\documentclass[aps,pra,superscriptaddress,twocolumn,floatfix,longbibliography]{revtex4-1}

\usepackage{minitoc}
\usepackage{tikz}
\usepackage{dcolumn}
\usepackage{graphicx,color}
\usepackage[utf8]{inputenc}
\usepackage{amsmath,amssymb,bm,amsfonts,dsfont,mathrsfs,amsthm}
\usepackage{braket}
\usepackage{txfonts,comment}
\usepackage[colorlinks=true,linkcolor=blue,citecolor=blue,urlcolor=blue]{hyperref}
\usepackage{soul}

\usepackage{booktabs}   
\usepackage{tabularx}

\begin{document}

\title{Stroboscopic Saturation of Multiparameter Quantum Limits in Distributed Quantum Sensing}

\author{Berihu Teklu}
\email{berihu.gebrehiwot@ku.ac.ae}
\affiliation{College of Computing and Mathematical Sciences, Department of Applied Mathematics and Sciences, Khalifa University of Science and Technology, 127788 Abu Dhabi, United Arab Emirates}
\affiliation{Center for Cyber-Physical Systems (C2PS), Khalifa University of Science and Technology, 127788, Abu Dhabi, United Arab Emirates}

\author{Victor Montenegro}
\email{victor.montenegro@ku.ac.ae}
\affiliation{College of Computing and Mathematical Sciences, Department of Applied Mathematics and Sciences, Khalifa University of Science and Technology, 127788 Abu Dhabi, United Arab Emirates}
\affiliation{Institute of Fundamental and Frontier Sciences, University of Electronic Science and Technology of China, Chengdu 611731, China}
\affiliation{Key Laboratory of Quantum Physics and Photonic Quantum Information, Ministry of Education, University of Electronic Science and Technology of China, Chengdu 611731, China}

\date{\today}

\begin{abstract}
High-precision sensors that exploit uniquely quantum phenomena have been shown to surpass the standard quantum limit of measurement precision. However, in the general scenario where multiple parameters are simultaneously encoded in a quantum probe, while surpassing the standard quantum limit is possible, its practical attainability is severely hindered. This difficulty arises due to the fundamental incompatibility among the optimal measurements required for estimating different parameters. A naturally multiparameter sensing scenario emerges when a network of quantum sensors is spatially distributed, with each individual sensor probing a distinct parameter of interest. The central goal in such a setting is twofold: first, to surpass the standard quantum limit in estimating global properties of the system---thereby achieving quantum-enhanced sensitivity for a given network size---and second, to explicitly identify the optimal measurement strategies necessary to practically attain this quantum advantage. Here, we analytically demonstrate quantum-enhanced sensitivity for a broad class of distributed quantum probes, including cases where the precision scales quadratically or quartically with the sensing resources. We construct the corresponding optimal measurement strategies that achieve the ultimate precision limits---namely, saturation of both the Holevo and quantum Cram\'{e}r-Rao bounds. We then apply our framework to two concrete scenarios: the simultaneous estimation of multiple gravitational accelerations (gravimetry) and coupling strengths across spatially separated locations. Feasibility analyses indicate that the proposed distributed quantum-enhanced sensing schemes are within reach of current experimental capabilities.
\end{abstract}

\maketitle

\section{Introduction}

Quantum sensing is a key pillar of quantum technologies, with rapidly growing theoretical development and experimental realizations~\cite{degen2017quantum, montenegro2025review, aslam2023quantum, ye2024essay, demille2024quantum}. At the heart of quantum sensing lies the exploitation of uniquely quantum effects to achieve sensing precision beyond the standard quantum limit, relative to a well-defined sensing resource~\cite{tan2021fisher, giovannetti2004quantum, giovannetti2006quantum, giovannetti2011advances}. Deep within the field, strong connections emerge with quantum information theory~\cite{braun2018quantum}, geometric approaches~\cite{braunstein1994statistical, sidhu2020geometric}, and quantum estimation theory~\cite{holevo2011probabilistic, vantrees2004detection, helstrom1967minimum, helstrom1976quantum, paris2009quantum}. The latter provides a natural framework to classify quantum sensing protocols into two broad classes: single-parameter estimation and multiparameter estimation. In the single-parameter case, the ultimate precision bounds are well established and known to be saturable with optimal measurement strategies~\cite{paris2009quantum}. In contrast, in the multiparameter setting, the symmetric logarithmic derivatives (SLDs) corresponding to different parameters generally do not commute~\cite{helstrom1974noncommuting, carollo2019quantumness}. This non-commutativity implies that the optimal measurements for different parameters are incompatible, preventing the existence of a single measurement that is simultaneously optimal for all parameters. Consequently, although the ultimate precision bounds---such as the Holevo or quantum Cram\'{e}r-Rao bounds---are formally established~\cite{liu2020quantum}, they are not guaranteed to be saturable in general. That is, except for special cases~\cite{pezze2017optimal}, one cannot construct a measurement basis that is both independent of the unknown parameters and capable of saturating the theoretical multiparameter lower bound on estimation precision.

Interestingly, the formalism of multiparameter quantum estimation naturally extends to scenarios involving spatially distributed networks of quantum sensors, giving rise to the framework of distributed quantum sensing~\cite{zhang2021distributed}. In this setting, the goal is to enable multiple sensors to collectively exploit shared entangled states in order to enhance the precision of estimating global properties of the system. Distributed quantum sensing has captured significant attention in recent years---developed extensively both theoretically~\cite{proctor2017networkedquantumsensing, hu2025optimalschemedistributedquantum, humphreys2013quantum, ge2018distributed, proctor2018multiparameter, zhuang2018distributed, agarwal2025saturation, eldredge2018optimal, rubio2020quantum, namkung2024optimal, zhang2017quantum, oh2020optimal} and experimentally~\cite{kim2024distributed, liu2021distributedquantum, guo2020distributed, kim2025distributed, xia2020demonstration, hong2021quantumenhanced, hong2025quantum, bate2025experimentaldistributedquantumsensing, polino2019femtosecond}. It has also been proposed for novel applications such as privacy-preserving quantum tasks~\cite{hassani2025privacy, bugalho2025privaterobuststates} and for probing quantum theory in curved spacetime~\cite{covey2025probing}. While the ultimate bounds on sensing precision in such distributed schemes are well established, distributed sensing inherits the fundamental challenges common to general multiparameter estimation problems: the typical absence of a universally optimal measurement capable of saturating the ultimate precision bound, and the inherent difficulty of demonstrating quantum-enhanced sensitivity for a fixed amount of sensing resources---further complicated by the spatial separation of the quantum probes, which introduces additional constraints on entanglement distribution and measurement implementation. This raises several critical questions: (i) Can one demonstrate the saturability of the ultimate lower bound for a broad class of quantum probes? (ii) Is it possible to construct an optimal, parameter-independent measurement basis that achieves this saturability? (iii) What are the fundamental limitations and necessary conditions for achieving quantum-enhanced sensitivity in networks of fixed size? (iv) Can these challenges be addressed using physically realizable quantum systems? 

In this work, we provide explicit answers to all four questions. (i) We analytically demonstrate quantum-enhanced sensitivity for a broad class of conditionally displaced quantum probes. We report cases where the precision scales quadratically or quartically with the sensing resources. These probes consist of a quantized electromagnetic field (here, mechanical oscillators prepared in thermal or coherent states) whose quadrature positions are conditionally displaced based on the state of an interacting subsystem. Remarkably, this class includes nonlinear quantum optomechanical systems~\cite{aspelmeyer2014cavity, qvarfort2018gravimetry}, solid-state spin systems~\cite{treutlein2014hybrid, rabl2009strong}, and hybrid dispersive tripartite architectures~\cite{restrepo2014single, restrepo2017fully}---making our analysis both general and experimentally relevant. The key mechanism enabling quantum-enhanced sensitivity is the disentanglement of the mechanical oscillators at full evolution cycles, which transfers all parameter-dependent information to the coupled subsystem. As a result, all global information is transferred to the subsystems, ensuring that optimal measurements can be performed exclusively on them---eliminating the need to measure the full composite system. (ii) Building on this, we show that the problem maps onto the estimation of relative phases~\cite{proctor2017networkedquantumsensing}. This mapping enables us to analytically construct an optimal, parameter-independent measurement basis via the Gram–Schmidt procedure~\cite{pezze2017optimal}. The validity of this construction is guaranteed by the weak incompatibility condition, which we prove rigorously~\cite{pezze2017optimal,liu2020quantum,carollo2019quantumness}. (iii) We further derive analytical conditions under which quantum-enhanced sensitivity can be achieved for a given network size, identifying how the precision enhancement competes with the available sensing resources---specifically, the initial number of probe excitations. (iv) Finally, having established the conditions for optimal and saturable quantum-enhanced sensing in this broad class of systems, we propose two concrete physical implementations: the simultaneous estimation of multiple gravitational accelerations (quantum gravimetry) and the estimation of coupling strengths across spatially separated locations. We also provide quantitative feasibility analyses demonstrating that these protocols are within reach of current experimental capabilities.

\section{Quantum Estimation Background}~\label{sec_QEB}

Consider a quantum state $\rho(\boldsymbol{\theta})$ that encodes $d$ unknown parameters $\boldsymbol{\theta}=(\theta_1,\dots,\theta_d)$. The fundamental lower bound on the precision of simultaneously estimating these parameters is given by the multiparameter quantum Cram\'{e}r-Rao inequality $\mathrm{Cov}[\check{\boldsymbol{\Theta}}] \geq [\mu F(\boldsymbol{\theta})]^{-1} \geq [\mu Q(\boldsymbol{\theta})]^{-1}$~\cite{liu2020quantum, albarelli2020perspective}, where $\mathrm{Cov}[\check{\boldsymbol{\Theta}}]_{ij}=\langle \check{\Theta}_i \check{\Theta}_j\rangle - \langle \check{\Theta}_i\rangle \langle \check{\Theta}_j\rangle$ are the entries of the covariance matrix of an unbiased estimator $\check{\boldsymbol{\Theta}}$ of $\boldsymbol{\theta}$, $\mu$ is the number of independent measurements, $F(\boldsymbol{\theta})$ is the classical Fisher information matrix (CFIM), and $Q(\boldsymbol{\theta})$ is the quantum Fisher information matrix (QFIM). Both Fisher matrices will be given explicitly later. Instead of working directly with matrix inequalities, in multiparameter quantum sensing it is often convenient to express the bounds in scalar form. To this end, one introduces a positive semidefinite weight matrix $\mathcal{W}$, leading to the scalar multiparameter Cram\'{e}r–Rao bounds $\mathrm{Tr}\left[\mathcal{W} \mathrm{Cov}[\check{\boldsymbol{\Theta}}] \right] \geq \frac{1}{\mu} \mathrm{Tr} \left[\mathcal{W} F(\boldsymbol{\theta})^{-1} \right] \geq \frac{1}{\mu} \mathrm{Tr} \left[\mathcal{W} Q(\boldsymbol{\theta})^{-1} \right]$. In this work, without loss of generality, we take $\mathcal{W}$ to be the identity matrix, which assigns uniform weight to all uncertainties in $\boldsymbol{\theta}$. Under this choice, the scalar bound becomes~\cite{liu2020quantum,albarelli2020perspective}
\begin{equation}
\mathrm{Tr}\left[\mathrm{Cov}[\check{\boldsymbol{\Theta}}] \right] \geq \frac{1}{\mu} \mathrm{Tr} \left[F(\boldsymbol{\theta})^{-1} \right] \geq \frac{1}{\mu} \mathrm{Tr} \left[Q(\boldsymbol{\theta})^{-1} \right],\label{eq_scalar_bound}
\end{equation}
where the CFIM has entries~\cite{liu2020quantum}
\begin{equation}
[F(\boldsymbol{\theta})]_{ij} = \int \frac{\partial_i p(x|\boldsymbol{\theta})\partial_j p(x|\boldsymbol{\theta})}{p(x|\boldsymbol{\theta})} dx, \hspace{1cm} \partial_i \equiv \partial/\partial \theta_i,
\end{equation}
with $p(x|\boldsymbol{\theta}) = \mathrm{Tr}[\rho(\boldsymbol{\theta}) \Pi_x]$ accounting for the conditional probability of obtaining outcome $x$ when measuring the state $\rho(\boldsymbol{\theta})$ with a positive operator-valued measure (POVM) $\{\Pi_x\}$~\cite{paris2009quantum}. This means that the CFIM determines the sensing precision for a given measurement strategy. On the other hand, the QFIM has entries~\cite{liu2020quantum, sidhu2020geometric}
\begin{equation}
[Q(\boldsymbol{\theta})]_{ij} = 2 \sum_{\substack{m,n \\ \varepsilon_m + \varepsilon_n\neq 0}}
\mathrm{Re} \left[\frac{\langle \varepsilon_m | \partial_i \rho(\boldsymbol{\theta}) | \varepsilon_n \rangle \langle \varepsilon_n | \partial_j \rho(\boldsymbol{\theta}) | \varepsilon_m \rangle }{\varepsilon_m + \varepsilon_n}\right],\label{eq_QFI_matrix_form}
\end{equation}
where $\rho(\boldsymbol{\theta}) = \sum_i \varepsilon_i(\boldsymbol{\theta})|\varepsilon_i(\boldsymbol{\theta})\rangle \langle \varepsilon_i(\boldsymbol{\theta})|$ is expressed in spectral decomposition with eigenvalues $\varepsilon_i(\boldsymbol{\theta})\equiv\varepsilon_i$ and eigenvectors $|\varepsilon_i(\boldsymbol{\theta})\rangle\equiv|\varepsilon_i\rangle$. As an important case, for pure states $\rho(\boldsymbol{\theta})=|\psi(\boldsymbol{\theta})\rangle\langle\psi(\boldsymbol{\theta})|$, Eq.~\eqref{eq_QFI_matrix_form} simplifies to~\cite{liu2020quantum, meyer2021fisher}:
\begin{equation}
[Q(\boldsymbol{\theta})]_{ij}{=}4 \mathrm{Re}\Big(\langle \partial_i \psi(\boldsymbol{\theta}) | \partial_j \psi(\boldsymbol{\theta}) \rangle{-}\langle \partial_i \psi(\boldsymbol{\theta}) | \psi(\boldsymbol{\theta}) \rangle \langle \psi(\boldsymbol{\theta}) | \partial_j \psi(\boldsymbol{\theta}) \rangle \Big).\label{eq_QFI_multi_pure}
\end{equation}

Ubiquitous to any quantum sensing protocol are two essential goals: (i) Demonstrate that $\mathrm{Tr}[\mathrm{Cov}[\check{\boldsymbol{\Theta}}]] \propto N_\mathrm{exc}^{-r}$, with $r > 1$. This scaling signifies quantum-enhanced sensitivity~\cite{giovannetti2004quantum, giovannetti2006quantum, giovannetti2011advances, montenegro2025review}, that is the total estimation uncertainty, quantified by the trace of the covariance matrix of the unbiased estimators $\check{\boldsymbol{\Theta}}$, decreases super-linearly with the sensing resource---here the number of initial excitations $N_\mathrm{exc}$. This means that by increasing the number of sensing resources, the estimation uncertainty decreases faster than the standard quantum limit $r = 1$~\cite{degen2017quantum, montenegro2025review}. This behavior is a hallmark of quantum metrology, where quantum features are used to surpass the standard quantum limit $r = 1$ and achieve quantum-enhanced scaling $r > 1$. In the context of distributed sensing, quantum-enhanced sensitivity should be demonstrated for a fixed number of sensors. (ii) Identifying the optimal measurement strategy that extracts the maximum precision allowed by the quantum probe, i.e. $\mathrm{Tr}[F(\boldsymbol{\theta})^{-1}] = \mathrm{Tr}[Q(\boldsymbol{\theta})^{-1}]$ in the multiparameter case, and $F(\theta) = Q(\theta)$ in the single-parameter case. The existence of a measurement that attains the quantum lower bound is what defines saturability. It is worth emphasizing that, unlike the single-parameter case where $\boldsymbol{\theta} = \theta$, in the multiparameter scenario the quantum Cram\'{e}r-Rao bound is generally not saturable~\cite{albarelli2020perspective}. This limitation arises because the optimal measurements corresponding to different parameters may be incompatible~\cite{carollo2019quantumness}. As a result, one cannot, in general, simultaneously achieve the precision limits for all parameters at once. This fundamental issue, stemming from non-commuting SLDs, places intrinsic constraints on the simultaneous estimation of multiple parameters~\cite{carollo2019quantumness, ragy2016compatibility, albarelli2019evaluating, szczykulska2016multiparameter, sidhu2020geometric, pezze2017optimal}.

\section{The Model}

\begin{figure}
    \centering
    \includegraphics[width=\linewidth]{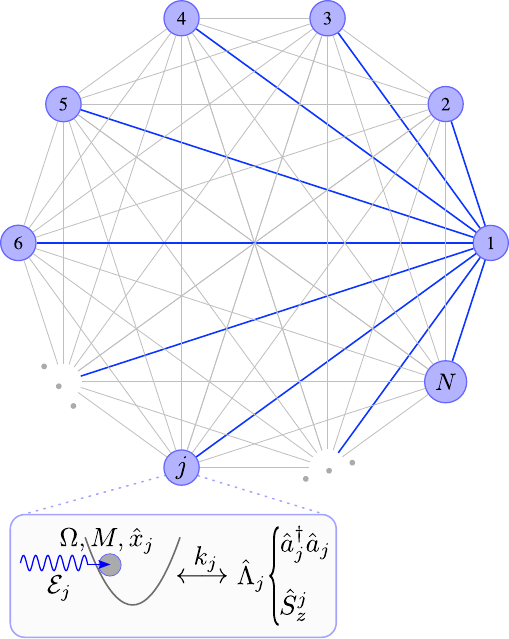}
    \caption{\textbf{Distributed quantum sensing model.} We consider a network of $N$ nodes. At each node $j$, the system consists of a mechanical oscillator of mass $M$ and frequency $\Omega$, driven at frequency $\mathcal{E}_j$. This oscillator interacts with a general quantum system $\hat{\Lambda}_j$. We propose two physical implementations in which $\hat{\Lambda}_j$ can be either the number operator $\hat{a}_j^\dagger \hat{a}_j$ or the higher-spin collective operator $\hat{S}_z^j$; see the main text for details. The sensing task is to simultaneously estimate the $N$ unknown parameters $k_j$ or $\mathcal{E}_j$ across the network.}
    \label{fig_schematics_model}
\end{figure}

We consider a network of $N$ quantum nodes. Each node, labeled by index $j$, is governed by the Hamiltonian
\begin{equation}
\hat{H}_j = \frac{1}{2M} \hat{p}_j^2 + \frac{M\Omega^2}{2} \hat{x}_j^2 - \hbar\frac{\tilde{k}_j}{x_0} \hat{\Lambda}_j \hat{x}_j + \hbar\frac{\tilde{\mathcal{E}}_j}{x_0} \hat{x}_j, \label{eq_generalHi}
\end{equation}
which describes a driven mechanical harmonic oscillator coupled to a general quantum system, see Fig.~\ref{fig_schematics_model}.
Specifically, the mechanical degree of freedom, with mass $M$, frequency $\Omega$, and driven amplitude $\tilde{\mathcal{E}}_j$, is described by canonical position and momentum operators $\hat{x}_j$ and $\hat{p}_j$, satisfying the commutation relation $[\hat{x}_j, \hat{p}_{j'}] = i\hbar\delta_{j,j'}$. The quantity $x_0 = \sqrt{\hbar/(2M\Omega)}$ denotes the zero-point fluctuation amplitude. The term proportional to $\tilde{k}_j$ captures a dispersive interaction between the mechanical position $\hat{x}_j$ and a general operator $\hat{\Lambda}_j$, obeying:
\begin{equation}
\hat{\Lambda}_j |\lambda_n\rangle_j{=}\lambda_{n,j}|\lambda_n\rangle_j; \hspace{0.5cm} _{j}\langle \lambda_{n} | \lambda_{n'} \rangle_{j'}{=}\delta_{n,n'}\delta_{j,j'}; \hspace{0.5cm} [\hat{\Lambda}_j{,}\hat{\Lambda}_{j'}]{=}0.\label{eq_requirements_Delta}
\end{equation}
where the index $n$ labels both the $n$th eigenvalue and its corresponding $n$th eigenvector $|\lambda_n\rangle_j$, which are associated with node $j$. Hence, the index $n$ runs over the spectrum of $\hat{\Lambda}_j$, whereas $1\leq j \leq N$. By introducing bosonic annihilation and creation operators defined as $\hat{b}_j = \tfrac{1}{2} \left( \tfrac{\hat{x}_j}{x_0} + i \frac{\hat{p}_j}{p_0} \right), \hat{b}_j^\dagger = \frac{1}{2} \left( \tfrac{\hat{x}_j}{x_0} - i \tfrac{\hat{p}_j}{p_0} \right)$, where $p_0 = \sqrt{\hbar M \Omega/2}$ and $[\hat{b}_j, \hat{b}_{j'}^\dagger] = \delta_{j,j'}$, the Hamiltonian in Eq.~\eqref{eq_generalHi} becomes
\begin{equation}
\frac{\hat{H}_j}{\Omega\hbar} = \hat{b}_j^\dagger \hat{b}_j - (k_j \hat{\Lambda}_j - \mathcal{E}_j)(\hat{b}_j^\dagger + \hat{b}_j),
\end{equation}
where we have defined the scaled quantities $k_j \equiv \frac{\tilde{k}_j}{\Omega}$ and $\mathcal{E}_j \equiv \frac{\tilde{\mathcal{E}}_j}{\Omega}$. The corresponding unitary time-evolution operator for the $j$th node can then be expressed in closed form as~\cite{bose1997preparation, qvarfort2018gravimetry, qvarfort2025solving}
\begin{equation}
\hat{u}_j(\tau) = e^{i (k_j \hat{\Lambda}_j - \mathcal{E}_j)^2 (\tau - \sin\tau)}e^{(k_j \hat{\Lambda}_j - \mathcal{E}_j)(\eta(\tau)\hat{b}_j^\dagger - \eta^*(\tau)\hat{b}_j)} e^{-i \hat{b}_j^\dagger \hat{b}_j \tau}. \label{eq_unitary_operator_tau}
\end{equation}
In the above expression, $\eta(\tau) \equiv 1 - e^{-i\tau}$, and $\tau \equiv \Omega t$ is the dimensionless time. Since $[\hat{H}_j, \hat{H}_{j'}] = 0$ for all $1 \leq (j,j') \leq N$, the total unitary operator for the network is given by the product
\begin{equation}
\hat{U}(\tau) = \prod_{j=1}^N \hat{u}_j(\tau). \label{eq_U_tau}
\end{equation}

In this work, we focus on a specific class of initial states that enable quantum-enhanced sensitivity across the entire network~\cite{maleki2022distributed}. Specifically, we consider states of the form:
\begin{equation}
    |\psi(0)\rangle = |\text{W}_N\rangle (|\alpha\rangle^{\otimes N}),
\end{equation}
where $|\alpha\rangle^{\otimes N}$ is a collection of $N$ coherent states with amplitudes $\alpha$, and $|W_N\rangle = \frac{1}{\sqrt{N}} \sum_{j=1}^N |j\rangle$ is the $W$-state~\cite{dur2000three, cao2003entanglement}. Each basis state $|j\rangle$ is defined as
\begin{equation}
|j\rangle = |\lambda'\rangle_1 |\lambda'\rangle_2 \ldots |\lambda\rangle_j \ldots |\lambda'\rangle_{N-1} |\lambda'\rangle_N,
\end{equation}
representing the configuration in which the eigenvalue $\lambda$ appears at site $j$, while all other sites are in the eigenstate $|\lambda'\rangle$. 

With the above definitions, it is straightforward to obtain the full wavefunction of the entire network at arbitrary time $\tau$, which takes the form:
\begin{equation}
    |\psi(\tau)\rangle = \frac{1}{\sqrt{N}}\left(|1\rangle|\alpha_1(\tau)\rangle+\sum_{j=2}^N e^{ \xi_j(\tau)}|j\rangle|\alpha_j(\tau)\rangle\right),\label{eq_wavefunction_arbitrary_time}
\end{equation}
where $\xi_j(\tau)\in\mathbb{C}$ is
\begin{multline}
    \xi_j(\tau)=\frac{1}{2}e^{-i\tau}(\lambda-\lambda')[\alpha (1-e^{2 i \tau})(k_1-k_j)-2[2\mathcal{E}_1 k_1-2\mathcal{E}_j k_j\\
 -(k_1-k_j)(k_1+k_j)(\lambda+\lambda')](\tau-\sin\tau)(\sin\tau-i\cos\tau)],
\end{multline}
and $|\alpha_j(\tau)\rangle=|\alpha e^{-i\tau}+(k_1\lambda'-\mathcal{E}_1)\eta(\tau)\rangle\cdots|\alpha e^{-i\tau}+(k_j\lambda-\mathcal{E}_j)\eta(\tau)\rangle\cdots|\alpha e^{-i\tau}+(k_N\lambda'-\mathcal{E}_N)\eta(\tau)\rangle$ denotes the $j$th-position of the eigenvalue $\lambda$ within a set of $N$ coherent states. See the detailed derivation in Sec.~\ref{sec_sm_analytical} of the Appendix.

\section{Dynamics of Quantum Entanglement}

To further characterize the quantum dynamics described by Eq.~\eqref{eq_wavefunction_arbitrary_time}, we evaluate the bipartite entanglement between the set of general systems $\{\hat{\Lambda}_j\}$ and the collection of mechanical oscillators. We quantify this entanglement using the linear entropy, see Sec.~\ref{sec_sm_entanglement} of the Appendix for its derivation:
\begin{equation}
    \mathcal{S}_L=\frac{2}{N^2}\left[\frac{N(N-1)}{2}{-}\sum_{1 \leq i < j \leq N} e^{2(k_i^2 + k_j^2)(\lambda-\lambda')^2 [\cos\tau -1]}\right].
\end{equation}

Three key insights follow from the above results. First, the linear entropy $\mathcal{S}_L$ is independent of both the initial mechanical coherent amplitude $\alpha$ and the mechanical driving amplitude $\mathcal{E}$. Second, at $\tau = \pi$, the system reaches a point of maximal correlation between its components. As a result, extracting the full information requires an entangled measurement across the entire composite system. Third, at $\tau = 2\pi$, these correlations vanish, and all parameter-dependent information becomes localized in the subsystems $\hat{\Lambda}_j$. Therefore, measurements performed solely on the subsystems are sufficient to recover all relevant information. Indeed, at the evolution time $\tau = 2\pi$ (and at integer multiples thereof), the unitary time-evolution operator reduces to:
\begin{equation}
\hat{u}_j(\tau = 2\pi) = e^{2\pi i (k_j \hat{\Lambda}_j - \mathcal{E}_j)^2} e^{-2\pi i \hat{b}_j^\dagger \hat{b}_j},\label{eq_unitary_operator}
\end{equation}
from which the full wavefunction of the entire network at $\tau = 2\pi$ takes the form
\begin{equation}
    |\psi(\tau=2\pi)\rangle = \frac{1}{\sqrt{N}}\left(|1\rangle + \sum_{j=2}^Ne^{i \beta_j \Phi_j}|j\rangle\right)(|\alpha\rangle^{\otimes N}).\label{eq_quantum_probe}
\end{equation}

In the above, $\beta_j$ denotes a multiplicative prefactor, while $\Phi_j$ represents the relative phase encoding the unknown parameters of interest. Note that the phases $\Phi_j$ are defined relative to a reference frequency at node $j=1$, such that they correspond to two-point phase differences, namely: $\Phi_j = \mathcal{E}_1 - \mathcal{E}_j$ or equivalently $\Phi_j = k_1 - k_j$. These relative phases are illustrated by the solid blue lines in Fig.~\ref{fig_schematics_model}. To provide $\beta_j$ and $\Phi_j$ explicit expressions, we distinguish between two distinct scenarios. 

In Case 1, all mechanical driving amplitudes are unknown and different, i.e. $\mathcal{E}_j \neq \mathcal{E}_{j'} \neq 0$, while all couplings strength are equal, $k_j = k$. In this situation, the relative phases in Eq.~\eqref{eq_quantum_probe} simplify to $\beta_j = 4\pi k(\lambda - \lambda')$, which is the same for all nodes (i.e. node-independent). The corresponding phase factors are $\Phi_j = \mathcal{E}_j^- = \mathcal{E}_1 - \mathcal{E}_j$. Thus, only the differences between $\mathcal{E}_1$ and each $\mathcal{E}_j$ appear explicitly. In Case 1, the estimation task is precisely to determine these relative unknown driving amplitudes $\mathcal{E}_j^-$. 

In Case 2, all mechanical drivings are set to zero, $\mathcal{E}_j = 0$, while the coupling strengths are different and nonzero, $k_j \neq k_{j'} \neq 0$. In this case, the relative phases in Eq.~\eqref{eq_quantum_probe} reduce to $\beta_j = 2\pi (k_1 + k_j)(\lambda' - \lambda)(\lambda + \lambda')$. The corresponding phase factors are $\Phi_j = k_j^- = k_1 - k_j$. Therefore, only the differences between $k_1$ and each $k_j$ appear explicitly. The estimation task in this Case 2 is to determine these relative unknown coupling strengths $k_j^-$. A concise overview of these two scenarios is provided in Table~\ref{tab_phases_cases}.

\begin{table}[t]
\centering
\renewcommand{\arraystretch}{1.6}
\begin{tabularx}{\columnwidth}{|c|c|c|X|}
\hline
Case & $\Phi_j$ & $\beta_j$ & Conditions \\
\hline
1 &
$\mathcal{E}_j^- = \mathcal{E}_1 - \mathcal{E}_j$ &
$4\pi k(\lambda - \lambda')$ &
$k_j = k$ and $\mathcal{E}_j \neq \mathcal{E}_{j'} \neq 0$.\\
\hline
2 &
$k_j^- = k_1 - k_j$ &
$2\pi k_j^+(\lambda' - \lambda)(\lambda + \lambda')$ &
$k_j \neq k_{j'} \neq 0$ and $\mathcal{E}_j = 0$.\\
\hline
\end{tabularx}\caption{\textbf{Distributed quantum sensing cases.} Case 1: The estimation task is to determine $\Phi_j = \mathcal{E}_j^- = \mathcal{E}_1 - \mathcal{E}_j$, under the conditions that all coupling strengths are identical $k_j = k$ for all $1 \leq j \leq N$, and all mechanical drivings are different and nonzero $\mathcal{E}_j \neq \mathcal{E}_{j'} \neq 0$ for all $1 \leq (j,j') \leq N$. Case 2: The estimation task is to determine $\Phi_j = k_j^- = k_1 - k_j$, under the conditions that all mechanical drivings vanish $\mathcal{E}_j = 0$ for all $1 \leq j \leq N$, while the coupling strengths are different and nonzero $k_j \neq k_{j'} \neq 0$ for all $1 \leq (j,j') \leq N$. We define $k_j^+ = k_1 + k_j$.}
\label{tab_phases_cases}
\end{table}

As a final observation, note that all mechanical oscillators return to their initial states $|\alpha\rangle$ at multiples of $\tau = 2\pi$ [see Eq.~\eqref{eq_quantum_probe}], while the subsystems associated with the operators $\{\hat{\Lambda}_j\}$ acquire distinct relative phases at each site $j$. This behavior suggests that the unknown relative parameters can be estimated solely by probing the system locally, namely measurements on $\{\hat{\Lambda}_j\}$. Indeed, the mechanical oscillators dynamically transfer the full information content to the general systems $\{\hat{\Lambda}_j\}$ at these specific times. At other times, e.g. at multiples of $\tau = \pi$, the system is maximally correlated, and hence a joint measurement across all modes and parties is required to extract the ultimate sensing capability of the network. Note that this disentangling behavior also holds when the mechanical oscillator is initially in a thermal state, thus avoiding initial ground state preparation of the mechanical object~\cite{oconnell2010quantum}. Indeed, a thermal state can be written in the Glauber–Sudarshan $P$-representation as $\rho_{\text{th}}{=}\int d^2\alpha P_{\text{th}}(\alpha) |\alpha\rangle\langle\alpha|$~\cite{orszag2024qoptics}. When this thermal state evolves under the action of $\hat{U}(\tau)$ of Eq.~\eqref{eq_U_tau}, the result is, in general, a conditionally displaced state of the form $\rho_{\text{th}}{\to}\int d^2\alpha P_{\text{th}}(\alpha)|\alpha_j(\tau)\rangle\langle\alpha_{j'}(\tau)|$, with displacements as in Eq.~\eqref{eq_wavefunction_arbitrary_time}. These displacements vanish at $\tau{=}2\pi$ since $\eta(\tau=2\pi){=}0$. As a result, the mechanical party returns to its initial state $\rho_\mathrm{th}$ at each full oscillation period. This mechanical disentanglement, followed by the transfer of information to all systems ${\hat{\Lambda}_j}$, constitutes the key dynamical feature that enables quantum-enhanced sensing across all nodes.

\section{Distributed Quantum-Enhanced Estimation}

With the quantum state of Eq.~\eqref{eq_quantum_probe} at hand, we now address the task of simultaneous multiparameter estimation across a network of size $N$. To this end, using Eq.~\eqref{eq_QFI_multi_pure}, we compute the trace of the inverse of the QFIM as, see details in Sec.~\ref{sec_sm_qfim} of the Appendix:
\begin{equation}
\mathrm{Tr}[Q^{-1}] = \frac{N}{2}\sum_{j=2}^N\frac{1}{\beta_j^2}.\label{eq_tr_inv_QFIM}
\end{equation}
The above expression demonstrates the achievement of quantum-enhanced sensitivity, as $\mathrm{Tr}[Q^{-1}]$ decreases quadratically with $\beta_j$, which itself scales linearly with the initial excitation amplitude $\beta_j \propto (\lambda - \lambda')$ in Case 1. Remarkably, in Case 2, the dependence becomes $\beta_j \propto (\lambda - \lambda')(\lambda + \lambda')$, leading to a quartic scaling of sensitivity with the initial excitations. To the best of our knowledge, this scaling has not been previously reported in the context of distributed quantum sensing schemes. This indicates that, in Case 2, the ultimate estimation precision improves at an even faster rate with increasing resources. The inverse-square and quartic scalings of the sensing uncertainty with the number of initial excitations (the sensing resource) provides a clear signature of quantum-enhanced sensing~\cite{giovannetti2004quantum, giovannetti2006quantum, giovannetti2011advances}. This is particularly evident in Case 1 of Table~\ref{tab_phases_cases}, where
\begin{equation}
\mathrm{Tr}[Q^{-1}] = \frac{N(N-1)}{32\pi^2 k^2(\lambda - \lambda')^2}, \hspace{1cm} \text{Case 1},
\end{equation}
Note that the lower bound is independent of the estimated parameters $\mathcal{E}_j^-$. Consequently, by imposing the condition $\mathrm{Tr}[Q^{-1}] < 1$, one obtains the following requirement for the considered sensing resources $(\lambda - \lambda') > \tfrac{N(N-1)}{32 \pi^2 k^2}$.

Remarkably, in Case 2 of Table~\ref{tab_phases_cases} one finds $\beta_j \propto (\lambda - \lambda')(\lambda + \lambda')$. Hence, in principle, a quartic scaling of $\mathrm{Tr}[Q^{-1}]$ of Eq.~\eqref{eq_tr_inv_QFIM} can be achieved when $\lambda' = 0$, that is:
\begin{equation}
    \mathrm{Tr}[Q^{-1}] = \frac{N}{8\pi^2\lambda^4}\sum_{j=2}^N \frac{1}{k_j^{+2}}, \hspace{1cm} \text{Case 2},
\end{equation}
In this case, the lower bound depends implicitly on the parameters $k_j^+ = k_1 + k_j$, which are, in principle, unknown. As a result, these parameters act as nuisances in the final estimation~\cite{albarelli2020perspective}. In the following, we propose and discuss the fundamental multiparameter precision limits of two physical implementations for realizing distributed quantum-enhanced sensing.

\section{Distributed Quantum Gravimetry: Case 1 Proposal}

To illustrate the fundamental distributed precision limits relevant to Case 1, see Table~\ref{tab_phases_cases}, we consider a network of $N$ nonlinear quantum optomechanical systems designed to estimate relative differences in gravitational acceleration encoded at spatially separated sites. This information is captured by the relative phases $\Phi_j = \mathcal{E}_1 - \mathcal{E}_j$, where $\mathcal{E}_j$ will be shown to explicitly depend on the local gravitational acceleration at site $j$. The aim therefore is to estimate these differences across the whole network at arbitrarily distances, thereby enabling distributed quantum-enhanced gravimetry. It is worth emphasizing that the estimation of a single gravitational acceleration parameter using nonlinear optomechanical interactions---known as gravimetry---has been recently explored~\cite{qvarfort2018gravimetry, armata2017quantum}. Optomechanical systems, which arise from the radiation-pressure interaction between a quantized electromagnetic field (hereafter referred to as the light field) and a mechanical oscillator, have been extensively studied both theoretically and experimentally~\cite{aspelmeyer2014cavity}, with applications ranging from quantum information tasks~\cite{montenegro2019enabling, neto2016quantum} to high-precision sensing~\cite{qvarfort2018gravimetry}.

Following our previous definitions, we describe a quantum optomechanical gravimetry network with $N$ nodes via the Hamiltonian at site $j$ as:
\begin{equation}
\hat{H}_j = \frac{\hat{p}_j^2}{2M} + \frac{M \Omega^2}{2} \hat{x}_j^2 - \hbar \frac{\tilde{k}}{x_0} \hat{a}^\dagger_j \hat{a}_j \hat{x}_j + \underbrace{M g_j \cos(\nu_j)\hat{x}_j}_{\Delta U_j^g}, \label{eq_nonlinear_probe}
\end{equation}
where $\hat{a}_j$ ($\hat{a}_j^\dagger$) denotes the annihilation (creation) operator of the light field, $\Delta U_j^g$ is the gravitational potential energy, $\nu_j$ is the rotation angle with respect to the free-fall direction (without loss of generality we take $\nu_j=0$, that is a vertical oriented gravimeter), and $g_j$ is the unknown gravitational acceleration. By comparing Eq.~\eqref{eq_nonlinear_probe} with Eq.~\eqref{eq_generalHi}, one directly identifies
\begin{equation}
\hat{\Lambda}_j = \hat{a}^\dagger_j \hat{a}_j,    
\end{equation}
which satisfies all the conditions specified in Eq.~\eqref{eq_requirements_Delta}. We consider the initial state
\begin{equation}
|\psi(0)\rangle = |\text{W}_N\rangle(|\alpha\rangle^{\otimes N}),
\end{equation}
where $|\text{W}_N\rangle = \tfrac{1}{\sqrt{N}}\big(|N_\mathrm{exc},0,\ldots,0\rangle + \cdots + |0,\ldots,0,N_\mathrm{exc}\rangle\big)$ is the $N$-partite W state with $N_\mathrm{exc}$ excitations. In this case, it is straightforward to identify $\lambda = N_\mathrm{exc}$ and $\lambda' = 0$. Since $\mathcal{E}_j=\tfrac{x_0Mg_j}{\Omega\hbar}$, then a simple use of chain rule yields the following QFIM in the International System of Units (SI)
\begin{equation}
\mathrm{Tr}[Q^{-1}] = \frac{N(N-1)}{32\pi^2k^2N_\mathrm{exc}^2}\left(\frac{\Omega\hbar}{x_0M}\right)^2.\label{eq_invQ_case1}
\end{equation}
Note that $k = \tilde{k}/\Omega$ is dimensionless, and $\big[\tfrac{\Omega \hbar}{x_0 M}\big] = \tfrac{m}{s^2}$. Hence, the units of $\mathrm{Tr}[Q^{-1}]$ are consistent with the covariance of the estimator. The final estimation results is:
\begin{equation}
    \mathrm{Tr}[\mathrm{Cov}[\boldsymbol{g}^-]]=\sum_{j=1}^N\mathrm{Var}[g_j^-]\geq \mu^{-1}\frac{N(N-1)}{32\pi^2k^2N_\mathrm{exc}^2}\left(\frac{\Omega\hbar}{x_0M}\right)^2.\label{eq_N_quantuoptomechanics}
\end{equation}

Note that in Eq.~\eqref{eq_N_quantuoptomechanics} the parameters satisfy $N \geq 2$ and $N_\mathrm{exc} \geq 1$, while $\Omega$, $M$, and $k$ are device dependent. For comparison, we consider three optomechanical architectures reported in Ref.~\cite{qvarfort2018gravimetry}: (i) a Fabry-Perot configuration with $\Omega = 10^{3}\text{Hz}$, $M = 10^{-6}\text{kg}$, and $k = 2.3$~\cite{bose1997preparation, pontin2018levitated}; (ii) a levitated micro-object confined in an ion trap coupled to an optical cavity with $\Omega = 10^{2}\text{Hz}$, $M = 10^{-14}\text{kg}$, and $k = 1963$~\cite{fonseca2016nonlinear, millen2015cavity}; and (iii) cold atoms in a cavity with $\Omega = 10^{2}\text{Hz}$, $M = 10^{-25}\text{kg}$, and $k = 2.3 \times 10^{6}$~\cite{ferdinand2008cavity}. 

\begin{figure}
    \centering
    \includegraphics[width=\linewidth]{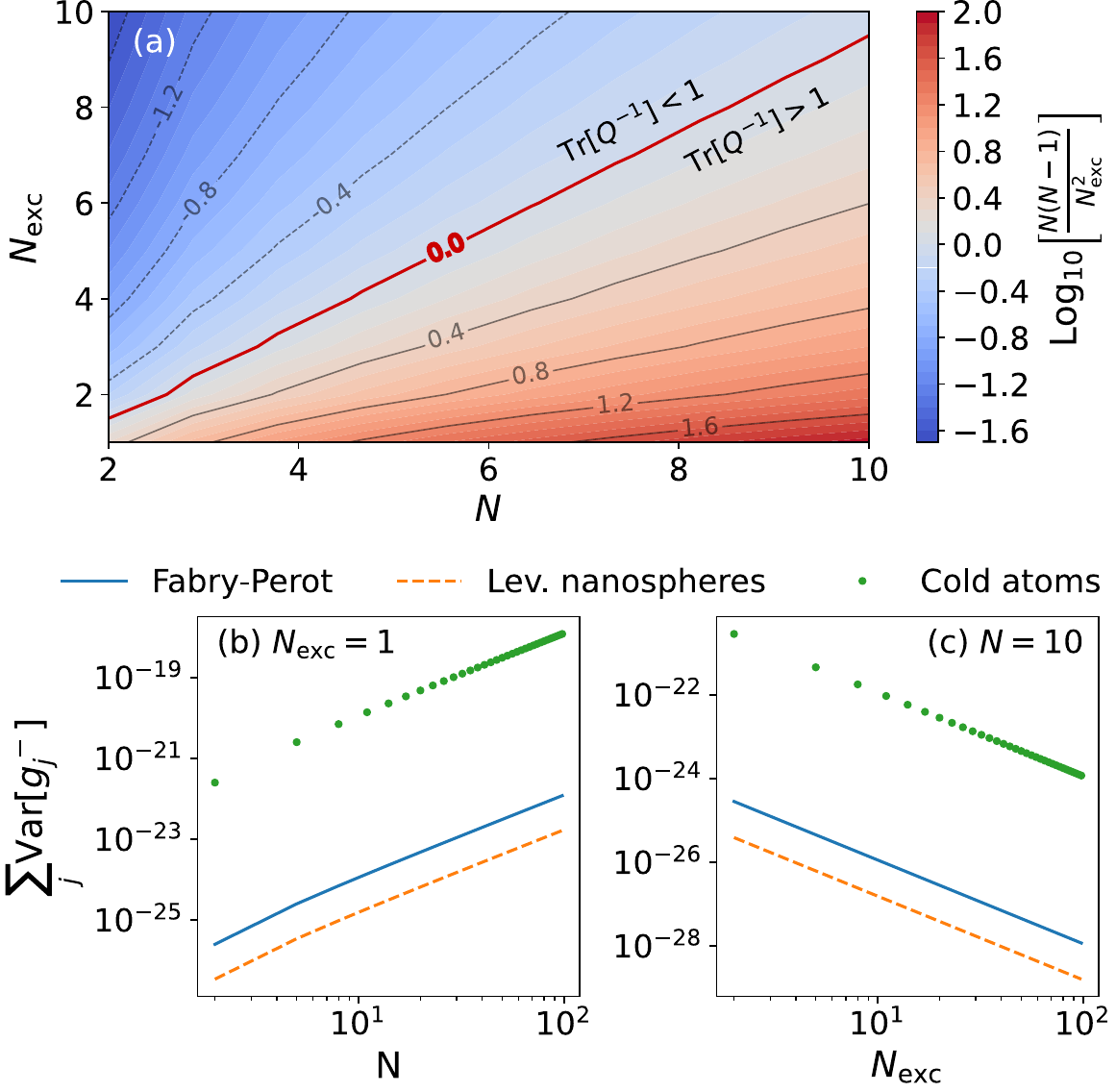}
    \caption{\textbf{Distributed quantum-enhanced sensing for Case 1.} (a) $\tfrac{N(N-1)}{N_\mathrm{exc}^2}$ as functions of network size $N$ and number of initial excitations $N_\mathrm{exc}$. (b) Sensitivity as a function of network size $N$ for a single excitation $N_\mathrm{exc} = 1$. (c) Sensitivity as a function of the number of excitations $N_\mathrm{exc}$ for a fixed network size $N = 10$. We consider $\mu=10^4$.}
    \label{fig_multiparameter_case1}
\end{figure}

To further illustrate the distributed quantum-enhanced sensitivity in Case 1, in Fig.~\ref{fig_multiparameter_case1}(a) we plot the quantity $\frac{N(N-1)}{N_\mathrm{exc}^2}$ from Eq.~\eqref{eq_invQ_case1} as functions of the network size $N$ and the number of excitations $N_\mathrm{exc}$. As the figure illustrates, the inverse-square dependence on $N_\mathrm{exc}$ competes directly with the $N$ growth of the network. Consequently, achieving lower sensing uncertainties imposes choosing $N_\mathrm{exc}$ at least on the order of $N$, that is $N_\mathrm{exc} \sim N$. Note that, for a given network size $N$, one indeed achieves the quadratic scaling of the QFIM with respect to our sensing resource, that is the initial probe excitations $N_\mathrm{exc}$.

In Fig.~\ref{fig_multiparameter_case1}(b), we show the distributed sensitivity as a function of $N$ for a single excitation $N_\mathrm{exc} = 1$. As seen in the figure, the sensitivity across the network increases quadratically with its size $N$ for a fixed $N_\mathrm{exc}$, with the levitated optomechanical system exhibiting the lowest overall uncertainties. This is readily evident from Eq.~\eqref{eq_N_quantuoptomechanics}. Moreover, the sensitivity (sum of variances) ranges from $\sim 10^{-27} \mathrm{m^2/s^4}$ for the levitated micro-sphere system to $\sim 10^{-18} \mathrm{m^2/s^4}$ for the cold atoms platform. For comparison with the single-parameter case, we define the root-mean-square total estimation error as $\Delta_{\mathrm{RMS}} \equiv \sqrt{\mathrm{Tr}[\mathrm{Cov}[\check{\boldsymbol\Theta}]]}$, which results in $10^{-15}\mathrm{m/s^2} \lesssim \Delta^{\text{Case 1}}_{\mathrm{RMS}} \lesssim 10^{-13}\mathrm{m/s^2}$ for the levitated optomechanical system, comparable to the minimal theoretical prediction for the standard deviation of $\Delta g \sim 10^{-15} \mathrm{m/s^2}$~\cite{qvarfort2018gravimetry}.

In Fig.~\ref{fig_multiparameter_case1}(c), we show the distributed quantum-enhanced sensitivity as a function of the number of excitations $N_\mathrm{exc}$ for a fixed network size $N = 10$. The figure illustrates that the hierarchy of optomechanical systems remains the same, with the overall sensitivity decreasing quadratically with $N_\mathrm{exc}$.

\subsection{Saturability of the Ultimate Quantum Precision Bound}

As discussed in Sec.~\ref{sec_QEB}, the saturation of the quantum Cram\'{e}r-Rao bound in multiparameter estimation is generally obstructed by the non-commutativity of the SLDs~\cite{carollo2019quantumness, liu2020quantum}. Hence, even if a multiparameter sensing scheme exhibits quantum-enhanced sensitivity, the absence of a measurement capable of saturating the bound renders the scheme fundamentally impractical. Remarkably, we show that our scheme saturates the quantum-enhanced precision bound when an appropriate projective measurement is applied. This demonstrates that the ultimate sensing limit can, in fact, be achieved through optimal measurements. 

The key observation is that our pure state probe satisfies the weak-commutativity condition for SLD-compatibility, namely $\langle \psi | [\mathcal{L}_k, \mathcal{L}_{k'}] | \psi \rangle = 0$ for all $k, k'$; see Sec.~\ref{sec_sm_weak} of the Appendix for a detailed proof. Under this condition, the SLD quantum Cram\'{e}r-Rao bound coincides with the attainable Holevo bound and is locally saturable~\cite{matsumoto2002approach, ragy2016compatibility}. For pure states, this weak-commutativity condition is both necessary and sufficient for local attainability of the SLD quantum Cram\'{e}r-Rao bound~\cite{matsumoto2002approach, ragy2016compatibility}. Building on this, Pezz\`{e} et al.~\cite{pezze2017optimal} derived necessary and sufficient conditions for projective measurements to saturate the QFIM in pure state models and provided an explicit Gram-Schmidt construction of the optimal projectors from the subspace spanned by the SLDs. We adopt that construction directly.

To illustrate the optimal measurement construction, consider the first non‑trivial network $N=3$ (the generalization to larger $N$ is direct) with projectors $\{\Pi_0,\Pi_1,\Pi_2\}$ onto the orthonormal kets with controllable known phases $\{{\vartheta_2,\vartheta_3}\}$
\begin{eqnarray}
\nonumber |\Sigma\rangle&=&\tfrac{1}{\sqrt{3}}\big(|1\rangle+e^{i\beta_2 \vartheta_2}|2\rangle+e^{i\beta_3 \vartheta_3}|3\rangle\big),\\
\nonumber |u_1\rangle&=&\tfrac{1}{\sqrt{2}}\big(-|1\rangle+e^{i\beta_2 \vartheta_2}|2\rangle\big),\\
|u_2\rangle&=&\tfrac{1}{\sqrt{6}}\big(-|1\rangle-e^{i\beta_2 \vartheta_2}|2\rangle+2e^{i\beta_3 \vartheta_3}|3\rangle\big), \label{eq_basis}
\end{eqnarray}
with $\Pi_0=|\Sigma\rangle\langle\Sigma|$, $\Pi_1=|u_1\rangle\langle u_1|$, and $\Pi_2=|u_2\rangle\langle u_2|$. For this measurement basis one computes the CFIM $\mathcal{F}(\vartheta_2,\vartheta_3)$ as a function of the controllable reference phases $\{\vartheta_2,\vartheta_3\}$. When the references are tuned to the true values $\{\Phi_2,\Phi_3\}$ (e.g., via adaptive procedures that first coarsely estimates the phases and then measures in the locally optimal basis~\cite{pezze2018quantummetrology, pezze2017optimal, mukhopadhyay2025current, montenegro2021global}), the CFIM equals the QFIM:
\begin{equation}
\lim_{{\vartheta_2,\vartheta_3}\to{\Phi_2,\Phi_3}}
\mathcal{F}(\vartheta_2,\vartheta_3)=Q(\Phi_2,\Phi_3)=\frac{4}{9}
\begin{pmatrix}
2 \beta_2^2 & - \beta_2 \beta_3 \\
\beta_2 \beta_3 & 2 \beta_3^2
\end{pmatrix}.
\end{equation}
This is precisely the saturability of the SLD-quantum Cram\'{e}r-Rao bound. The same Gram-Schmidt construction of the optimal measurement basis can be generalized immediately to any number of network size $N$~\cite{pezze2017optimal}.

\section{Distributed Coupling-Strength Estimation: Case 2 Proposal}

We now consider the fundamental distributed precision limits for Case 2, see Table~\ref{tab_phases_cases}, where all coupling strengths are unknown and the mechanical oscillators are undriven. For this scenario, we propose two physical implementations: (i) the optomechanical architectures described above, and (ii) conditionally displaced spin–mechanical systems~\cite{treutlein2014hybrid}. Conditional displacement Hamiltonians describe interactions in which the energy separation between the emitter's internal states (e.g., a spin system) is strongly influenced by local electrostatic or magnetic fields~\cite{treutlein2014hybrid}. These local fields can be engineered using, for example, oscillating electrodes or vibrating magnetic tips attached to a mechanical oscillator. This implies, equivalently, that the displacement of the mechanical oscillator becomes conditional on the emitter's quantum state~\cite{treutlein2014hybrid}. Such Hamiltonians have been extensively studied both theoretically~\cite{khosla2018displacemon, rabl2010quantum, rabl2009strong, rabl2010cooling, braccini2023large, rao2016heralded, montenegro2017macroscopic, montenegro2018ground, tufarelli2011oscillator, scala2013matterwave, yin2013large, spiller2006quantum, kumar2017magnetometrty, tufarelli2012reconstructing, montenegro2014nonlinearity} and experimentally~\cite{treutlein2014hybrid}. The paradigmatic form is given by
\begin{equation}
\hat{H}_j = \frac{\hat{p}_j^2}{2M} + \frac{M\Omega^2}{2}\hat{x}_j^2 - \hbar\frac{\tilde{k}_j}{x_0} \hat{S}^{j}_z \hat{x}_j,
\end{equation}
where $\tilde{k}_j$ is the unknown coupling strength at site $j$. $\hat{S}^{j}_\beta = \sum_{i=1}^{N_\mathrm{exc}} \frac{\hat{\sigma}_{i,j}^\beta}{2}$ with $(\beta = x, y, z)$ denotes the collective spin operator at site $j$, composed of $N_\mathrm{exc}$ spin-$\tfrac{1}{2}$ particles indexed by $i$, where $\hat{\sigma}_{i,j}^\beta$ are the Pauli matrices along the $\beta$ direction. Note that the coupling constant $\tilde{k}_j$ is system-dependent and can encode multiple unknown physical properties of the system, including magnetic fields, magnetic field gradients, circulating currents, deformation potential constants, gate capacitances, and gate charges (see Ref.~\cite{treutlein2014hybrid} for a detailed list of relevant coupling constants). Consequently, estimating these distributed coupling constants across the sensor network enables one to infer valuable information about the physical properties of disparate systems. Furthermore, for consistency with Case 1, we denote by $N_\mathrm{exc}$ the total number of individual spins coupled to the mechanical oscillator. The operators $\hat{S}^{j}_\beta$ satisfy the usual commutation relations
\begin{eqnarray}
[\hat{S}^{j}_n, \hat{S}^{j'}_m]&=&i\hbar \varepsilon_{nml} \delta_{j,j'}\hat{S}^{j}_l,\\
(\hat{S}^{j})^2|s,m\rangle&=&\hbar^2 s(s+1)|s,m\rangle,\\
\hat{S}^{j}_z|s,m\rangle&=&\hbar m|s,m\rangle,
\end{eqnarray}
where $\varepsilon_{nml}$ is the Levi-Civita symbol, $s \leq N_\mathrm{exc}/2$ is the spin quantum number, and $-s \leq m \leq s$ labels the angular momentum projection along the $z$-axis. We focus in the symmetric subspace $2s=N_\mathrm{exc}$, with $N_\mathrm{exc}$ assumed even. Analogously to the previous case, we consider the initial state
\begin{equation}
|\psi(0)\rangle = |\text{W}_N\rangle (|\alpha\rangle^{\otimes N}),
\end{equation}
where
\begin{equation}
|\text{W}_N\rangle = \frac{1}{\sqrt{N}}\Big(|\pm \frac{N_\mathrm{exc}}{2},0,\ldots,0\rangle + \cdots + |0,\ldots,0,\pm \frac{N_\mathrm{exc}}{2}\rangle\Big),
\end{equation}
for $N_\mathrm{exc}\geq 2$, is the $N$-partite W state. Note that for spin-1/2, that is $N_\mathrm{exc}=1$, implies $\lambda=\pm \tfrac{N_\mathrm{exc}}{2}$ and $\lambda'=\mp \tfrac{N_\mathrm{exc}}{2}$. Thus, $\beta_j\propto(\lambda-\lambda')(\lambda+\lambda')=0$, resulting in singular Fisher matrix, see Eq.~\eqref{eq_tr_inv_QFIM}. For the case of $N_\mathrm{exc}\geq 2$, the parameters are properly identified as $\lambda = \pm \tfrac{N_\mathrm{exc}}{2}$ and $\lambda' = 0$. While other combinations are possible, maximal sensitivity is achieved when $\lambda'-\lambda$ is largest, that is, for $\lambda$ considering $m = \pm s = \pm \tfrac{N_\mathrm{exc}}{2}$ and thus $\lambda'$ considering $m = 0$. Using the chain rule as before, one gets:
\begin{equation}
    \sum_{j=1}^N\mathrm{Var}[\tilde{k}_j^-]\geq \frac{1}{\mu}\frac{2\Omega^2 N}{\pi^2N_\mathrm{exc}^4}\sum_{j=2}^N [k_j^+]^{-2}.
\end{equation}
As seen from the above expression, the additional terms effectively act as nuisance parameters in the sensing protocol, in the sense that they contribute unwanted dependencies which degrade the estimation precision~\cite{albarelli2020perspective}. To further simplify the above expression, we assume that the effective couplings satisfy $k_j^+ \sim 2k$, meaning they are all of the same order of magnitude. This places us in the worst-case scenario, where no coupling dominates and sensitivity is evenly distributed across the parameters. Equivalently, one can adopt a stepwise estimation protocol. First performing a coarse calibration of the nuisance parameters $k_j^+$, followed by a refined estimation of the target parameters $k_j^-$~\cite{mukhopadhyay2025beatingjointquantumestimation}. With this simplification, one gets:
\begin{equation}
\sum_{j=1}^N\mathrm{Var}[\tilde{k}_j^-]\geq \frac{1}{\mu}\frac{2\Omega^2 N}{\pi^2N_\mathrm{exc}^4}\sum_{j=2}^N [k_j^+]^{-2}\sim \frac{1}{\mu}\frac{\Omega^2 N(N-1)}{2\pi^2 k^2 N_\mathrm{exc}^4}.\label{eq_case2_sm_om}
\end{equation}

\begin{figure}
    \centering
    \includegraphics[width=\linewidth]{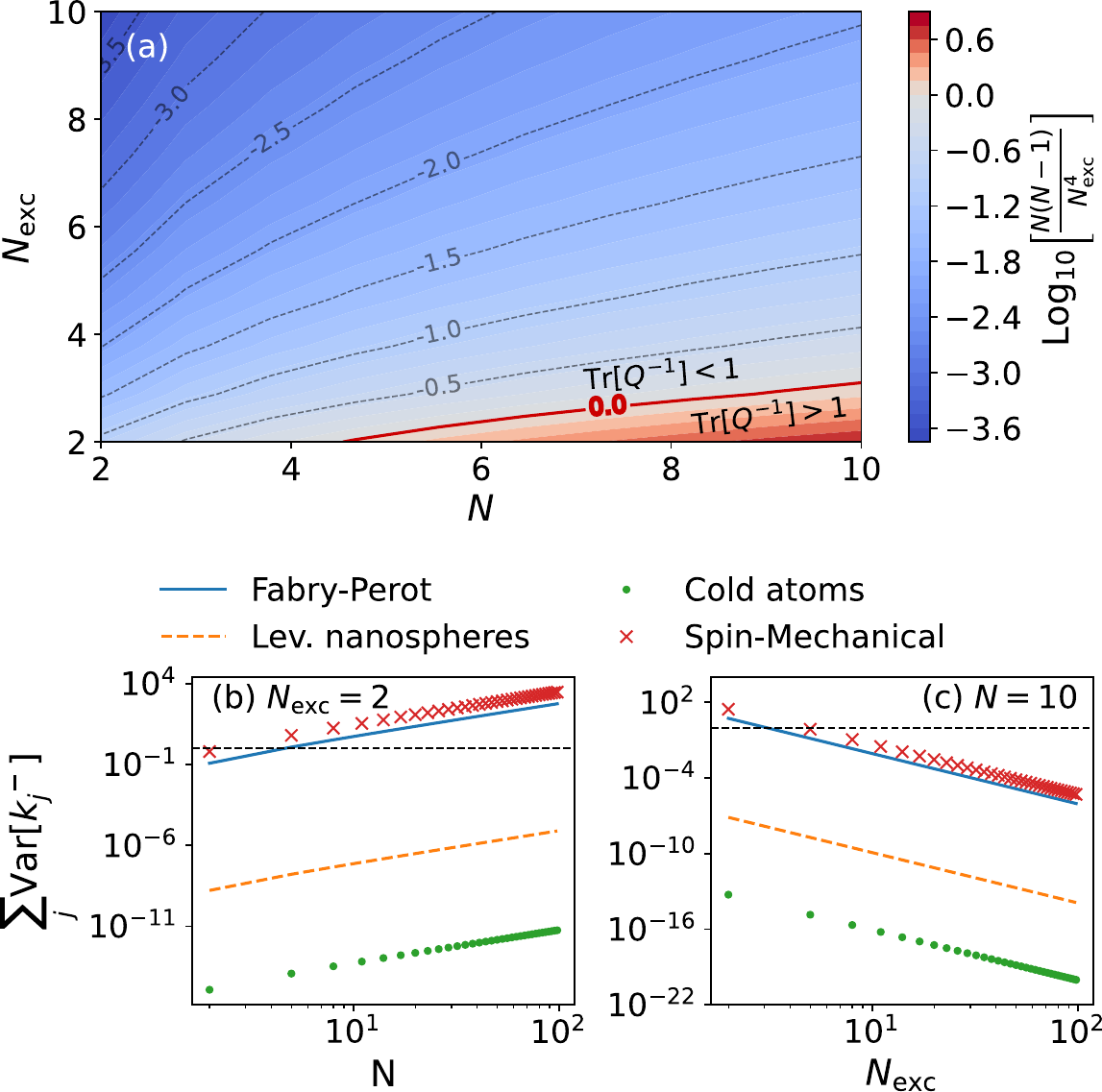}
    \caption{\textbf{Distributed quantum-enhanced sensing for Case 2.} (a) Dependence of the network size $N$ and the number of excitations $N_\mathrm{exc}$ in the right-hand side of Eq.~\eqref{eq_case2_sm_om}. (b) Sensitivity as a function of network size $N$ for a single excitation $N_\mathrm{exc} = 2$. (c) Sensitivity as a function of the number of excitations $N_\mathrm{exc}$ for a fixed network size $N = 10$. We consider $\mu=10^4$.}
    \label{fig_overall_sensitivity_case2}
\end{figure}

To illustrate the distributed quantum-enhanced sensitivity in Case 2, Fig.~\ref{fig_overall_sensitivity_case2}(a) shows the dependence of the sensing precision on both the network size $N$ and the number of initial probe excitations $N_\mathrm{exc}$ as given by the right-hand side of Eq.~\eqref{eq_case2_sm_om}. In contrast to Case 1, here, the inverse-quartic scaling $N_\mathrm{exc}^4$ competes with the $N^2$ scaling of the network size. As a result, achieving lower sensing uncertainty requires only $N_\mathrm{exc} \sim \sqrt{N}$, marking a favorable resource tradeoff in distributed quantum-enhanced metrology.

In Fig.~\ref{fig_overall_sensitivity_case2}(b), we show the distributed sensitivity as a function of $N$ for fixed $N_\mathrm{exc} = 1$. In contrast to the previous case, the hierarchy of optomechanical sensitivities is inverted, with the cold atoms platform achieving the lowest distributed multiparameter uncertainty for estimating the relative interaction coupling strengths---here, $10^{-9}\mathrm{Hz} \lesssim \Delta^{\text{Case 2}}_\mathrm{RMS} \lesssim 10^{-7}\mathrm{Hz}$. As predicted by Eq.~\eqref{eq_case2_sm_om}, the overall uncertainty increases quadratically with $N$ for a given $N_\mathrm{exc}$. Moreover, the Fabry–Perot configuration and the spin–mechanical system exhibit the worst scenarios with the highest overall uncertainties. For reference, the black dashed line in Fig.~\ref{fig_overall_sensitivity_case2}(b) indicates $\sum_j \mathrm{Var}[k_j^-] = 1$. For the spin–mechanical system, we consider mechanical frequencies as low as $\Omega \sim 10^3\mathrm{Hz}$, with higher frequencies further degrading the sensitivity.

In Fig.~\ref{fig_overall_sensitivity_case2}(c), we show the distributed quantum-enhanced sensitivity as a function of the number of excitations $N_\mathrm{exc}$ for a fixed network size $N = 10$. As the figure shows, the cold atoms system achieves the lowest distributed uncertainty $
10^{-12} \mathrm{Hz} \lesssim \Delta^\text{Case 2}_\mathrm{RMS} \lesssim 10^{-9}\mathrm{Hz}$. Furthermore, a higher number of excitations (or spin number) is required to reach lower sensitivities.

\section{Distributed Quantum-Enhanced Single-Parameter Estimation}

A particularly important case arises when only one relative parameter $\Phi_j$ remains unknown, with all others assumed known. This defines a distributed single-parameter scenario. Within our general multiparameter framework, the single-parameter precision limits is computed directly from the $(j,j)$ entry of the QFIM, that is $[Q]_{jj}\equiv Q^{\mathrm{single}}(\Phi_j)$, corresponding to the precision limits of the unknown parameter difference between node 1 and node $j$; see Sec.~\ref{sec_sm_qfim} of the Appendix for details:
\begin{equation}
Q^{\mathrm{single}}(\Phi_j) =
\begin{cases} \sim \frac{(N-1)}{N^2} k^2 N_\mathrm{exc}^2, & \text{for Case 1},\\
\sim \frac{(N-1)}{N^2}k^{+ 2}_jN_\mathrm{exc}^4, & \text{for Case 2}.
\end{cases}\label{eq_single_nonuisance}
\end{equation}
This reveals a clear competition between the network size $N$ and the number of initial excitations $N_\mathrm{exc}$, analogous to the case where all $\Phi_j$ are unknown throughout the network.

The central importance of single-parameter quantum estimation rests on three key advantages. First, in the limit of large data sets, it is possible to construct unbiased and efficient estimators that asymptotically saturate the quantum Cram\'er–Rao bound~\cite{meyer2021fisher}. Second, an optimal measurement that achieves the ultimate sensitivity always exists for single-parameter sensing scenarios~\cite{paris2009quantum}. Third, because such an optimal measurement is guaranteed, the quantum Cram\'er–Rao bound serves as a rigorous and operationally meaningful benchmark for experiments, against which any practical (non-optimal) measurement strategy can be directly compared~\cite{montenegro2025review, paris2009quantum}. The optimal measurement basis is obtained from the eigenbasis of the SLD operator $\mathcal{L}(\theta)$, which is defined for pure states as:
\begin{equation}
   \mathcal{L}(\theta) = 2 \left( |\partial_\theta \psi_\theta \rangle \langle \psi_\theta| + |\psi_\theta \rangle \langle \partial_\theta \psi_\theta| \right). \label{eq_SLDs_pure}
\end{equation}
Using the expression for the SLD given in Eq.~\eqref{eq_SLDs_pure}, we can explicitly construct the optimal measurement basis for the quantum probe state defined in Eq.~\eqref{eq_quantum_probe} for estimating $\Phi_k$ as---see Sec.~\ref{sec_sm_optimal_measurement_single} of the Appendix for a detailed derivation:
\begin{equation}
\Pi_\pm^{(k)}=|v_k^{\pm}\rangle\langle v_k^{\pm}|,
\end{equation}
where
\begin{equation}
    |v_k^{\pm}\rangle=\frac{1}{\sqrt2}\left[\Big(1\mp \frac{i}{\sqrt{N-1}}\Big)|\psi\rangle \pm i \sqrt{\frac{N}{N-1}}e^{i\beta_k \Phi_k}|k\rangle\right].
\end{equation}

Finally, another important case occurs when one aims to estimate a single-parameter $\Phi_j$ while all the other parameters $\{\Phi_k\}_{k\neq j}$ are also unknown. In this nuisance scenario~\cite{albarelli2020perspective}, the precision is bounded by $\mathrm{Var}[\Phi_j] \geq \tfrac{1}{\mu} (Q^{-1})_{jj} = \tfrac{1}{\mu} \tfrac{N}{2 \beta_j^2}$. As expected, the uncertainty is degraded by a factor of $\tfrac{N}{2(N-1)}$ compared to the case shown in Eq.~\eqref{eq_single_nonuisance} where all other parameters are known.

\section{Discussion}

Our proposal has both advantages and limitations.

On the advantages side: First, a key strength is that we provide a fully analytical derivation for estimating multiple unknown parameters distributed over $N$ spatially separated nodes, applicable to a broad and physically relevant class of quantum systems. Second, the quantum-enhanced sensitivity in our protocol relies crucially on stroboscopic disentangling generated by the conditional‑displacement interaction. At times $\Omega t = 2\pi$, all parameter dependence is guaranteed to be imprinted solely on the probe (e.g., light modes or collective spin systems), while the mechanical oscillator returns to---and decouples in---its initial state. As a result, all relevant information about the unknown parameters can be extracted through local measurements on the probe only, using an optimally constructed measurement basis~\cite{pezze2017optimal}. Furthermore, since the probe state is a $W$-type state, the measurement yields a discrete set of outcomes. This structure enables the use of efficient estimators, such as maximum-likelihood and Bayesian inference methods, which can be practically employed to saturate the precision bound~\cite{rubio2020bayesian, rubio2019quantummetrology}. Third, the protocol exhibits quadratic and quartic scaling of the Fisher information with respect to the number of initial probe excitations---a hallmark of quantum-enhanced performance. Fourth, the sensing nodes can be realized across a wide range of opto- and spin-mechanical platforms. In all case studies considered, the overall root-mean-square error closely tracks the single-node quantum limit (see Cases 1–2)~\cite{qvarfort2018gravimetry, montenegro2025heisenberg}. This shows that the distributed protocol does not incur a significant penalty compared to the best single-node strategy, while enabling spatially distributed field sensing. Fifth, promising physical implementations include levitated micro- and nano-spheres for gravimetry, and cold-atom optomechanical architectures for estimating distributed coupling strengths. More broadly, a nonlinear optomechanical interaction and a conditional-displacement spin-mechanical coupling can be combined into a hybrid interaction of the form
\begin{equation}
\hat{H}_\mathrm{hybrid} \propto (\hat{a}^\dagger \hat{a} + \tilde{\Omega} \hat{S}_z^j)(\hat{b}^\dagger + \hat{b}),
\end{equation}
where $\tilde{\Omega}$ is an dummy parameter. This effective interaction maps directly onto our model and, as a result, inherits the same precision guarantees~\cite{restrepo2014single, treutlein2014hybrid, wang2025remote}. This provides a concrete and scalable route to extend the present scheme to a variety of hybrid quantum architectures.

Three practical challenges merit discussion: First, preparing and verifying the required entangled resource states at scale is nontrivial~\cite{guhne2009entanglement, polino2020photonic}. Nevertheless, the generation of $W$-states has been extensively studied---both theoretically~\cite{dur2000three, cao2003entanglement, guo2002scheme, neto2017steady, chobari2025generationteleportationparticlew} and experimentally~\cite{park2025entangled, pu2018experimental, eibl2004experimental, haefner2005scalable, mattar2017experimental}. In the specific nonlinear optomechanical setup used in Case 1, the relevant entangled resource reduces to a multipartite N00N-type state. Here too, high-$N$ N00N states have been experimentally realized~\cite{mitchell2004super, walther2004debroglie, afek2010highnoon, hong2021quantumenhanced, polino2020photonic} and theoretically proposed~\cite{zhang2018scalable}. These developments collectively point to viable routes for both state preparation and readout. However, the associated overheads---especially as the number of nodes $N$ increases---remain a key bottleneck. Second, to keep our results analytically transparent and broadly accessible, we assume ideal unitary dynamics and do not fully model decoherence channels. However, an analytical approximation of the conditional-displacement dynamics under dissipation can, in principle, be incorporated by following the approach of Ref.~\cite{bose1997preparation}. This method alternates between unitary evolution and nonunitary damping steps, resulting in a probe state with explicitly damped off-diagonal complex coefficients. This effectively captures the impact of decoherence without resorting to a full quantum master equation treatment. Third, the scheme relies on operating at stroboscopic times $\Omega t = 2\pi$. Deviations from these times lead to residual entanglement between the probe and the mechanical oscillator, which degrades decoupling and thus reduces sensing performance. This sensitivity to timing errors is documented in stroboscopic gravimetric opto‑ and spin‑mechanical protocols~\cite{montenegro2025heisenberg}. While such imperfections can partially degrade the ideal quantum-enhanced scaling, the protocol can still maintain an advantage over classical sensing strategies under experimentally realistic levels of noise and timing deviations.

\section{Conclusions}

In this work, we present four key contributions. First, we analytically derive the ultimate multiparameter precision limits for a broad and physically relevant class of quantum probes. Our results demonstrate distributed quantum-enhanced sensitivity across the entire sensing network. Specifically, for a network of size $N$, we show that the achievable precision scales quadratically or quartically with respect to the sensing resource $N_\mathrm{exc}$. This scaling behavior is a defining signature of quantum-enhanced sensing. To the best of our knowledge, the quartic scaling has not been previously reported in the context of distributed quantum sensing schemes. Second, we rigorously prove that our quantum probes satisfy the weak-incompatibility condition. This condition ensures that both the Holevo bound and the quantum Cram\'{e}r-Rao bound are saturable---guaranteeing that the ultimate multiparameter precision limits can, in principle, be achieved. Third, building on this result, we explicitly construct an optimal measurement basis that does not depend on the unknown parameters and remains valid across the entire sensing network. Fourth, we apply our general results to two physically realizable systems: (i) nonlinear cavity quantum optomechanical systems, and (ii) solid-state spin systems coupled to mechanical oscillators---both of which are extendable to hybrid spin-optomechanical platforms~\cite{restrepo2017fully}. The first system enables the estimation of unknown gravitational accelerations at different spatial locations. The second allows for the estimation of coupling strengths across spatially separated points. Finally, we provide concrete experimental parameters showing that these protocols are readily implementable with current technology.

\section*{Acknowledgements}
This work was supported by Khalifa University of Science and Technology through Project No. 8474000739 (RIG-2024-033). V.M. thanks support from the National Natural Science Foundation of China Grants No. 12374482 and No. W2432005.

\bibliography{Distributed}

\begin{thebibliography}{108}%
\makeatletter
\providecommand \@ifxundefined [1]{%
 \@ifx{#1\undefined}
}%
\providecommand \@ifnum [1]{%
 \ifnum #1\expandafter \@firstoftwo
 \else \expandafter \@secondoftwo
 \fi
}%
\providecommand \@ifx [1]{%
 \ifx #1\expandafter \@firstoftwo
 \else \expandafter \@secondoftwo
 \fi
}%
\providecommand \natexlab [1]{#1}%
\providecommand \enquote  [1]{``#1''}%
\providecommand \bibnamefont  [1]{#1}%
\providecommand \bibfnamefont [1]{#1}%
\providecommand \citenamefont [1]{#1}%
\providecommand \href@noop [0]{\@secondoftwo}%
\providecommand \href [0]{\begingroup \@sanitize@url \@href}%
\providecommand \@href[1]{\@@startlink{#1}\@@href}%
\providecommand \@@href[1]{\endgroup#1\@@endlink}%
\providecommand \@sanitize@url [0]{\catcode `\\12\catcode `\$12\catcode
  `\&12\catcode `\#12\catcode `\^12\catcode `\_12\catcode `\%12\relax}%
\providecommand \@@startlink[1]{}%
\providecommand \@@endlink[0]{}%
\providecommand \url  [0]{\begingroup\@sanitize@url \@url }%
\providecommand \@url [1]{\endgroup\@href {#1}{\urlprefix }}%
\providecommand \urlprefix  [0]{URL }%
\providecommand \Eprint [0]{\href }%
\providecommand \doibase [0]{http://dx.doi.org/}%
\providecommand \selectlanguage [0]{\@gobble}%
\providecommand \bibinfo  [0]{\@secondoftwo}%
\providecommand \bibfield  [0]{\@secondoftwo}%
\providecommand \translation [1]{[#1]}%
\providecommand \BibitemOpen [0]{}%
\providecommand \bibitemStop [0]{}%
\providecommand \bibitemNoStop [0]{.\EOS\space}%
\providecommand \EOS [0]{\spacefactor3000\relax}%
\providecommand \BibitemShut  [1]{\csname bibitem#1\endcsname}%
\let\auto@bib@innerbib\@empty
\bibitem [{\citenamefont {Degen}\ \emph {et~al.}(2017)\citenamefont {Degen},
  \citenamefont {Reinhard},\ and\ \citenamefont
  {Cappellaro}}]{degen2017quantum}%
  \BibitemOpen
  \bibfield  {author} {\bibinfo {author} {\bibfnamefont {Christian~L}\
  \bibnamefont {Degen}}, \bibinfo {author} {\bibfnamefont {Friedemann}\
  \bibnamefont {Reinhard}}, \ and\ \bibinfo {author} {\bibfnamefont {Paola}\
  \bibnamefont {Cappellaro}},\ }\bibfield  {title} {\enquote {\bibinfo {title}
  {Quantum sensing},}\ }\href {https://doi.org/10.1103/RevModPhys.89.035002}
  {\bibfield  {journal} {\bibinfo  {journal} {Rev. Mod. Phys.}\ }\textbf
  {\bibinfo {volume} {89}},\ \bibinfo {pages} {035002} (\bibinfo {year}
  {2017})}\BibitemShut {NoStop}%
\bibitem [{\citenamefont {Montenegro}\ \emph {et~al.}(2025)\citenamefont
  {Montenegro}, \citenamefont {Mukhopadhyay}, \citenamefont {Yousefjani},
  \citenamefont {Sarkar}, \citenamefont {Mishra}, \citenamefont {Paris},\ and\
  \citenamefont {Bayat}}]{montenegro2025review}%
  \BibitemOpen
  \bibfield  {author} {\bibinfo {author} {\bibfnamefont {Victor}\ \bibnamefont
  {Montenegro}}, \bibinfo {author} {\bibfnamefont {Chiranjib}\ \bibnamefont
  {Mukhopadhyay}}, \bibinfo {author} {\bibfnamefont {Rozhin}\ \bibnamefont
  {Yousefjani}}, \bibinfo {author} {\bibfnamefont {Saubhik}\ \bibnamefont
  {Sarkar}}, \bibinfo {author} {\bibfnamefont {Utkarsh}\ \bibnamefont
  {Mishra}}, \bibinfo {author} {\bibfnamefont {Matteo~G.A.}\ \bibnamefont
  {Paris}}, \ and\ \bibinfo {author} {\bibfnamefont {Abolfazl}\ \bibnamefont
  {Bayat}},\ }\bibfield  {title} {\enquote {\bibinfo {title} {Review: Quantum
  metrology and sensing with many-body systems},}\ }\href {\doibase
  https://doi.org/10.1016/j.physrep.2025.05.005} {\bibfield  {journal}
  {\bibinfo  {journal} {Physics Reports}\ }\textbf {\bibinfo {volume} {1134}},\
  \bibinfo {pages} {1--62} (\bibinfo {year} {2025})},\ \bibinfo {note} {review:
  Quantum metrology and sensing with many-body systems}\BibitemShut {NoStop}%
\bibitem [{\citenamefont {Aslam}\ \emph {et~al.}(2023)\citenamefont {Aslam},
  \citenamefont {Zhou}, \citenamefont {Urbach}, \citenamefont {Turner},
  \citenamefont {Walsworth}, \citenamefont {Lukin},\ and\ \citenamefont
  {Park}}]{aslam2023quantum}%
  \BibitemOpen
  \bibfield  {author} {\bibinfo {author} {\bibfnamefont {Nabeel}\ \bibnamefont
  {Aslam}}, \bibinfo {author} {\bibfnamefont {Hengyun}\ \bibnamefont {Zhou}},
  \bibinfo {author} {\bibfnamefont {Elana~K}\ \bibnamefont {Urbach}}, \bibinfo
  {author} {\bibfnamefont {Matthew~J}\ \bibnamefont {Turner}}, \bibinfo
  {author} {\bibfnamefont {Ronald~L}\ \bibnamefont {Walsworth}}, \bibinfo
  {author} {\bibfnamefont {Mikhail~D}\ \bibnamefont {Lukin}}, \ and\ \bibinfo
  {author} {\bibfnamefont {Hongkun}\ \bibnamefont {Park}},\ }\bibfield  {title}
  {\enquote {\bibinfo {title} {Quantum sensors for biomedical applications},}\
  }\href {https://doi.org/10.1038/s42254-023-00558-3} {\bibfield  {journal}
  {\bibinfo  {journal} {Nat. Rev. Phys.}\ }\textbf {\bibinfo {volume} {5}},\
  \bibinfo {pages} {157--169} (\bibinfo {year} {2023})}\BibitemShut {NoStop}%
\bibitem [{\citenamefont {Ye}\ and\ \citenamefont
  {Zoller}(2024)}]{ye2024essay}%
  \BibitemOpen
  \bibfield  {author} {\bibinfo {author} {\bibfnamefont {Jun}\ \bibnamefont
  {Ye}}\ and\ \bibinfo {author} {\bibfnamefont {Peter}\ \bibnamefont
  {Zoller}},\ }\bibfield  {title} {\enquote {\bibinfo {title} {Essay: Quantum
  sensing with atomic, molecular, and optical platforms for fundamental
  physics},}\ }\href {\doibase 10.1103/PhysRevLett.132.190001} {\bibfield
  {journal} {\bibinfo  {journal} {Phys. Rev. Lett.}\ }\textbf {\bibinfo
  {volume} {132}},\ \bibinfo {pages} {190001} (\bibinfo {year}
  {2024})}\BibitemShut {NoStop}%
\bibitem [{\citenamefont {DeMille}\ \emph {et~al.}(2024)\citenamefont
  {DeMille}, \citenamefont {Hutzler}, \citenamefont {Rey},\ and\ \citenamefont
  {Zelevinsky}}]{demille2024quantum}%
  \BibitemOpen
  \bibfield  {author} {\bibinfo {author} {\bibfnamefont {David}\ \bibnamefont
  {DeMille}}, \bibinfo {author} {\bibfnamefont {Nicholas~R.}\ \bibnamefont
  {Hutzler}}, \bibinfo {author} {\bibfnamefont {Ana~Maria}\ \bibnamefont
  {Rey}}, \ and\ \bibinfo {author} {\bibfnamefont {Tanya}\ \bibnamefont
  {Zelevinsky}},\ }\bibfield  {title} {\enquote {\bibinfo {title} {Quantum
  sensing and metrology for fundamental physics with molecules},}\ }\href
  {\doibase 10.1038/s41567-024-02499-9} {\bibfield  {journal} {\bibinfo
  {journal} {Nature Physics}\ }\textbf {\bibinfo {volume} {20}},\ \bibinfo
  {pages} {741--749} (\bibinfo {year} {2024})}\BibitemShut {NoStop}%
\bibitem [{\citenamefont {Tan}\ \emph {et~al.}(2021)\citenamefont {Tan},
  \citenamefont {Narasimhachar},\ and\ \citenamefont {Regula}}]{tan2021fisher}%
  \BibitemOpen
  \bibfield  {author} {\bibinfo {author} {\bibfnamefont {Kok~Chuan}\
  \bibnamefont {Tan}}, \bibinfo {author} {\bibfnamefont {Varun}\ \bibnamefont
  {Narasimhachar}}, \ and\ \bibinfo {author} {\bibfnamefont {Bartosz}\
  \bibnamefont {Regula}},\ }\bibfield  {title} {\enquote {\bibinfo {title}
  {Fisher information universally identifies quantum resources},}\ }\href
  {\doibase 10.1103/PhysRevLett.127.200402} {\bibfield  {journal} {\bibinfo
  {journal} {Phys. Rev. Lett.}\ }\textbf {\bibinfo {volume} {127}},\ \bibinfo
  {pages} {200402} (\bibinfo {year} {2021})}\BibitemShut {NoStop}%
\bibitem [{\citenamefont {Giovannetti}\ \emph {et~al.}(2004)\citenamefont
  {Giovannetti}, \citenamefont {Lloyd},\ and\ \citenamefont
  {Maccone}}]{giovannetti2004quantum}%
  \BibitemOpen
  \bibfield  {author} {\bibinfo {author} {\bibfnamefont {Vittorio}\
  \bibnamefont {Giovannetti}}, \bibinfo {author} {\bibfnamefont {Seth}\
  \bibnamefont {Lloyd}}, \ and\ \bibinfo {author} {\bibfnamefont {Lorenzo}\
  \bibnamefont {Maccone}},\ }\bibfield  {title} {\enquote {\bibinfo {title}
  {Quantum-enhanced measurements: beating the standard quantum limit},}\
  }\href@noop {} {\bibfield  {journal} {\bibinfo  {journal} {Science}\ }\textbf
  {\bibinfo {volume} {306}},\ \bibinfo {pages} {1330--1336} (\bibinfo {year}
  {2004})}\BibitemShut {NoStop}%
\bibitem [{\citenamefont {Giovannetti}\ \emph {et~al.}(2006)\citenamefont
  {Giovannetti}, \citenamefont {Lloyd},\ and\ \citenamefont
  {Maccone}}]{giovannetti2006quantum}%
  \BibitemOpen
  \bibfield  {author} {\bibinfo {author} {\bibfnamefont {Vittorio}\
  \bibnamefont {Giovannetti}}, \bibinfo {author} {\bibfnamefont {Seth}\
  \bibnamefont {Lloyd}}, \ and\ \bibinfo {author} {\bibfnamefont {Lorenzo}\
  \bibnamefont {Maccone}},\ }\bibfield  {title} {\enquote {\bibinfo {title}
  {Quantum metrology},}\ }\href {\doibase 10.1103/PhysRevLett.96.010401}
  {\bibfield  {journal} {\bibinfo  {journal} {Phys. Rev. Lett.}\ }\textbf
  {\bibinfo {volume} {96}},\ \bibinfo {pages} {010401} (\bibinfo {year}
  {2006})}\BibitemShut {NoStop}%
\bibitem [{\citenamefont {Giovannetti}\ \emph {et~al.}(2011)\citenamefont
  {Giovannetti}, \citenamefont {Lloyd},\ and\ \citenamefont
  {Maccone}}]{giovannetti2011advances}%
  \BibitemOpen
  \bibfield  {author} {\bibinfo {author} {\bibfnamefont {Vittorio}\
  \bibnamefont {Giovannetti}}, \bibinfo {author} {\bibfnamefont {Seth}\
  \bibnamefont {Lloyd}}, \ and\ \bibinfo {author} {\bibfnamefont {Lorenzo}\
  \bibnamefont {Maccone}},\ }\bibfield  {title} {\enquote {\bibinfo {title}
  {Advances in quantum metrology},}\ }\href@noop {} {\bibfield  {journal}
  {\bibinfo  {journal} {Nature photonics}\ }\textbf {\bibinfo {volume} {5}},\
  \bibinfo {pages} {222--229} (\bibinfo {year} {2011})}\BibitemShut {NoStop}%
\bibitem [{\citenamefont {Braun}\ \emph {et~al.}(2018)\citenamefont {Braun},
  \citenamefont {Adesso}, \citenamefont {Benatti}, \citenamefont {Floreanini},
  \citenamefont {Marzolino}, \citenamefont {Mitchell},\ and\ \citenamefont
  {Pirandola}}]{braun2018quantum}%
  \BibitemOpen
  \bibfield  {author} {\bibinfo {author} {\bibfnamefont {Daniel}\ \bibnamefont
  {Braun}}, \bibinfo {author} {\bibfnamefont {Gerardo}\ \bibnamefont {Adesso}},
  \bibinfo {author} {\bibfnamefont {Fabio}\ \bibnamefont {Benatti}}, \bibinfo
  {author} {\bibfnamefont {Roberto}\ \bibnamefont {Floreanini}}, \bibinfo
  {author} {\bibfnamefont {Ugo}\ \bibnamefont {Marzolino}}, \bibinfo {author}
  {\bibfnamefont {Morgan~W.}\ \bibnamefont {Mitchell}}, \ and\ \bibinfo
  {author} {\bibfnamefont {Stefano}\ \bibnamefont {Pirandola}},\ }\bibfield
  {title} {\enquote {\bibinfo {title} {Quantum-enhanced measurements without
  entanglement},}\ }\href {\doibase 10.1103/RevModPhys.90.035006} {\bibfield
  {journal} {\bibinfo  {journal} {Rev. Mod. Phys.}\ }\textbf {\bibinfo {volume}
  {90}},\ \bibinfo {pages} {035006} (\bibinfo {year} {2018})}\BibitemShut
  {NoStop}%
\bibitem [{\citenamefont {Braunstein}\ and\ \citenamefont
  {Caves}(1994)}]{braunstein1994statistical}%
  \BibitemOpen
  \bibfield  {author} {\bibinfo {author} {\bibfnamefont {Samuel~L.}\
  \bibnamefont {Braunstein}}\ and\ \bibinfo {author} {\bibfnamefont
  {Carlton~M.}\ \bibnamefont {Caves}},\ }\bibfield  {title} {\enquote {\bibinfo
  {title} {Statistical distance and the geometry of quantum states},}\ }\href
  {\doibase 10.1103/PhysRevLett.72.3439} {\bibfield  {journal} {\bibinfo
  {journal} {Phys. Rev. Lett.}\ }\textbf {\bibinfo {volume} {72}},\ \bibinfo
  {pages} {3439--3443} (\bibinfo {year} {1994})}\BibitemShut {NoStop}%
\bibitem [{\citenamefont {Sidhu}\ and\ \citenamefont
  {Kok}(2020)}]{sidhu2020geometric}%
  \BibitemOpen
  \bibfield  {author} {\bibinfo {author} {\bibfnamefont {Jasminder~S}\
  \bibnamefont {Sidhu}}\ and\ \bibinfo {author} {\bibfnamefont {Pieter}\
  \bibnamefont {Kok}},\ }\bibfield  {title} {\enquote {\bibinfo {title}
  {Geometric perspective on quantum parameter estimation},}\ }\href
  {https://doi.org/10.1116/1.5119961} {\bibfield  {journal} {\bibinfo
  {journal} {AVS Quantum Science}\ }\textbf {\bibinfo {volume} {2}},\ \bibinfo
  {pages} {014701} (\bibinfo {year} {2020})}\BibitemShut {NoStop}%
\bibitem [{\citenamefont {Holevo}(2011)}]{holevo2011probabilistic}%
  \BibitemOpen
  \bibfield  {author} {\bibinfo {author} {\bibfnamefont {A.S.}\ \bibnamefont
  {Holevo}},\ }\href@noop {} {\emph {\bibinfo {title} {Probabilistic and
  Statistical Aspects of Quantum Theory}}}\ (\bibinfo  {publisher} {Edizioni
  della Normale, Pisa},\ \bibinfo {year} {2011})\BibitemShut {NoStop}%
\bibitem [{\citenamefont {Van~Trees}(2004)}]{vantrees2004detection}%
  \BibitemOpen
  \bibfield  {author} {\bibinfo {author} {\bibfnamefont {Harry~L.}\
  \bibnamefont {Van~Trees}},\ }\href {\doibase 10.1002/0471221082} {\emph
  {\bibinfo {title} {Detection, Estimation, and Modulation Theory, Part I}}},\
  \bibinfo {edition} {2nd}\ ed.\ (\bibinfo  {publisher} {Wiley-Interscience},\
  \bibinfo {year} {2004})\BibitemShut {NoStop}%
\bibitem [{\citenamefont {Helstrom}(1967)}]{helstrom1967minimum}%
  \BibitemOpen
  \bibfield  {author} {\bibinfo {author} {\bibfnamefont {C.W.}\ \bibnamefont
  {Helstrom}},\ }\bibfield  {title} {\enquote {\bibinfo {title} {Minimum
  mean-squared error of estimates in quantum statistics},}\ }\href
  {https://doi.org/10.1016/0375-9601(67)90366-0} {\bibfield  {journal}
  {\bibinfo  {journal} {Phys. Lett. A}\ }\textbf {\bibinfo {volume} {25}},\
  \bibinfo {pages} {101--102} (\bibinfo {year} {1967})}\BibitemShut {NoStop}%
\bibitem [{\citenamefont {Helstrom}(1976)}]{helstrom1976quantum}%
  \BibitemOpen
  \bibfield  {author} {\bibinfo {author} {\bibfnamefont {Carl~W}\ \bibnamefont
  {Helstrom}},\ }\href@noop {} {\emph {\bibinfo {title} {Quantum Detection and
  Estimation Theory}}}\ (\bibinfo  {publisher} {Academic Press},\ \bibinfo
  {year} {1976})\BibitemShut {NoStop}%
\bibitem [{\citenamefont {Paris}(2009)}]{paris2009quantum}%
  \BibitemOpen
  \bibfield  {author} {\bibinfo {author} {\bibfnamefont {Matteo~GA}\
  \bibnamefont {Paris}},\ }\bibfield  {title} {\enquote {\bibinfo {title}
  {Quantum estimation for quantum technology},}\ }\href
  {https://doi.org/10.1142/S0219749909004839} {\bibfield  {journal} {\bibinfo
  {journal} {Int. J. Quantum Inf.}\ }\textbf {\bibinfo {volume} {7}},\ \bibinfo
  {pages} {125--137} (\bibinfo {year} {2009})}\BibitemShut {NoStop}%
\bibitem [{\citenamefont {Helstrom}\ and\ \citenamefont
  {Kennedy}(1974)}]{helstrom1974noncommuting}%
  \BibitemOpen
  \bibfield  {author} {\bibinfo {author} {\bibfnamefont {C.}~\bibnamefont
  {Helstrom}}\ and\ \bibinfo {author} {\bibfnamefont {R.}~\bibnamefont
  {Kennedy}},\ }\bibfield  {title} {\enquote {\bibinfo {title} {Noncommuting
  observables in quantum detection and estimation theory},}\ }\href {\doibase
  10.1109/TIT.1974.1055173} {\bibfield  {journal} {\bibinfo  {journal} {IEEE
  Trans. Inf. Theory.}\ }\textbf {\bibinfo {volume} {20}},\ \bibinfo {pages}
  {16--24} (\bibinfo {year} {1974})}\BibitemShut {NoStop}%
\bibitem [{\citenamefont {Carollo}\ \emph {et~al.}(2019)\citenamefont
  {Carollo}, \citenamefont {Spagnolo}, \citenamefont {Dubkov},\ and\
  \citenamefont {Valenti}}]{carollo2019quantumness}%
  \BibitemOpen
  \bibfield  {author} {\bibinfo {author} {\bibfnamefont {Angelo}\ \bibnamefont
  {Carollo}}, \bibinfo {author} {\bibfnamefont {Bernardo}\ \bibnamefont
  {Spagnolo}}, \bibinfo {author} {\bibfnamefont {Alexander~A}\ \bibnamefont
  {Dubkov}}, \ and\ \bibinfo {author} {\bibfnamefont {Davide}\ \bibnamefont
  {Valenti}},\ }\bibfield  {title} {\enquote {\bibinfo {title} {On quantumness
  in multi-parameter quantum estimation},}\ }\href
  {https://doi.org/10.1088/1742-5468/ab3ccb} {\bibfield  {journal} {\bibinfo
  {journal} {J. Stat. Mech.}\ }\textbf {\bibinfo {volume} {2019}},\ \bibinfo
  {pages} {094010} (\bibinfo {year} {2019})}\BibitemShut {NoStop}%
\bibitem [{\citenamefont {Liu}\ \emph {et~al.}(2020)\citenamefont {Liu},
  \citenamefont {Yuan}, \citenamefont {Lu},\ and\ \citenamefont
  {Wang}}]{liu2020quantum}%
  \BibitemOpen
  \bibfield  {author} {\bibinfo {author} {\bibfnamefont {Jing}\ \bibnamefont
  {Liu}}, \bibinfo {author} {\bibfnamefont {Haidong}\ \bibnamefont {Yuan}},
  \bibinfo {author} {\bibfnamefont {Xiao-Ming}\ \bibnamefont {Lu}}, \ and\
  \bibinfo {author} {\bibfnamefont {Xiaoguang}\ \bibnamefont {Wang}},\
  }\bibfield  {title} {\enquote {\bibinfo {title} {Quantum fisher information
  matrix and multiparameter estimation},}\ }\href
  {https://iopscience.iop.org/article/10.1088/1751-8121/ab5d4d} {\bibfield
  {journal} {\bibinfo  {journal} {J. Phys. A: Math. Theor.}\ }\textbf {\bibinfo
  {volume} {53}},\ \bibinfo {pages} {023001} (\bibinfo {year}
  {2020})}\BibitemShut {NoStop}%
\bibitem [{\citenamefont {Pezz\`e}\ \emph {et~al.}(2017)\citenamefont
  {Pezz\`e}, \citenamefont {Ciampini}, \citenamefont {Spagnolo}, \citenamefont
  {Humphreys}, \citenamefont {Datta}, \citenamefont {Walmsley}, \citenamefont
  {Barbieri}, \citenamefont {Sciarrino},\ and\ \citenamefont
  {Smerzi}}]{pezze2017optimal}%
  \BibitemOpen
  \bibfield  {author} {\bibinfo {author} {\bibfnamefont {Luca}\ \bibnamefont
  {Pezz\`e}}, \bibinfo {author} {\bibfnamefont {Mario~A.}\ \bibnamefont
  {Ciampini}}, \bibinfo {author} {\bibfnamefont {Nicol\`o}\ \bibnamefont
  {Spagnolo}}, \bibinfo {author} {\bibfnamefont {Peter~C.}\ \bibnamefont
  {Humphreys}}, \bibinfo {author} {\bibfnamefont {Animesh}\ \bibnamefont
  {Datta}}, \bibinfo {author} {\bibfnamefont {Ian~A.}\ \bibnamefont
  {Walmsley}}, \bibinfo {author} {\bibfnamefont {Marco}\ \bibnamefont
  {Barbieri}}, \bibinfo {author} {\bibfnamefont {Fabio}\ \bibnamefont
  {Sciarrino}}, \ and\ \bibinfo {author} {\bibfnamefont {Augusto}\ \bibnamefont
  {Smerzi}},\ }\bibfield  {title} {\enquote {\bibinfo {title} {Optimal
  measurements for simultaneous quantum estimation of multiple phases},}\
  }\href {\doibase 10.1103/PhysRevLett.119.130504} {\bibfield  {journal}
  {\bibinfo  {journal} {Phys. Rev. Lett.}\ }\textbf {\bibinfo {volume} {119}},\
  \bibinfo {pages} {130504} (\bibinfo {year} {2017})}\BibitemShut {NoStop}%
\bibitem [{\citenamefont {Zhang}\ and\ \citenamefont
  {Zhuang}(2021)}]{zhang2021distributed}%
  \BibitemOpen
  \bibfield  {author} {\bibinfo {author} {\bibfnamefont {Zheshen}\ \bibnamefont
  {Zhang}}\ and\ \bibinfo {author} {\bibfnamefont {Quntao}\ \bibnamefont
  {Zhuang}},\ }\bibfield  {title} {\enquote {\bibinfo {title} {Distributed
  quantum sensing},}\ }\href {\doibase 10.1088/2058-9565/abd4c3} {\bibfield
  {journal} {\bibinfo  {journal} {Quantum Science and Technology}\ }\textbf
  {\bibinfo {volume} {6}},\ \bibinfo {pages} {043001} (\bibinfo {year}
  {2021})}\BibitemShut {NoStop}%
\bibitem [{\citenamefont {Proctor}\ \emph {et~al.}(2017)\citenamefont
  {Proctor}, \citenamefont {Knott},\ and\ \citenamefont
  {Dunningham}}]{proctor2017networkedquantumsensing}%
  \BibitemOpen
  \bibfield  {author} {\bibinfo {author} {\bibfnamefont {T.~J.}\ \bibnamefont
  {Proctor}}, \bibinfo {author} {\bibfnamefont {P.~A.}\ \bibnamefont {Knott}},
  \ and\ \bibinfo {author} {\bibfnamefont {J.~A.}\ \bibnamefont {Dunningham}},\
  }\href {https://arxiv.org/abs/1702.04271} {\enquote {\bibinfo {title}
  {Networked quantum sensing},}\ } (\bibinfo {year} {2017}),\ \Eprint
  {http://arxiv.org/abs/1702.04271} {arXiv:1702.04271 [quant-ph]} \BibitemShut
  {NoStop}%
\bibitem [{\citenamefont {Hu}\ \emph {et~al.}(2025)\citenamefont {Hu},
  \citenamefont {Zang}, \citenamefont {Wang}, \citenamefont {Zhong},
  \citenamefont {Yuan}, \citenamefont {Jiang},\ and\ \citenamefont
  {Saleem}}]{hu2025optimalschemedistributedquantum}%
  \BibitemOpen
  \bibfield  {author} {\bibinfo {author} {\bibfnamefont {Zhiyao}\ \bibnamefont
  {Hu}}, \bibinfo {author} {\bibfnamefont {Allen}\ \bibnamefont {Zang}},
  \bibinfo {author} {\bibfnamefont {Jianwei}\ \bibnamefont {Wang}}, \bibinfo
  {author} {\bibfnamefont {Tian}\ \bibnamefont {Zhong}}, \bibinfo {author}
  {\bibfnamefont {Haidong}\ \bibnamefont {Yuan}}, \bibinfo {author}
  {\bibfnamefont {Liang}\ \bibnamefont {Jiang}}, \ and\ \bibinfo {author}
  {\bibfnamefont {Zain~H.}\ \bibnamefont {Saleem}},\ }\href
  {https://arxiv.org/abs/2509.18334} {\enquote {\bibinfo {title} {Optimal
  scheme for distributed quantum metrology},}\ } (\bibinfo {year} {2025}),\
  \Eprint {http://arxiv.org/abs/2509.18334} {arXiv:2509.18334 [quant-ph]}
  \BibitemShut {NoStop}%
\bibitem [{\citenamefont {Humphreys}\ \emph {et~al.}(2013)\citenamefont
  {Humphreys}, \citenamefont {Barbieri}, \citenamefont {Datta},\ and\
  \citenamefont {Walmsley}}]{humphreys2013quantum}%
  \BibitemOpen
  \bibfield  {author} {\bibinfo {author} {\bibfnamefont {Peter~C.}\
  \bibnamefont {Humphreys}}, \bibinfo {author} {\bibfnamefont {Marco}\
  \bibnamefont {Barbieri}}, \bibinfo {author} {\bibfnamefont {Animesh}\
  \bibnamefont {Datta}}, \ and\ \bibinfo {author} {\bibfnamefont {Ian~A.}\
  \bibnamefont {Walmsley}},\ }\bibfield  {title} {\enquote {\bibinfo {title}
  {Quantum enhanced multiple phase estimation},}\ }\href {\doibase
  10.1103/PhysRevLett.111.070403} {\bibfield  {journal} {\bibinfo  {journal}
  {Phys. Rev. Lett.}\ }\textbf {\bibinfo {volume} {111}},\ \bibinfo {pages}
  {070403} (\bibinfo {year} {2013})}\BibitemShut {NoStop}%
\bibitem [{\citenamefont {Ge}\ \emph {et~al.}(2018)\citenamefont {Ge},
  \citenamefont {Jacobs}, \citenamefont {Eldredge}, \citenamefont {Gorshkov},\
  and\ \citenamefont {Foss-Feig}}]{ge2018distributed}%
  \BibitemOpen
  \bibfield  {author} {\bibinfo {author} {\bibfnamefont {Wenchao}\ \bibnamefont
  {Ge}}, \bibinfo {author} {\bibfnamefont {Kurt}\ \bibnamefont {Jacobs}},
  \bibinfo {author} {\bibfnamefont {Zachary}\ \bibnamefont {Eldredge}},
  \bibinfo {author} {\bibfnamefont {Alexey~V.}\ \bibnamefont {Gorshkov}}, \
  and\ \bibinfo {author} {\bibfnamefont {Michael}\ \bibnamefont {Foss-Feig}},\
  }\bibfield  {title} {\enquote {\bibinfo {title} {Distributed quantum
  metrology with linear networks and separable inputs},}\ }\href {\doibase
  10.1103/PhysRevLett.121.043604} {\bibfield  {journal} {\bibinfo  {journal}
  {Phys. Rev. Lett.}\ }\textbf {\bibinfo {volume} {121}},\ \bibinfo {pages}
  {043604} (\bibinfo {year} {2018})}\BibitemShut {NoStop}%
\bibitem [{\citenamefont {Proctor}\ \emph {et~al.}(2018)\citenamefont
  {Proctor}, \citenamefont {Knott},\ and\ \citenamefont
  {Dunningham}}]{proctor2018multiparameter}%
  \BibitemOpen
  \bibfield  {author} {\bibinfo {author} {\bibfnamefont {Timothy~J.}\
  \bibnamefont {Proctor}}, \bibinfo {author} {\bibfnamefont {Paul~A.}\
  \bibnamefont {Knott}}, \ and\ \bibinfo {author} {\bibfnamefont {Jacob~A.}\
  \bibnamefont {Dunningham}},\ }\bibfield  {title} {\enquote {\bibinfo {title}
  {Multiparameter estimation in networked quantum sensors},}\ }\href {\doibase
  10.1103/PhysRevLett.120.080501} {\bibfield  {journal} {\bibinfo  {journal}
  {Phys. Rev. Lett.}\ }\textbf {\bibinfo {volume} {120}},\ \bibinfo {pages}
  {080501} (\bibinfo {year} {2018})}\BibitemShut {NoStop}%
\bibitem [{\citenamefont {Zhuang}\ \emph {et~al.}(2018)\citenamefont {Zhuang},
  \citenamefont {Zhang},\ and\ \citenamefont
  {Shapiro}}]{zhuang2018distributed}%
  \BibitemOpen
  \bibfield  {author} {\bibinfo {author} {\bibfnamefont {Quntao}\ \bibnamefont
  {Zhuang}}, \bibinfo {author} {\bibfnamefont {Zheshen}\ \bibnamefont {Zhang}},
  \ and\ \bibinfo {author} {\bibfnamefont {Jeffrey~H.}\ \bibnamefont
  {Shapiro}},\ }\bibfield  {title} {\enquote {\bibinfo {title} {Distributed
  quantum sensing using continuous-variable multipartite entanglement},}\
  }\href {\doibase 10.1103/PhysRevA.97.032329} {\bibfield  {journal} {\bibinfo
  {journal} {Phys. Rev. A}\ }\textbf {\bibinfo {volume} {97}},\ \bibinfo
  {pages} {032329} (\bibinfo {year} {2018})}\BibitemShut {NoStop}%
\bibitem [{\citenamefont {Agarwal}(2025)}]{agarwal2025saturation}%
  \BibitemOpen
  \bibfield  {author} {\bibinfo {author} {\bibfnamefont {Girish~S.}\
  \bibnamefont {Agarwal}},\ }\bibfield  {title} {\enquote {\bibinfo {title}
  {Saturation of the quantum cram\'er-rao bound for distributed sensing via
  error sensitivity in su(1,1)-su($m$) interferometry},}\ }\href {\doibase
  10.1103/h2x6-dz96} {\bibfield  {journal} {\bibinfo  {journal} {Phys. Rev. A}\
  }\textbf {\bibinfo {volume} {112}},\ \bibinfo {pages} {032439} (\bibinfo
  {year} {2025})}\BibitemShut {NoStop}%
\bibitem [{\citenamefont {Eldredge}\ \emph {et~al.}(2018)\citenamefont
  {Eldredge}, \citenamefont {Foss-Feig}, \citenamefont {Gross}, \citenamefont
  {Rolston},\ and\ \citenamefont {Gorshkov}}]{eldredge2018optimal}%
  \BibitemOpen
  \bibfield  {author} {\bibinfo {author} {\bibfnamefont {Zachary}\ \bibnamefont
  {Eldredge}}, \bibinfo {author} {\bibfnamefont {Michael}\ \bibnamefont
  {Foss-Feig}}, \bibinfo {author} {\bibfnamefont {Jonathan~A.}\ \bibnamefont
  {Gross}}, \bibinfo {author} {\bibfnamefont {S.~L.}\ \bibnamefont {Rolston}},
  \ and\ \bibinfo {author} {\bibfnamefont {Alexey~V.}\ \bibnamefont
  {Gorshkov}},\ }\bibfield  {title} {\enquote {\bibinfo {title} {Optimal and
  secure measurement protocols for quantum sensor networks},}\ }\href {\doibase
  10.1103/PhysRevA.97.042337} {\bibfield  {journal} {\bibinfo  {journal} {Phys.
  Rev. A}\ }\textbf {\bibinfo {volume} {97}},\ \bibinfo {pages} {042337}
  (\bibinfo {year} {2018})}\BibitemShut {NoStop}%
\bibitem [{\citenamefont {Rubio}\ \emph {et~al.}(2020)\citenamefont {Rubio},
  \citenamefont {Knott}, \citenamefont {Proctor},\ and\ \citenamefont
  {Dunningham}}]{rubio2020quantum}%
  \BibitemOpen
  \bibfield  {author} {\bibinfo {author} {\bibfnamefont {Jesús}\ \bibnamefont
  {Rubio}}, \bibinfo {author} {\bibfnamefont {Paul~A}\ \bibnamefont {Knott}},
  \bibinfo {author} {\bibfnamefont {Timothy~J}\ \bibnamefont {Proctor}}, \ and\
  \bibinfo {author} {\bibfnamefont {Jacob~A}\ \bibnamefont {Dunningham}},\
  }\bibfield  {title} {\enquote {\bibinfo {title} {Quantum sensing networks for
  the estimation of linear functions},}\ }\href {\doibase
  10.1088/1751-8121/ab9d46} {\bibfield  {journal} {\bibinfo  {journal} {Journal
  of Physics A: Mathematical and Theoretical}\ }\textbf {\bibinfo {volume}
  {53}},\ \bibinfo {pages} {344001} (\bibinfo {year} {2020})}\BibitemShut
  {NoStop}%
\bibitem [{\citenamefont {Namkung}\ \emph {et~al.}(2024)\citenamefont
  {Namkung}, \citenamefont {Kim}, \citenamefont {Hong}, \citenamefont {Kim},
  \citenamefont {Lee},\ and\ \citenamefont {Lim}}]{namkung2024optimal}%
  \BibitemOpen
  \bibfield  {author} {\bibinfo {author} {\bibfnamefont {Min}\ \bibnamefont
  {Namkung}}, \bibinfo {author} {\bibfnamefont {Dong-Hyun}\ \bibnamefont
  {Kim}}, \bibinfo {author} {\bibfnamefont {Seongjin}\ \bibnamefont {Hong}},
  \bibinfo {author} {\bibfnamefont {Yong-Su}\ \bibnamefont {Kim}}, \bibinfo
  {author} {\bibfnamefont {Changhyoup}\ \bibnamefont {Lee}}, \ and\ \bibinfo
  {author} {\bibfnamefont {Hyang-Tag}\ \bibnamefont {Lim}},\ }\bibfield
  {title} {\enquote {\bibinfo {title} {Optimal multiple-phase estimation with
  multi-mode noon states against photon loss},}\ }\href {\doibase
  10.1088/1367-2630/ad5eaf} {\bibfield  {journal} {\bibinfo  {journal} {New
  Journal of Physics}\ }\textbf {\bibinfo {volume} {26}},\ \bibinfo {pages}
  {073028} (\bibinfo {year} {2024})}\BibitemShut {NoStop}%
\bibitem [{\citenamefont {Zhang}\ and\ \citenamefont
  {Chan}(2017)}]{zhang2017quantum}%
  \BibitemOpen
  \bibfield  {author} {\bibinfo {author} {\bibfnamefont {Lu}~\bibnamefont
  {Zhang}}\ and\ \bibinfo {author} {\bibfnamefont {Kam Wai~Clifford}\
  \bibnamefont {Chan}},\ }\bibfield  {title} {\enquote {\bibinfo {title}
  {Quantum multiparameter estimation with generalized balanced multimode
  noon-like states},}\ }\href {\doibase 10.1103/PhysRevA.95.032321} {\bibfield
  {journal} {\bibinfo  {journal} {Phys. Rev. A}\ }\textbf {\bibinfo {volume}
  {95}},\ \bibinfo {pages} {032321} (\bibinfo {year} {2017})}\BibitemShut
  {NoStop}%
\bibitem [{\citenamefont {Oh}\ \emph {et~al.}(2020)\citenamefont {Oh},
  \citenamefont {Lee}, \citenamefont {Lie},\ and\ \citenamefont
  {Jeong}}]{oh2020optimal}%
  \BibitemOpen
  \bibfield  {author} {\bibinfo {author} {\bibfnamefont {Changhun}\
  \bibnamefont {Oh}}, \bibinfo {author} {\bibfnamefont {Changhyoup}\
  \bibnamefont {Lee}}, \bibinfo {author} {\bibfnamefont {Seok~Hyung}\
  \bibnamefont {Lie}}, \ and\ \bibinfo {author} {\bibfnamefont {Hyunseok}\
  \bibnamefont {Jeong}},\ }\bibfield  {title} {\enquote {\bibinfo {title}
  {Optimal distributed quantum sensing using gaussian states},}\ }\href
  {\doibase 10.1103/PhysRevResearch.2.023030} {\bibfield  {journal} {\bibinfo
  {journal} {Phys. Rev. Res.}\ }\textbf {\bibinfo {volume} {2}},\ \bibinfo
  {pages} {023030} (\bibinfo {year} {2020})}\BibitemShut {NoStop}%
\bibitem [{\citenamefont {Kim}\ \emph {et~al.}(2024)\citenamefont {Kim},
  \citenamefont {Hong}, \citenamefont {Kim}, \citenamefont {Kim}, \citenamefont
  {Lee}, \citenamefont {Pooser}, \citenamefont {Oh}, \citenamefont {Lee},
  \citenamefont {Lee},\ and\ \citenamefont {Lim}}]{kim2024distributed}%
  \BibitemOpen
  \bibfield  {author} {\bibinfo {author} {\bibfnamefont {Dong-Hyun}\
  \bibnamefont {Kim}}, \bibinfo {author} {\bibfnamefont {Seongjin}\
  \bibnamefont {Hong}}, \bibinfo {author} {\bibfnamefont {Yong-Su}\
  \bibnamefont {Kim}}, \bibinfo {author} {\bibfnamefont {Yosep}\ \bibnamefont
  {Kim}}, \bibinfo {author} {\bibfnamefont {Seung-Woo}\ \bibnamefont {Lee}},
  \bibinfo {author} {\bibfnamefont {Raphael~C.}\ \bibnamefont {Pooser}},
  \bibinfo {author} {\bibfnamefont {Kyunghwan}\ \bibnamefont {Oh}}, \bibinfo
  {author} {\bibfnamefont {Su-Yong}\ \bibnamefont {Lee}}, \bibinfo {author}
  {\bibfnamefont {Changhyoup}\ \bibnamefont {Lee}}, \ and\ \bibinfo {author}
  {\bibfnamefont {Hyang-Tag}\ \bibnamefont {Lim}},\ }\bibfield  {title}
  {\enquote {\bibinfo {title} {Distributed quantum sensing of multiple phases
  with fewer photons},}\ }\href {\doibase 10.1038/s41467-023-44204-z}
  {\bibfield  {journal} {\bibinfo  {journal} {Nature Communications}\ }\textbf
  {\bibinfo {volume} {15}},\ \bibinfo {pages} {266} (\bibinfo {year}
  {2024})}\BibitemShut {NoStop}%
\bibitem [{\citenamefont {Liu}\ \emph {et~al.}(2021)\citenamefont {Liu},
  \citenamefont {Zhang}, \citenamefont {Li}, \citenamefont {Zhang},
  \citenamefont {Yin}, \citenamefont {Fei}, \citenamefont {Li}, \citenamefont
  {Liu}, \citenamefont {Xu}, \citenamefont {Chen},\ and\ \citenamefont
  {Pan}}]{liu2021distributedquantum}%
  \BibitemOpen
  \bibfield  {author} {\bibinfo {author} {\bibfnamefont {Li-Zheng}\
  \bibnamefont {Liu}}, \bibinfo {author} {\bibfnamefont {Yu-Zhe}\ \bibnamefont
  {Zhang}}, \bibinfo {author} {\bibfnamefont {Zheng-Da}\ \bibnamefont {Li}},
  \bibinfo {author} {\bibfnamefont {Rui}\ \bibnamefont {Zhang}}, \bibinfo
  {author} {\bibfnamefont {Xu-Fei}\ \bibnamefont {Yin}}, \bibinfo {author}
  {\bibfnamefont {Yue-Yang}\ \bibnamefont {Fei}}, \bibinfo {author}
  {\bibfnamefont {Li}~\bibnamefont {Li}}, \bibinfo {author} {\bibfnamefont
  {Nai-Le}\ \bibnamefont {Liu}}, \bibinfo {author} {\bibfnamefont {Feihu}\
  \bibnamefont {Xu}}, \bibinfo {author} {\bibfnamefont {Yu-Ao}\ \bibnamefont
  {Chen}}, \ and\ \bibinfo {author} {\bibfnamefont {Jian-Wei}\ \bibnamefont
  {Pan}},\ }\bibfield  {title} {\enquote {\bibinfo {title} {Distributed quantum
  phase estimation with entangled photons},}\ }\href {\doibase
  10.1038/s41566-020-00718-2} {\bibfield  {journal} {\bibinfo  {journal}
  {Nature Photonics}\ }\textbf {\bibinfo {volume} {15}},\ \bibinfo {pages}
  {137--142} (\bibinfo {year} {2021})}\BibitemShut {NoStop}%
\bibitem [{\citenamefont {Guo}\ \emph {et~al.}(2020)\citenamefont {Guo},
  \citenamefont {Breum}, \citenamefont {Borregaard}, \citenamefont {Izumi},
  \citenamefont {Larsen}, \citenamefont {Gehring}, \citenamefont {Christandl},
  \citenamefont {Neergaard-Nielsen},\ and\ \citenamefont
  {Andersen}}]{guo2020distributed}%
  \BibitemOpen
  \bibfield  {author} {\bibinfo {author} {\bibfnamefont {Xueshi}\ \bibnamefont
  {Guo}}, \bibinfo {author} {\bibfnamefont {Casper~R.}\ \bibnamefont {Breum}},
  \bibinfo {author} {\bibfnamefont {Johannes}\ \bibnamefont {Borregaard}},
  \bibinfo {author} {\bibfnamefont {Shuro}\ \bibnamefont {Izumi}}, \bibinfo
  {author} {\bibfnamefont {Mikkel~V.}\ \bibnamefont {Larsen}}, \bibinfo
  {author} {\bibfnamefont {Tobias}\ \bibnamefont {Gehring}}, \bibinfo {author}
  {\bibfnamefont {Matthias}\ \bibnamefont {Christandl}}, \bibinfo {author}
  {\bibfnamefont {Jonas~S.}\ \bibnamefont {Neergaard-Nielsen}}, \ and\ \bibinfo
  {author} {\bibfnamefont {Ulrik~L.}\ \bibnamefont {Andersen}},\ }\bibfield
  {title} {\enquote {\bibinfo {title} {Distributed quantum sensing in a
  continuous-variable entangled network},}\ }\href {\doibase
  10.1038/s41567-019-0743-x} {\bibfield  {journal} {\bibinfo  {journal} {Nature
  Physics}\ }\textbf {\bibinfo {volume} {16}},\ \bibinfo {pages} {281--284}
  (\bibinfo {year} {2020})}\BibitemShut {NoStop}%
\bibitem [{\citenamefont {Kim}\ \emph {et~al.}(2025)\citenamefont {Kim},
  \citenamefont {Hong}, \citenamefont {Kim}, \citenamefont {Oh}, \citenamefont
  {Lee}, \citenamefont {Lee},\ and\ \citenamefont {Lim}}]{kim2025distributed}%
  \BibitemOpen
  \bibfield  {author} {\bibinfo {author} {\bibfnamefont {Dong-Hyun}\
  \bibnamefont {Kim}}, \bibinfo {author} {\bibfnamefont {Seongjin}\
  \bibnamefont {Hong}}, \bibinfo {author} {\bibfnamefont {Yong-Su}\
  \bibnamefont {Kim}}, \bibinfo {author} {\bibfnamefont {Kyunghwan}\
  \bibnamefont {Oh}}, \bibinfo {author} {\bibfnamefont {Su-Yong}\ \bibnamefont
  {Lee}}, \bibinfo {author} {\bibfnamefont {Changhyoup}\ \bibnamefont {Lee}}, \
  and\ \bibinfo {author} {\bibfnamefont {Hyang-Tag}\ \bibnamefont {Lim}},\
  }\bibfield  {title} {\enquote {\bibinfo {title} {Distributed quantum sensing
  with multimode $n00n$ states},}\ }\href {\doibase 10.1103/4vdx-7224}
  {\bibfield  {journal} {\bibinfo  {journal} {Phys. Rev. Lett.}\ }\textbf
  {\bibinfo {volume} {135}},\ \bibinfo {pages} {050802} (\bibinfo {year}
  {2025})}\BibitemShut {NoStop}%
\bibitem [{\citenamefont {Xia}\ \emph {et~al.}(2020)\citenamefont {Xia},
  \citenamefont {Li}, \citenamefont {Clark}, \citenamefont {Hart},
  \citenamefont {Zhuang},\ and\ \citenamefont {Zhang}}]{xia2020demonstration}%
  \BibitemOpen
  \bibfield  {author} {\bibinfo {author} {\bibfnamefont {Yi}~\bibnamefont
  {Xia}}, \bibinfo {author} {\bibfnamefont {Wei}\ \bibnamefont {Li}}, \bibinfo
  {author} {\bibfnamefont {William}\ \bibnamefont {Clark}}, \bibinfo {author}
  {\bibfnamefont {Darlene}\ \bibnamefont {Hart}}, \bibinfo {author}
  {\bibfnamefont {Quntao}\ \bibnamefont {Zhuang}}, \ and\ \bibinfo {author}
  {\bibfnamefont {Zheshen}\ \bibnamefont {Zhang}},\ }\bibfield  {title}
  {\enquote {\bibinfo {title} {Demonstration of a reconfigurable entangled
  radio-frequency photonic sensor network},}\ }\href {\doibase
  10.1103/PhysRevLett.124.150502} {\bibfield  {journal} {\bibinfo  {journal}
  {Phys. Rev. Lett.}\ }\textbf {\bibinfo {volume} {124}},\ \bibinfo {pages}
  {150502} (\bibinfo {year} {2020})}\BibitemShut {NoStop}%
\bibitem [{\citenamefont {Hong}\ \emph {et~al.}(2021)\citenamefont {Hong},
  \citenamefont {ur~Rehman}, \citenamefont {Kim}, \citenamefont {Cho},
  \citenamefont {Lee}, \citenamefont {Jung}, \citenamefont {Moon},
  \citenamefont {Han},\ and\ \citenamefont {Lim}}]{hong2021quantumenhanced}%
  \BibitemOpen
  \bibfield  {author} {\bibinfo {author} {\bibfnamefont {Seongjin}\
  \bibnamefont {Hong}}, \bibinfo {author} {\bibfnamefont {Junaid}\ \bibnamefont
  {ur~Rehman}}, \bibinfo {author} {\bibfnamefont {Yong-Su}\ \bibnamefont
  {Kim}}, \bibinfo {author} {\bibfnamefont {Young-Wook}\ \bibnamefont {Cho}},
  \bibinfo {author} {\bibfnamefont {Seung-Woo}\ \bibnamefont {Lee}}, \bibinfo
  {author} {\bibfnamefont {Hojoong}\ \bibnamefont {Jung}}, \bibinfo {author}
  {\bibfnamefont {Sung}\ \bibnamefont {Moon}}, \bibinfo {author} {\bibfnamefont
  {Sang-Wook}\ \bibnamefont {Han}}, \ and\ \bibinfo {author} {\bibfnamefont
  {Hyang-Tag}\ \bibnamefont {Lim}},\ }\bibfield  {title} {\enquote {\bibinfo
  {title} {Quantum enhanced multiple-phase estimation with multi-mode n00n
  states},}\ }\href {\doibase 10.1038/s41467-021-25451-4} {\bibfield  {journal}
  {\bibinfo  {journal} {Nature Communications}\ }\textbf {\bibinfo {volume}
  {12}},\ \bibinfo {pages} {5211} (\bibinfo {year} {2021})}\BibitemShut
  {NoStop}%
\bibitem [{\citenamefont {Hong}\ \emph {et~al.}(2025)\citenamefont {Hong},
  \citenamefont {Feldman}, \citenamefont {Marvinney}, \citenamefont {Lee},
  \citenamefont {Lee}, \citenamefont {Febbraro}, \citenamefont {Marino},\ and\
  \citenamefont {Pooser}}]{hong2025quantum}%
  \BibitemOpen
  \bibfield  {author} {\bibinfo {author} {\bibfnamefont {Seongjin}\
  \bibnamefont {Hong}}, \bibinfo {author} {\bibfnamefont {Matthew~A.}\
  \bibnamefont {Feldman}}, \bibinfo {author} {\bibfnamefont {Claire~E.}\
  \bibnamefont {Marvinney}}, \bibinfo {author} {\bibfnamefont {Donghwa}\
  \bibnamefont {Lee}}, \bibinfo {author} {\bibfnamefont {Changhyoup}\
  \bibnamefont {Lee}}, \bibinfo {author} {\bibfnamefont {Michael~T.}\
  \bibnamefont {Febbraro}}, \bibinfo {author} {\bibfnamefont {Alberto~M.}\
  \bibnamefont {Marino}}, \ and\ \bibinfo {author} {\bibfnamefont {Raphael~C.}\
  \bibnamefont {Pooser}},\ }\bibfield  {title} {\enquote {\bibinfo {title}
  {Quantum-enhanced distributed phase sensing with a truncated su(1,1)
  interferometer},}\ }\href {\doibase 10.1103/PhysRevResearch.7.023231}
  {\bibfield  {journal} {\bibinfo  {journal} {Phys. Rev. Res.}\ }\textbf
  {\bibinfo {volume} {7}},\ \bibinfo {pages} {023231} (\bibinfo {year}
  {2025})}\BibitemShut {NoStop}%
\bibitem [{\citenamefont {Bate}\ \emph {et~al.}(2025)\citenamefont {Bate},
  \citenamefont {Hamann}, \citenamefont {Canteri}, \citenamefont {Winkler},
  \citenamefont {Koong}, \citenamefont {Krutyanskiy}, \citenamefont {Dür},\
  and\ \citenamefont {Lanyon}}]{bate2025experimentaldistributedquantumsensing}%
  \BibitemOpen
  \bibfield  {author} {\bibinfo {author} {\bibfnamefont {James}\ \bibnamefont
  {Bate}}, \bibinfo {author} {\bibfnamefont {Arne}\ \bibnamefont {Hamann}},
  \bibinfo {author} {\bibfnamefont {Marco}\ \bibnamefont {Canteri}}, \bibinfo
  {author} {\bibfnamefont {Armin}\ \bibnamefont {Winkler}}, \bibinfo {author}
  {\bibfnamefont {Zhe~Xian}\ \bibnamefont {Koong}}, \bibinfo {author}
  {\bibfnamefont {Victor}\ \bibnamefont {Krutyanskiy}}, \bibinfo {author}
  {\bibfnamefont {Wolfgang}\ \bibnamefont {Dür}}, \ and\ \bibinfo {author}
  {\bibfnamefont {Benjamin~Peter}\ \bibnamefont {Lanyon}},\ }\href
  {https://arxiv.org/abs/2501.08940} {\enquote {\bibinfo {title} {Experimental
  distributed quantum sensing in a noisy environment},}\ } (\bibinfo {year}
  {2025}),\ \Eprint {http://arxiv.org/abs/2501.08940} {arXiv:2501.08940
  [quant-ph]} \BibitemShut {NoStop}%
\bibitem [{\citenamefont {Polino}\ \emph {et~al.}(2019)\citenamefont {Polino},
  \citenamefont {Riva}, \citenamefont {Valeri}, \citenamefont {Silvestri},
  \citenamefont {Corrielli}, \citenamefont {Crespi}, \citenamefont {Spagnolo},
  \citenamefont {Osellame},\ and\ \citenamefont
  {Sciarrino}}]{polino2019femtosecond}%
  \BibitemOpen
  \bibfield  {author} {\bibinfo {author} {\bibfnamefont {Emanuele}\
  \bibnamefont {Polino}}, \bibinfo {author} {\bibfnamefont {Martina}\
  \bibnamefont {Riva}}, \bibinfo {author} {\bibfnamefont {Mauro}\ \bibnamefont
  {Valeri}}, \bibinfo {author} {\bibfnamefont {Raffaele}\ \bibnamefont
  {Silvestri}}, \bibinfo {author} {\bibfnamefont {Giacomo}\ \bibnamefont
  {Corrielli}}, \bibinfo {author} {\bibfnamefont {Andrea}\ \bibnamefont
  {Crespi}}, \bibinfo {author} {\bibfnamefont {Nicol\`{o}}\ \bibnamefont
  {Spagnolo}}, \bibinfo {author} {\bibfnamefont {Roberto}\ \bibnamefont
  {Osellame}}, \ and\ \bibinfo {author} {\bibfnamefont {Fabio}\ \bibnamefont
  {Sciarrino}},\ }\bibfield  {title} {\enquote {\bibinfo {title} {Experimental
  multiphase estimation on a chip},}\ }\href {\doibase 10.1364/OPTICA.6.000288}
  {\bibfield  {journal} {\bibinfo  {journal} {Optica}\ }\textbf {\bibinfo
  {volume} {6}},\ \bibinfo {pages} {288--295} (\bibinfo {year}
  {2019})}\BibitemShut {NoStop}%
\bibitem [{\citenamefont {Hassani}\ \emph {et~al.}(2025)\citenamefont
  {Hassani}, \citenamefont {Scheiner}, \citenamefont {Paris},\ and\
  \citenamefont {Markham}}]{hassani2025privacy}%
  \BibitemOpen
  \bibfield  {author} {\bibinfo {author} {\bibfnamefont {Majid}\ \bibnamefont
  {Hassani}}, \bibinfo {author} {\bibfnamefont {Santiago}\ \bibnamefont
  {Scheiner}}, \bibinfo {author} {\bibfnamefont {Matteo G.~A.}\ \bibnamefont
  {Paris}}, \ and\ \bibinfo {author} {\bibfnamefont {Damian}\ \bibnamefont
  {Markham}},\ }\bibfield  {title} {\enquote {\bibinfo {title} {Privacy in
  networks of quantum sensors},}\ }\href {\doibase
  10.1103/PhysRevLett.134.030802} {\bibfield  {journal} {\bibinfo  {journal}
  {Phys. Rev. Lett.}\ }\textbf {\bibinfo {volume} {134}},\ \bibinfo {pages}
  {030802} (\bibinfo {year} {2025})}\BibitemShut {NoStop}%
\bibitem [{\citenamefont {Bugalho}\ \emph {et~al.}(2025)\citenamefont
  {Bugalho}, \citenamefont {Hassani}, \citenamefont {Omar},\ and\ \citenamefont
  {Markham}}]{bugalho2025privaterobuststates}%
  \BibitemOpen
  \bibfield  {author} {\bibinfo {author} {\bibfnamefont {Lu{\'{i}}s}\
  \bibnamefont {Bugalho}}, \bibinfo {author} {\bibfnamefont {Majid}\
  \bibnamefont {Hassani}}, \bibinfo {author} {\bibfnamefont {Yasser}\
  \bibnamefont {Omar}}, \ and\ \bibinfo {author} {\bibfnamefont {Damian}\
  \bibnamefont {Markham}},\ }\bibfield  {title} {\enquote {\bibinfo {title}
  {Private and {R}obust {S}tates for {D}istributed {Q}uantum {S}ensing},}\
  }\href {\doibase 10.22331/q-2025-01-15-1596} {\bibfield  {journal} {\bibinfo
  {journal} {{Quantum}}\ }\textbf {\bibinfo {volume} {9}},\ \bibinfo {pages}
  {1596} (\bibinfo {year} {2025})}\BibitemShut {NoStop}%
\bibitem [{\citenamefont {Covey}\ \emph {et~al.}(2025)\citenamefont {Covey},
  \citenamefont {Pikovski},\ and\ \citenamefont
  {Borregaard}}]{covey2025probing}%
  \BibitemOpen
  \bibfield  {author} {\bibinfo {author} {\bibfnamefont {Jacob~P.}\
  \bibnamefont {Covey}}, \bibinfo {author} {\bibfnamefont {Igor}\ \bibnamefont
  {Pikovski}}, \ and\ \bibinfo {author} {\bibfnamefont {Johannes}\ \bibnamefont
  {Borregaard}},\ }\bibfield  {title} {\enquote {\bibinfo {title} {Probing
  curved spacetime with a distributed atomic processor clock},}\ }\href
  {\doibase 10.1103/q188-b1cr} {\bibfield  {journal} {\bibinfo  {journal} {PRX
  Quantum}\ }\textbf {\bibinfo {volume} {6}},\ \bibinfo {pages} {030310}
  (\bibinfo {year} {2025})}\BibitemShut {NoStop}%
\bibitem [{\citenamefont {Aspelmeyer}\ \emph {et~al.}(2014)\citenamefont
  {Aspelmeyer}, \citenamefont {Kippenberg},\ and\ \citenamefont
  {Marquardt}}]{aspelmeyer2014cavity}%
  \BibitemOpen
  \bibfield  {author} {\bibinfo {author} {\bibfnamefont {Markus}\ \bibnamefont
  {Aspelmeyer}}, \bibinfo {author} {\bibfnamefont {Tobias~J}\ \bibnamefont
  {Kippenberg}}, \ and\ \bibinfo {author} {\bibfnamefont {Florian}\
  \bibnamefont {Marquardt}},\ }\bibfield  {title} {\enquote {\bibinfo {title}
  {Cavity optomechanics},}\ }\href {https://doi.org/10.1103/RevModPhys.86.1391}
  {\bibfield  {journal} {\bibinfo  {journal} {Reviews of Modern Physics}\
  }\textbf {\bibinfo {volume} {86}},\ \bibinfo {pages} {1391--1452} (\bibinfo
  {year} {2014})}\BibitemShut {NoStop}%
\bibitem [{\citenamefont {Qvarfort}\ \emph {et~al.}(2018)\citenamefont
  {Qvarfort}, \citenamefont {Serafini}, \citenamefont {Barker},\ and\
  \citenamefont {Bose}}]{qvarfort2018gravimetry}%
  \BibitemOpen
  \bibfield  {author} {\bibinfo {author} {\bibfnamefont {Sofia}\ \bibnamefont
  {Qvarfort}}, \bibinfo {author} {\bibfnamefont {Alessio}\ \bibnamefont
  {Serafini}}, \bibinfo {author} {\bibfnamefont {P.~F.}\ \bibnamefont
  {Barker}}, \ and\ \bibinfo {author} {\bibfnamefont {Sougato}\ \bibnamefont
  {Bose}},\ }\bibfield  {title} {\enquote {\bibinfo {title} {Gravimetry through
  non-linear optomechanics},}\ }\href {\doibase 10.1038/s41467-018-06037-z}
  {\bibfield  {journal} {\bibinfo  {journal} {Nat. Commun.}\ }\textbf {\bibinfo
  {volume} {9}},\ \bibinfo {pages} {3690} (\bibinfo {year} {2018})}\BibitemShut
  {NoStop}%
\bibitem [{\citenamefont {Treutlein}\ \emph {et~al.}(2014)\citenamefont
  {Treutlein}, \citenamefont {Genes}, \citenamefont {Hammerer}, \citenamefont
  {Poggio},\ and\ \citenamefont {Rabl}}]{treutlein2014hybrid}%
  \BibitemOpen
  \bibfield  {author} {\bibinfo {author} {\bibfnamefont {Philipp}\ \bibnamefont
  {Treutlein}}, \bibinfo {author} {\bibfnamefont {Claudiu}\ \bibnamefont
  {Genes}}, \bibinfo {author} {\bibfnamefont {Klemens}\ \bibnamefont
  {Hammerer}}, \bibinfo {author} {\bibfnamefont {Martino}\ \bibnamefont
  {Poggio}}, \ and\ \bibinfo {author} {\bibfnamefont {Peter}\ \bibnamefont
  {Rabl}},\ }\enquote {\bibinfo {title} {Hybrid mechanical systems},}\ in\
  \href {\doibase 10.1007/978-3-642-55312-7_14} {\emph {\bibinfo {booktitle}
  {Cavity Optomechanics}}}\ (\bibinfo  {publisher} {Springer Berlin
  Heidelberg},\ \bibinfo {year} {2014})\ p.\ \bibinfo {pages}
  {327–351}\BibitemShut {NoStop}%
\bibitem [{\citenamefont {Rabl}\ \emph {et~al.}(2009)\citenamefont {Rabl},
  \citenamefont {Cappellaro}, \citenamefont {Dutt}, \citenamefont {Jiang},
  \citenamefont {Maze},\ and\ \citenamefont {Lukin}}]{rabl2009strong}%
  \BibitemOpen
  \bibfield  {author} {\bibinfo {author} {\bibfnamefont {P.}~\bibnamefont
  {Rabl}}, \bibinfo {author} {\bibfnamefont {P.}~\bibnamefont {Cappellaro}},
  \bibinfo {author} {\bibfnamefont {M.~V.~Gurudev}\ \bibnamefont {Dutt}},
  \bibinfo {author} {\bibfnamefont {L.}~\bibnamefont {Jiang}}, \bibinfo
  {author} {\bibfnamefont {J.~R.}\ \bibnamefont {Maze}}, \ and\ \bibinfo
  {author} {\bibfnamefont {M.~D.}\ \bibnamefont {Lukin}},\ }\bibfield  {title}
  {\enquote {\bibinfo {title} {Strong magnetic coupling between an electronic
  spin qubit and a mechanical resonator},}\ }\href {\doibase
  10.1103/PhysRevB.79.041302} {\bibfield  {journal} {\bibinfo  {journal} {Phys.
  Rev. B}\ }\textbf {\bibinfo {volume} {79}},\ \bibinfo {pages} {041302}
  (\bibinfo {year} {2009})}\BibitemShut {NoStop}%
\bibitem [{\citenamefont {Restrepo}\ \emph {et~al.}(2014)\citenamefont
  {Restrepo}, \citenamefont {Ciuti},\ and\ \citenamefont
  {Favero}}]{restrepo2014single}%
  \BibitemOpen
  \bibfield  {author} {\bibinfo {author} {\bibfnamefont {Juan}\ \bibnamefont
  {Restrepo}}, \bibinfo {author} {\bibfnamefont {Cristiano}\ \bibnamefont
  {Ciuti}}, \ and\ \bibinfo {author} {\bibfnamefont {Ivan}\ \bibnamefont
  {Favero}},\ }\bibfield  {title} {\enquote {\bibinfo {title} {Single-polariton
  optomechanics},}\ }\href {\doibase 10.1103/PhysRevLett.112.013601} {\bibfield
   {journal} {\bibinfo  {journal} {Physical Review Letters}\ }\textbf {\bibinfo
  {volume} {112}},\ \bibinfo {pages} {013601} (\bibinfo {year}
  {2014})}\BibitemShut {NoStop}%
\bibitem [{\citenamefont {Restrepo}\ \emph {et~al.}(2017)\citenamefont
  {Restrepo}, \citenamefont {Favero},\ and\ \citenamefont
  {Ciuti}}]{restrepo2017fully}%
  \BibitemOpen
  \bibfield  {author} {\bibinfo {author} {\bibfnamefont {Juan}\ \bibnamefont
  {Restrepo}}, \bibinfo {author} {\bibfnamefont {Ivan}\ \bibnamefont {Favero}},
  \ and\ \bibinfo {author} {\bibfnamefont {Cristiano}\ \bibnamefont {Ciuti}},\
  }\bibfield  {title} {\enquote {\bibinfo {title} {Fully coupled hybrid cavity
  optomechanics: Quantum interferences and correlations},}\ }\href {\doibase
  10.1103/PhysRevA.95.023832} {\bibfield  {journal} {\bibinfo  {journal} {Phys.
  Rev. A}\ }\textbf {\bibinfo {volume} {95}},\ \bibinfo {pages} {023832}
  (\bibinfo {year} {2017})}\BibitemShut {NoStop}%
\bibitem [{\citenamefont {Albarelli}\ \emph {et~al.}(2020)\citenamefont
  {Albarelli}, \citenamefont {Barbieri}, \citenamefont {Genoni},\ and\
  \citenamefont {Gianani}}]{albarelli2020perspective}%
  \BibitemOpen
  \bibfield  {author} {\bibinfo {author} {\bibfnamefont {Francesco}\
  \bibnamefont {Albarelli}}, \bibinfo {author} {\bibfnamefont {Marco}\
  \bibnamefont {Barbieri}}, \bibinfo {author} {\bibfnamefont {Marco~G}\
  \bibnamefont {Genoni}}, \ and\ \bibinfo {author} {\bibfnamefont {Ilaria}\
  \bibnamefont {Gianani}},\ }\bibfield  {title} {\enquote {\bibinfo {title} {A
  perspective on multiparameter quantum metrology: From theoretical tools to
  applications in quantum imaging},}\ }\href
  {https://doi.org/10.1016/j.physleta.2020.126311} {\bibfield  {journal}
  {\bibinfo  {journal} {Phys. Lett. A}\ }\textbf {\bibinfo {volume} {384}},\
  \bibinfo {pages} {126311} (\bibinfo {year} {2020})}\BibitemShut {NoStop}%
\bibitem [{\citenamefont {Meyer}(2021)}]{meyer2021fisher}%
  \BibitemOpen
  \bibfield  {author} {\bibinfo {author} {\bibfnamefont {Johannes~Jakob}\
  \bibnamefont {Meyer}},\ }\bibfield  {title} {\enquote {\bibinfo {title}
  {Fisher information in noisy intermediate-scale quantum applications},}\
  }\href {https://doi.org/10.22331/q-2021-09-09-539} {\bibfield  {journal}
  {\bibinfo  {journal} {Quantum}\ }\textbf {\bibinfo {volume} {5}},\ \bibinfo
  {pages} {539} (\bibinfo {year} {2021})}\BibitemShut {NoStop}%
\bibitem [{\citenamefont {Ragy}\ \emph {et~al.}(2016)\citenamefont {Ragy},
  \citenamefont {Jarzyna},\ and\ \citenamefont
  {Demkowicz-Dobrza{\'n}ski}}]{ragy2016compatibility}%
  \BibitemOpen
  \bibfield  {author} {\bibinfo {author} {\bibfnamefont {Sammy}\ \bibnamefont
  {Ragy}}, \bibinfo {author} {\bibfnamefont {Marcin}\ \bibnamefont {Jarzyna}},
  \ and\ \bibinfo {author} {\bibfnamefont {Rafa{\l}}\ \bibnamefont
  {Demkowicz-Dobrza{\'n}ski}},\ }\bibfield  {title} {\enquote {\bibinfo {title}
  {Compatibility in multiparameter quantum metrology},}\ }\href {\doibase
  10.1103/PhysRevA.94.052108} {\bibfield  {journal} {\bibinfo  {journal}
  {Physical Review A}\ }\textbf {\bibinfo {volume} {94}},\ \bibinfo {pages}
  {052108} (\bibinfo {year} {2016})}\BibitemShut {NoStop}%
\bibitem [{\citenamefont {Albarelli}\ \emph {et~al.}(2019)\citenamefont
  {Albarelli}, \citenamefont {Friel},\ and\ \citenamefont
  {Datta}}]{albarelli2019evaluating}%
  \BibitemOpen
  \bibfield  {author} {\bibinfo {author} {\bibfnamefont {Francesco}\
  \bibnamefont {Albarelli}}, \bibinfo {author} {\bibfnamefont {Jamie~F.}\
  \bibnamefont {Friel}}, \ and\ \bibinfo {author} {\bibfnamefont {Animesh}\
  \bibnamefont {Datta}},\ }\bibfield  {title} {\enquote {\bibinfo {title}
  {Evaluating the holevo cram{\'e}r–rao bound for multiparameter quantum
  metrology},}\ }\href {\doibase 10.1103/PhysRevLett.123.200503} {\bibfield
  {journal} {\bibinfo  {journal} {Physical Review Letters}\ }\textbf {\bibinfo
  {volume} {123}},\ \bibinfo {pages} {200503} (\bibinfo {year}
  {2019})}\BibitemShut {NoStop}%
\bibitem [{\citenamefont {Szczykulska}\ \emph {et~al.}(2016)\citenamefont
  {Szczykulska}, \citenamefont {Baumgratz},\ and\ \citenamefont
  {Datta}}]{szczykulska2016multiparameter}%
  \BibitemOpen
  \bibfield  {author} {\bibinfo {author} {\bibfnamefont {Magdalena}\
  \bibnamefont {Szczykulska}}, \bibinfo {author} {\bibfnamefont {Tillmann}\
  \bibnamefont {Baumgratz}}, \ and\ \bibinfo {author} {\bibfnamefont {Animesh}\
  \bibnamefont {Datta}},\ }\bibfield  {title} {\enquote {\bibinfo {title}
  {Multi-parameter quantum metrology},}\ }\href {\doibase
  10.1080/23746149.2016.1230476} {\bibfield  {journal} {\bibinfo  {journal}
  {Advances in Physics: X}\ }\textbf {\bibinfo {volume} {1}},\ \bibinfo {pages}
  {621--639} (\bibinfo {year} {2016})}\BibitemShut {NoStop}%
\bibitem [{\citenamefont {Bose}\ \emph {et~al.}(1997)\citenamefont {Bose},
  \citenamefont {Jacobs},\ and\ \citenamefont {Knight}}]{bose1997preparation}%
  \BibitemOpen
  \bibfield  {author} {\bibinfo {author} {\bibfnamefont {S.}~\bibnamefont
  {Bose}}, \bibinfo {author} {\bibfnamefont {K.}~\bibnamefont {Jacobs}}, \ and\
  \bibinfo {author} {\bibfnamefont {P.~L.}\ \bibnamefont {Knight}},\ }\bibfield
   {title} {\enquote {\bibinfo {title} {Preparation of nonclassical states in
  cavities with a moving mirror},}\ }\href {\doibase 10.1103/PhysRevA.56.4175}
  {\bibfield  {journal} {\bibinfo  {journal} {Phys. Rev. A}\ }\textbf {\bibinfo
  {volume} {56}},\ \bibinfo {pages} {4175--4186} (\bibinfo {year}
  {1997})}\BibitemShut {NoStop}%
\bibitem [{\citenamefont {Qvarfort}\ and\ \citenamefont
  {Pikovski}(2025)}]{qvarfort2025solving}%
  \BibitemOpen
  \bibfield  {author} {\bibinfo {author} {\bibfnamefont {Sofia}\ \bibnamefont
  {Qvarfort}}\ and\ \bibinfo {author} {\bibfnamefont {Igor}\ \bibnamefont
  {Pikovski}},\ }\bibfield  {title} {\enquote {\bibinfo {title} {Solving
  quantum dynamics with a lie-algebra decoupling method},}\ }\href {\doibase
  10.1103/PRXQuantum.6.010201} {\bibfield  {journal} {\bibinfo  {journal} {PRX
  Quantum}\ }\textbf {\bibinfo {volume} {6}},\ \bibinfo {pages} {010201}
  (\bibinfo {year} {2025})}\BibitemShut {NoStop}%
\bibitem [{\citenamefont {Maleki}\ and\ \citenamefont
  {Zubairy}(2022)}]{maleki2022distributed}%
  \BibitemOpen
  \bibfield  {author} {\bibinfo {author} {\bibfnamefont {Yusef}\ \bibnamefont
  {Maleki}}\ and\ \bibinfo {author} {\bibfnamefont {M.~Suhail}\ \bibnamefont
  {Zubairy}},\ }\bibfield  {title} {\enquote {\bibinfo {title} {Distributed
  phase estimation and networked quantum sensors with $w$-type quantum
  probes},}\ }\href {\doibase 10.1103/PhysRevA.105.032428} {\bibfield
  {journal} {\bibinfo  {journal} {Phys. Rev. A}\ }\textbf {\bibinfo {volume}
  {105}},\ \bibinfo {pages} {032428} (\bibinfo {year} {2022})}\BibitemShut
  {NoStop}%
\bibitem [{\citenamefont {D{"u}r}\ \emph {et~al.}(2000)\citenamefont {D{"u}r},
  \citenamefont {Vidal},\ and\ \citenamefont {Cirac}}]{dur2000three}%
  \BibitemOpen
  \bibfield  {author} {\bibinfo {author} {\bibfnamefont {Wolfgang}\
  \bibnamefont {D{"u}r}}, \bibinfo {author} {\bibfnamefont {Guifr{'e}}\
  \bibnamefont {Vidal}}, \ and\ \bibinfo {author} {\bibfnamefont {J.~I.}\
  \bibnamefont {Cirac}},\ }\bibfield  {title} {\enquote {\bibinfo {title}
  {Three qubits can be entangled in two inequivalent ways},}\ }\href {\doibase
  10.1103/PhysRevA.62.062314} {\bibfield  {journal} {\bibinfo  {journal}
  {Physical Review A}\ }\textbf {\bibinfo {volume} {62}},\ \bibinfo {pages}
  {062314} (\bibinfo {year} {2000})}\BibitemShut {NoStop}%
\bibitem [{\citenamefont {Cao}\ and\ \citenamefont
  {Yang}(2003)}]{cao2003entanglement}%
  \BibitemOpen
  \bibfield  {author} {\bibinfo {author} {\bibfnamefont {Zhuo-Liang}\
  \bibnamefont {Cao}}\ and\ \bibinfo {author} {\bibfnamefont {Ming}\
  \bibnamefont {Yang}},\ }\bibfield  {title} {\enquote {\bibinfo {title}
  {Entanglement distillation for three-particle w class states},}\ }\href
  {\doibase 10.1088/0953-4075/36/21/005} {\bibfield  {journal} {\bibinfo
  {journal} {Journal of Physics B: Atomic, Molecular and Optical Physics}\
  }\textbf {\bibinfo {volume} {36}},\ \bibinfo {pages} {4245} (\bibinfo {year}
  {2003})}\BibitemShut {NoStop}%
\bibitem [{\citenamefont {O'Connell}\ \emph {et~al.}(2010)\citenamefont
  {O'Connell}, \citenamefont {Hofheinz}, \citenamefont {Ansmann}, \citenamefont
  {Bialczak}, \citenamefont {Lenander}, \citenamefont {Lucero}, \citenamefont
  {Neeley}, \citenamefont {Sank}, \citenamefont {Wang}, \citenamefont {Weides},
  \citenamefont {Wenner}, \citenamefont {Martinis},\ and\ \citenamefont
  {Cleland}}]{oconnell2010quantum}%
  \BibitemOpen
  \bibfield  {author} {\bibinfo {author} {\bibfnamefont {A.~D.}\ \bibnamefont
  {O'Connell}}, \bibinfo {author} {\bibfnamefont {M.}~\bibnamefont {Hofheinz}},
  \bibinfo {author} {\bibfnamefont {M.}~\bibnamefont {Ansmann}}, \bibinfo
  {author} {\bibfnamefont {Radoslaw~C.}\ \bibnamefont {Bialczak}}, \bibinfo
  {author} {\bibfnamefont {M.}~\bibnamefont {Lenander}}, \bibinfo {author}
  {\bibfnamefont {Erik}\ \bibnamefont {Lucero}}, \bibinfo {author}
  {\bibfnamefont {M.}~\bibnamefont {Neeley}}, \bibinfo {author} {\bibfnamefont
  {D.}~\bibnamefont {Sank}}, \bibinfo {author} {\bibfnamefont {H.}~\bibnamefont
  {Wang}}, \bibinfo {author} {\bibfnamefont {M.}~\bibnamefont {Weides}},
  \bibinfo {author} {\bibfnamefont {J.}~\bibnamefont {Wenner}}, \bibinfo
  {author} {\bibfnamefont {John~M.}\ \bibnamefont {Martinis}}, \ and\ \bibinfo
  {author} {\bibfnamefont {A.~N.}\ \bibnamefont {Cleland}},\ }\bibfield
  {title} {\enquote {\bibinfo {title} {Quantum ground state and single-phonon
  control of a mechanical resonator},}\ }\href {\doibase 10.1038/nature08967}
  {\bibfield  {journal} {\bibinfo  {journal} {Nature}\ }\textbf {\bibinfo
  {volume} {464}},\ \bibinfo {pages} {697--703} (\bibinfo {year}
  {2010})}\BibitemShut {NoStop}%
\bibitem [{\citenamefont {Orszag}(2024)}]{orszag2024qoptics}%
  \BibitemOpen
  \bibfield  {author} {\bibinfo {author} {\bibfnamefont {Miguel}\ \bibnamefont
  {Orszag}},\ }\href@noop {} {\emph {\bibinfo {title} {Quantum Optics}}},\
  \bibinfo {edition} {4th}\ ed.\ (\bibinfo  {publisher} {Springer Nature
  Switzerland},\ \bibinfo {year} {2024})\BibitemShut {NoStop}%
\bibitem [{\citenamefont {Armata}\ \emph {et~al.}(2017)\citenamefont {Armata},
  \citenamefont {Latmiral}, \citenamefont {Plato},\ and\ \citenamefont
  {Kim}}]{armata2017quantum}%
  \BibitemOpen
  \bibfield  {author} {\bibinfo {author} {\bibfnamefont {F.}~\bibnamefont
  {Armata}}, \bibinfo {author} {\bibfnamefont {L.}~\bibnamefont {Latmiral}},
  \bibinfo {author} {\bibfnamefont {A.~D.~K.}\ \bibnamefont {Plato}}, \ and\
  \bibinfo {author} {\bibfnamefont {M.~S.}\ \bibnamefont {Kim}},\ }\bibfield
  {title} {\enquote {\bibinfo {title} {Quantum limits to gravity estimation
  with optomechanics},}\ }\href {\doibase 10.1103/PhysRevA.96.043824}
  {\bibfield  {journal} {\bibinfo  {journal} {Phys. Rev. A}\ }\textbf {\bibinfo
  {volume} {96}},\ \bibinfo {pages} {043824} (\bibinfo {year}
  {2017})}\BibitemShut {NoStop}%
\bibitem [{\citenamefont {Montenegro}\ \emph {et~al.}(2019)\citenamefont
  {Montenegro}, \citenamefont {Ferraro},\ and\ \citenamefont
  {Bose}}]{montenegro2019enabling}%
  \BibitemOpen
  \bibfield  {author} {\bibinfo {author} {\bibfnamefont {Victor}\ \bibnamefont
  {Montenegro}}, \bibinfo {author} {\bibfnamefont {Alessandro}\ \bibnamefont
  {Ferraro}}, \ and\ \bibinfo {author} {\bibfnamefont {Sougato}\ \bibnamefont
  {Bose}},\ }\bibfield  {title} {\enquote {\bibinfo {title} {Enabling
  entanglement distillation via optomechanics},}\ }\href {\doibase
  10.1103/PhysRevA.100.042310} {\bibfield  {journal} {\bibinfo  {journal}
  {Phys. Rev. A}\ }\textbf {\bibinfo {volume} {100}},\ \bibinfo {pages}
  {042310} (\bibinfo {year} {2019})}\BibitemShut {NoStop}%
\bibitem [{\citenamefont {de~Moraes~Neto}\ \emph {et~al.}(2016)\citenamefont
  {de~Moraes~Neto}, \citenamefont {Andrade}, \citenamefont {Montenegro},\ and\
  \citenamefont {Bose}}]{neto2016quantum}%
  \BibitemOpen
  \bibfield  {author} {\bibinfo {author} {\bibfnamefont {G.~D.}\ \bibnamefont
  {de~Moraes~Neto}}, \bibinfo {author} {\bibfnamefont {F.~M.}\ \bibnamefont
  {Andrade}}, \bibinfo {author} {\bibfnamefont {V.}~\bibnamefont {Montenegro}},
  \ and\ \bibinfo {author} {\bibfnamefont {S.}~\bibnamefont {Bose}},\
  }\bibfield  {title} {\enquote {\bibinfo {title} {Quantum state transfer in
  optomechanical arrays},}\ }\href {\doibase 10.1103/PhysRevA.93.062339}
  {\bibfield  {journal} {\bibinfo  {journal} {Phys. Rev. A}\ }\textbf {\bibinfo
  {volume} {93}},\ \bibinfo {pages} {062339} (\bibinfo {year}
  {2016})}\BibitemShut {NoStop}%
\bibitem [{\citenamefont {Pontin}\ \emph {et~al.}(2018)\citenamefont {Pontin},
  \citenamefont {Mourounas}, \citenamefont {Geraci},\ and\ \citenamefont
  {Barker}}]{pontin2018levitated}%
  \BibitemOpen
  \bibfield  {author} {\bibinfo {author} {\bibfnamefont {A}~\bibnamefont
  {Pontin}}, \bibinfo {author} {\bibfnamefont {L~S}\ \bibnamefont {Mourounas}},
  \bibinfo {author} {\bibfnamefont {A~A}\ \bibnamefont {Geraci}}, \ and\
  \bibinfo {author} {\bibfnamefont {P~F}\ \bibnamefont {Barker}},\ }\bibfield
  {title} {\enquote {\bibinfo {title} {Levitated optomechanics with a fiber
  fabry–perot interferometer},}\ }\href {\doibase 10.1088/1367-2630/aaa71c}
  {\bibfield  {journal} {\bibinfo  {journal} {New Journal of Physics}\ }\textbf
  {\bibinfo {volume} {20}},\ \bibinfo {pages} {023017} (\bibinfo {year}
  {2018})}\BibitemShut {NoStop}%
\bibitem [{\citenamefont {Fonseca}\ \emph {et~al.}(2016)\citenamefont
  {Fonseca}, \citenamefont {Aranas}, \citenamefont {Millen}, \citenamefont
  {Monteiro},\ and\ \citenamefont {Barker}}]{fonseca2016nonlinear}%
  \BibitemOpen
  \bibfield  {author} {\bibinfo {author} {\bibfnamefont {P.~Z.~G.}\
  \bibnamefont {Fonseca}}, \bibinfo {author} {\bibfnamefont {E.~B.}\
  \bibnamefont {Aranas}}, \bibinfo {author} {\bibfnamefont {J.}~\bibnamefont
  {Millen}}, \bibinfo {author} {\bibfnamefont {T.~S.}\ \bibnamefont
  {Monteiro}}, \ and\ \bibinfo {author} {\bibfnamefont {P.~F.}\ \bibnamefont
  {Barker}},\ }\bibfield  {title} {\enquote {\bibinfo {title} {Nonlinear
  dynamics and strong cavity cooling of levitated nanoparticles},}\ }\href
  {\doibase 10.1103/PhysRevLett.117.173602} {\bibfield  {journal} {\bibinfo
  {journal} {Phys. Rev. Lett.}\ }\textbf {\bibinfo {volume} {117}},\ \bibinfo
  {pages} {173602} (\bibinfo {year} {2016})}\BibitemShut {NoStop}%
\bibitem [{\citenamefont {Millen}\ \emph {et~al.}(2015)\citenamefont {Millen},
  \citenamefont {Fonseca}, \citenamefont {Mavrogordatos}, \citenamefont
  {Monteiro},\ and\ \citenamefont {Barker}}]{millen2015cavity}%
  \BibitemOpen
  \bibfield  {author} {\bibinfo {author} {\bibfnamefont {J.}~\bibnamefont
  {Millen}}, \bibinfo {author} {\bibfnamefont {P.~Z.~G.}\ \bibnamefont
  {Fonseca}}, \bibinfo {author} {\bibfnamefont {T.}~\bibnamefont
  {Mavrogordatos}}, \bibinfo {author} {\bibfnamefont {T.~S.}\ \bibnamefont
  {Monteiro}}, \ and\ \bibinfo {author} {\bibfnamefont {P.~F.}\ \bibnamefont
  {Barker}},\ }\bibfield  {title} {\enquote {\bibinfo {title} {Cavity cooling a
  single charged levitated nanosphere},}\ }\href {\doibase
  10.1103/PhysRevLett.114.123602} {\bibfield  {journal} {\bibinfo  {journal}
  {Phys. Rev. Lett.}\ }\textbf {\bibinfo {volume} {114}},\ \bibinfo {pages}
  {123602} (\bibinfo {year} {2015})}\BibitemShut {NoStop}%
\bibitem [{\citenamefont {Brennecke}\ \emph {et~al.}(2008)\citenamefont
  {Brennecke}, \citenamefont {Ritter}, \citenamefont {Donner},\ and\
  \citenamefont {Esslinger}}]{ferdinand2008cavity}%
  \BibitemOpen
  \bibfield  {author} {\bibinfo {author} {\bibfnamefont {Ferdinand}\
  \bibnamefont {Brennecke}}, \bibinfo {author} {\bibfnamefont {Stephan}\
  \bibnamefont {Ritter}}, \bibinfo {author} {\bibfnamefont {Tobias}\
  \bibnamefont {Donner}}, \ and\ \bibinfo {author} {\bibfnamefont {Tilman}\
  \bibnamefont {Esslinger}},\ }\bibfield  {title} {\enquote {\bibinfo {title}
  {Cavity optomechanics with a bose-einstein condensate},}\ }\href {\doibase
  10.1126/science.1163218} {\bibfield  {journal} {\bibinfo  {journal}
  {Science}\ }\textbf {\bibinfo {volume} {322}},\ \bibinfo {pages} {235--238}
  (\bibinfo {year} {2008})}\BibitemShut {NoStop}%
\bibitem [{\citenamefont {Matsumoto}(2002)}]{matsumoto2002approach}%
  \BibitemOpen
  \bibfield  {author} {\bibinfo {author} {\bibfnamefont {Keiji}\ \bibnamefont
  {Matsumoto}},\ }\bibfield  {title} {\enquote {\bibinfo {title} {A new
  approach to the cram{\'e}r--rao-type bound of the pure-state model},}\ }\href
  {\doibase 10.1088/0305-4470/35/13/307} {\bibfield  {journal} {\bibinfo
  {journal} {Journal of Physics A: Mathematical and General}\ }\textbf
  {\bibinfo {volume} {35}},\ \bibinfo {pages} {3111--3124} (\bibinfo {year}
  {2002})}\BibitemShut {NoStop}%
\bibitem [{\citenamefont {Pezz\`e}\ \emph {et~al.}(2018)\citenamefont
  {Pezz\`e}, \citenamefont {Smerzi}, \citenamefont {Oberthaler}, \citenamefont
  {Schmied},\ and\ \citenamefont {Treutlein}}]{pezze2018quantummetrology}%
  \BibitemOpen
  \bibfield  {author} {\bibinfo {author} {\bibfnamefont {Luca}\ \bibnamefont
  {Pezz\`e}}, \bibinfo {author} {\bibfnamefont {Augusto}\ \bibnamefont
  {Smerzi}}, \bibinfo {author} {\bibfnamefont {Markus~K.}\ \bibnamefont
  {Oberthaler}}, \bibinfo {author} {\bibfnamefont {Roman}\ \bibnamefont
  {Schmied}}, \ and\ \bibinfo {author} {\bibfnamefont {Philipp}\ \bibnamefont
  {Treutlein}},\ }\bibfield  {title} {\enquote {\bibinfo {title} {Quantum
  metrology with nonclassical states of atomic ensembles},}\ }\href {\doibase
  10.1103/RevModPhys.90.035005} {\bibfield  {journal} {\bibinfo  {journal}
  {Reviews of Modern Physics}\ }\textbf {\bibinfo {volume} {90}},\ \bibinfo
  {pages} {035005} (\bibinfo {year} {2018})}\BibitemShut {NoStop}%
\bibitem [{\citenamefont {Mukhopadhyay}\ \emph
  {et~al.}(2025{\natexlab{a}})\citenamefont {Mukhopadhyay}, \citenamefont
  {Montenegro},\ and\ \citenamefont {Bayat}}]{mukhopadhyay2025current}%
  \BibitemOpen
  \bibfield  {author} {\bibinfo {author} {\bibfnamefont {Chiranjib}\
  \bibnamefont {Mukhopadhyay}}, \bibinfo {author} {\bibfnamefont {Victor}\
  \bibnamefont {Montenegro}}, \ and\ \bibinfo {author} {\bibfnamefont
  {Abolfazl}\ \bibnamefont {Bayat}},\ }\bibfield  {title} {\enquote {\bibinfo
  {title} {Current trends in global quantum metrology},}\ }\href {\doibase
  10.1088/1751-8121/adb112} {\bibfield  {journal} {\bibinfo  {journal} {Journal
  of Physics A: Mathematical and Theoretical}\ }\textbf {\bibinfo {volume}
  {58}},\ \bibinfo {pages} {063001} (\bibinfo {year}
  {2025}{\natexlab{a}})}\BibitemShut {NoStop}%
\bibitem [{\citenamefont {Montenegro}\ \emph {et~al.}(2021)\citenamefont
  {Montenegro}, \citenamefont {Mishra},\ and\ \citenamefont
  {Bayat}}]{montenegro2021global}%
  \BibitemOpen
  \bibfield  {author} {\bibinfo {author} {\bibfnamefont {Victor}\ \bibnamefont
  {Montenegro}}, \bibinfo {author} {\bibfnamefont {Utkarsh}\ \bibnamefont
  {Mishra}}, \ and\ \bibinfo {author} {\bibfnamefont {Abolfazl}\ \bibnamefont
  {Bayat}},\ }\bibfield  {title} {\enquote {\bibinfo {title} {Global sensing
  and its impact for quantum many-body probes with criticality},}\ }\href
  {\doibase 10.1103/PhysRevLett.126.200501} {\bibfield  {journal} {\bibinfo
  {journal} {Phys. Rev. Lett.}\ }\textbf {\bibinfo {volume} {126}},\ \bibinfo
  {pages} {200501} (\bibinfo {year} {2021})}\BibitemShut {NoStop}%
\bibitem [{\citenamefont {Khosla}\ \emph {et~al.}(2018)\citenamefont {Khosla},
  \citenamefont {Vanner}, \citenamefont {Ares},\ and\ \citenamefont
  {Laird}}]{khosla2018displacemon}%
  \BibitemOpen
  \bibfield  {author} {\bibinfo {author} {\bibfnamefont {K.~E.}\ \bibnamefont
  {Khosla}}, \bibinfo {author} {\bibfnamefont {M.~R.}\ \bibnamefont {Vanner}},
  \bibinfo {author} {\bibfnamefont {N.}~\bibnamefont {Ares}}, \ and\ \bibinfo
  {author} {\bibfnamefont {E.~A.}\ \bibnamefont {Laird}},\ }\bibfield  {title}
  {\enquote {\bibinfo {title} {Displacemon electromechanics: How to detect
  quantum interference in a nanomechanical resonator},}\ }\href {\doibase
  10.1103/PhysRevX.8.021052} {\bibfield  {journal} {\bibinfo  {journal} {Phys.
  Rev. X}\ }\textbf {\bibinfo {volume} {8}},\ \bibinfo {pages} {021052}
  (\bibinfo {year} {2018})}\BibitemShut {NoStop}%
\bibitem [{\citenamefont {Rabl}\ \emph {et~al.}(2010)\citenamefont {Rabl},
  \citenamefont {Kolkowitz}, \citenamefont {Koppens}, \citenamefont {Harris},
  \citenamefont {Zoller},\ and\ \citenamefont {Lukin}}]{rabl2010quantum}%
  \BibitemOpen
  \bibfield  {author} {\bibinfo {author} {\bibfnamefont {P.}~\bibnamefont
  {Rabl}}, \bibinfo {author} {\bibfnamefont {S.~J.}\ \bibnamefont {Kolkowitz}},
  \bibinfo {author} {\bibfnamefont {F.~H.~L.}\ \bibnamefont {Koppens}},
  \bibinfo {author} {\bibfnamefont {J.~G.~E.}\ \bibnamefont {Harris}}, \bibinfo
  {author} {\bibfnamefont {P.}~\bibnamefont {Zoller}}, \ and\ \bibinfo {author}
  {\bibfnamefont {M.~D.}\ \bibnamefont {Lukin}},\ }\bibfield  {title} {\enquote
  {\bibinfo {title} {A quantum spin transducer based on nanoelectromechanical
  resonator arrays},}\ }\href {\doibase 10.1038/nphys1679} {\bibfield
  {journal} {\bibinfo  {journal} {Nature Physics}\ }\textbf {\bibinfo {volume}
  {6}},\ \bibinfo {pages} {602–608} (\bibinfo {year} {2010})}\BibitemShut
  {NoStop}%
\bibitem [{\citenamefont {Rabl}(2010)}]{rabl2010cooling}%
  \BibitemOpen
  \bibfield  {author} {\bibinfo {author} {\bibfnamefont {P.}~\bibnamefont
  {Rabl}},\ }\bibfield  {title} {\enquote {\bibinfo {title} {Cooling of
  mechanical motion with a two-level system: The high-temperature regime},}\
  }\href {\doibase 10.1103/PhysRevB.82.165320} {\bibfield  {journal} {\bibinfo
  {journal} {Phys. Rev. B}\ }\textbf {\bibinfo {volume} {82}},\ \bibinfo
  {pages} {165320} (\bibinfo {year} {2010})}\BibitemShut {NoStop}%
\bibitem [{\citenamefont {Braccini}\ \emph {et~al.}(2023)\citenamefont
  {Braccini}, \citenamefont {Schut}, \citenamefont {Serafini}, \citenamefont
  {Mazumdar},\ and\ \citenamefont {Bose}}]{braccini2023large}%
  \BibitemOpen
  \bibfield  {author} {\bibinfo {author} {\bibfnamefont {Lorenzo}\ \bibnamefont
  {Braccini}}, \bibinfo {author} {\bibfnamefont {Martine}\ \bibnamefont
  {Schut}}, \bibinfo {author} {\bibfnamefont {Alessio}\ \bibnamefont
  {Serafini}}, \bibinfo {author} {\bibfnamefont {Anupam}\ \bibnamefont
  {Mazumdar}}, \ and\ \bibinfo {author} {\bibfnamefont {Sougato}\ \bibnamefont
  {Bose}},\ }\href@noop {} {\enquote {\bibinfo {title} {Large spin
  stern-gerlach interferometry for gravitational entanglement},}\ } (\bibinfo
  {year} {2023}),\ \Eprint {http://arxiv.org/abs/2312.05170} {arXiv:2312.05170
  [quant-ph]} \BibitemShut {NoStop}%
\bibitem [{\citenamefont {Rao}\ \emph {et~al.}(2016)\citenamefont {Rao},
  \citenamefont {Momenzadeh},\ and\ \citenamefont
  {Wrachtrup}}]{rao2016heralded}%
  \BibitemOpen
  \bibfield  {author} {\bibinfo {author} {\bibfnamefont {D.~D.~Bhaktavatsala}\
  \bibnamefont {Rao}}, \bibinfo {author} {\bibfnamefont {S.~Ali}\ \bibnamefont
  {Momenzadeh}}, \ and\ \bibinfo {author} {\bibfnamefont {J\"org}\ \bibnamefont
  {Wrachtrup}},\ }\bibfield  {title} {\enquote {\bibinfo {title} {Heralded
  control of mechanical motion by single spins},}\ }\href {\doibase
  10.1103/PhysRevLett.117.077203} {\bibfield  {journal} {\bibinfo  {journal}
  {Phys. Rev. Lett.}\ }\textbf {\bibinfo {volume} {117}},\ \bibinfo {pages}
  {077203} (\bibinfo {year} {2016})}\BibitemShut {NoStop}%
\bibitem [{\citenamefont {Montenegro}\ \emph {et~al.}(2017)\citenamefont
  {Montenegro}, \citenamefont {Coto}, \citenamefont {Eremeev},\ and\
  \citenamefont {Orszag}}]{montenegro2017macroscopic}%
  \BibitemOpen
  \bibfield  {author} {\bibinfo {author} {\bibfnamefont {V\'{\i}ctor}\
  \bibnamefont {Montenegro}}, \bibinfo {author} {\bibfnamefont {Ra\'ul}\
  \bibnamefont {Coto}}, \bibinfo {author} {\bibfnamefont {Vitalie}\
  \bibnamefont {Eremeev}}, \ and\ \bibinfo {author} {\bibfnamefont {Miguel}\
  \bibnamefont {Orszag}},\ }\bibfield  {title} {\enquote {\bibinfo {title}
  {Macroscopic nonclassical-state preparation via postselection},}\ }\href
  {\doibase 10.1103/PhysRevA.96.053851} {\bibfield  {journal} {\bibinfo
  {journal} {Phys. Rev. A}\ }\textbf {\bibinfo {volume} {96}},\ \bibinfo
  {pages} {053851} (\bibinfo {year} {2017})}\BibitemShut {NoStop}%
\bibitem [{\citenamefont {Montenegro}\ \emph {et~al.}(2018)\citenamefont
  {Montenegro}, \citenamefont {Coto}, \citenamefont {Eremeev},\ and\
  \citenamefont {Orszag}}]{montenegro2018ground}%
  \BibitemOpen
  \bibfield  {author} {\bibinfo {author} {\bibfnamefont {V\'{\i}ctor}\
  \bibnamefont {Montenegro}}, \bibinfo {author} {\bibfnamefont {Ra\'ul}\
  \bibnamefont {Coto}}, \bibinfo {author} {\bibfnamefont {Vitalie}\
  \bibnamefont {Eremeev}}, \ and\ \bibinfo {author} {\bibfnamefont {Miguel}\
  \bibnamefont {Orszag}},\ }\bibfield  {title} {\enquote {\bibinfo {title}
  {Ground-state cooling of a nanomechanical oscillator with $n$ spins},}\
  }\href {\doibase 10.1103/PhysRevA.98.053837} {\bibfield  {journal} {\bibinfo
  {journal} {Phys. Rev. A}\ }\textbf {\bibinfo {volume} {98}},\ \bibinfo
  {pages} {053837} (\bibinfo {year} {2018})}\BibitemShut {NoStop}%
\bibitem [{\citenamefont {Tufarelli}\ \emph {et~al.}(2011)\citenamefont
  {Tufarelli}, \citenamefont {Kim},\ and\ \citenamefont
  {Bose}}]{tufarelli2011oscillator}%
  \BibitemOpen
  \bibfield  {author} {\bibinfo {author} {\bibfnamefont {Tommaso}\ \bibnamefont
  {Tufarelli}}, \bibinfo {author} {\bibfnamefont {M.~S.}\ \bibnamefont {Kim}},
  \ and\ \bibinfo {author} {\bibfnamefont {Sougato}\ \bibnamefont {Bose}},\
  }\bibfield  {title} {\enquote {\bibinfo {title} {Oscillator state
  reconstruction via tunable qubit coupling in markovian environments},}\
  }\href {\doibase 10.1103/PhysRevA.83.062120} {\bibfield  {journal} {\bibinfo
  {journal} {Phys. Rev. A}\ }\textbf {\bibinfo {volume} {83}},\ \bibinfo
  {pages} {062120} (\bibinfo {year} {2011})}\BibitemShut {NoStop}%
\bibitem [{\citenamefont {Scala}\ \emph {et~al.}(2013)\citenamefont {Scala},
  \citenamefont {Kim}, \citenamefont {Morley}, \citenamefont {Barker},\ and\
  \citenamefont {Bose}}]{scala2013matterwave}%
  \BibitemOpen
  \bibfield  {author} {\bibinfo {author} {\bibfnamefont {M.}~\bibnamefont
  {Scala}}, \bibinfo {author} {\bibfnamefont {M.~S.}\ \bibnamefont {Kim}},
  \bibinfo {author} {\bibfnamefont {G.~W.}\ \bibnamefont {Morley}}, \bibinfo
  {author} {\bibfnamefont {P.~F.}\ \bibnamefont {Barker}}, \ and\ \bibinfo
  {author} {\bibfnamefont {S.}~\bibnamefont {Bose}},\ }\bibfield  {title}
  {\enquote {\bibinfo {title} {Matter-wave interferometry of a levitated
  thermal nano-oscillator induced and probed by a spin},}\ }\href {\doibase
  10.1103/PhysRevLett.111.180403} {\bibfield  {journal} {\bibinfo  {journal}
  {Phys. Rev. Lett.}\ }\textbf {\bibinfo {volume} {111}},\ \bibinfo {pages}
  {180403} (\bibinfo {year} {2013})}\BibitemShut {NoStop}%
\bibitem [{\citenamefont {Yin}\ \emph {et~al.}(2013)\citenamefont {Yin},
  \citenamefont {Li}, \citenamefont {Zhang},\ and\ \citenamefont
  {Duan}}]{yin2013large}%
  \BibitemOpen
  \bibfield  {author} {\bibinfo {author} {\bibfnamefont {Zhang-qi}\
  \bibnamefont {Yin}}, \bibinfo {author} {\bibfnamefont {Tongcang}\
  \bibnamefont {Li}}, \bibinfo {author} {\bibfnamefont {Xiang}\ \bibnamefont
  {Zhang}}, \ and\ \bibinfo {author} {\bibfnamefont {L.~M.}\ \bibnamefont
  {Duan}},\ }\bibfield  {title} {\enquote {\bibinfo {title} {Large quantum
  superpositions of a levitated nanodiamond through spin-optomechanical
  coupling},}\ }\href@noop {} {\bibfield  {journal} {\bibinfo  {journal}
  {Physical Review A}\ }\textbf {\bibinfo {volume} {88}} (\bibinfo {year}
  {2013})}\BibitemShut {NoStop}%
\bibitem [{\citenamefont {Spiller}\ \emph {et~al.}(2006)\citenamefont
  {Spiller}, \citenamefont {Nemoto}, \citenamefont {Braunstein}, \citenamefont
  {Munro}, \citenamefont {Loock},\ and\ \citenamefont
  {Milburn}}]{spiller2006quantum}%
  \BibitemOpen
  \bibfield  {author} {\bibinfo {author} {\bibfnamefont {T~P}\ \bibnamefont
  {Spiller}}, \bibinfo {author} {\bibfnamefont {Kae}\ \bibnamefont {Nemoto}},
  \bibinfo {author} {\bibfnamefont {Samuel~L}\ \bibnamefont {Braunstein}},
  \bibinfo {author} {\bibfnamefont {W~J}\ \bibnamefont {Munro}}, \bibinfo
  {author} {\bibfnamefont {P~van}\ \bibnamefont {Loock}}, \ and\ \bibinfo
  {author} {\bibfnamefont {G~J}\ \bibnamefont {Milburn}},\ }\bibfield  {title}
  {\enquote {\bibinfo {title} {Quantum computation by communication},}\ }\href
  {\doibase 10.1088/1367-2630/8/2/030} {\bibfield  {journal} {\bibinfo
  {journal} {New Journal of Physics}\ }\textbf {\bibinfo {volume} {8}},\
  \bibinfo {pages} {30–30} (\bibinfo {year} {2006})}\BibitemShut {NoStop}%
\bibitem [{\citenamefont {Kumar}\ and\ \citenamefont
  {Bhattacharya}(2017)}]{kumar2017magnetometrty}%
  \BibitemOpen
  \bibfield  {author} {\bibinfo {author} {\bibfnamefont {Pardeep}\ \bibnamefont
  {Kumar}}\ and\ \bibinfo {author} {\bibfnamefont {M.}~\bibnamefont
  {Bhattacharya}},\ }\bibfield  {title} {\enquote {\bibinfo {title}
  {Magnetometry via spin-mechanical coupling in levitated optomechanics},}\
  }\href {\doibase 10.1364/OE.25.019568} {\bibfield  {journal} {\bibinfo
  {journal} {Opt. Express}\ }\textbf {\bibinfo {volume} {25}},\ \bibinfo
  {pages} {19568--19582} (\bibinfo {year} {2017})}\BibitemShut {NoStop}%
\bibitem [{\citenamefont {Tufarelli}\ \emph {et~al.}(2012)\citenamefont
  {Tufarelli}, \citenamefont {Ferraro}, \citenamefont {Kim},\ and\
  \citenamefont {Bose}}]{tufarelli2012reconstructing}%
  \BibitemOpen
  \bibfield  {author} {\bibinfo {author} {\bibfnamefont {Tommaso}\ \bibnamefont
  {Tufarelli}}, \bibinfo {author} {\bibfnamefont {Alessandro}\ \bibnamefont
  {Ferraro}}, \bibinfo {author} {\bibfnamefont {M.~S.}\ \bibnamefont {Kim}}, \
  and\ \bibinfo {author} {\bibfnamefont {Sougato}\ \bibnamefont {Bose}},\
  }\bibfield  {title} {\enquote {\bibinfo {title} {Reconstructing the quantum
  state of oscillator networks with a single qubit},}\ }\href {\doibase
  10.1103/PhysRevA.85.032334} {\bibfield  {journal} {\bibinfo  {journal} {Phys.
  Rev. A}\ }\textbf {\bibinfo {volume} {85}},\ \bibinfo {pages} {032334}
  (\bibinfo {year} {2012})}\BibitemShut {NoStop}%
\bibitem [{\citenamefont {Montenegro}\ \emph {et~al.}(2014)\citenamefont
  {Montenegro}, \citenamefont {Ferraro},\ and\ \citenamefont
  {Bose}}]{montenegro2014nonlinearity}%
  \BibitemOpen
  \bibfield  {author} {\bibinfo {author} {\bibfnamefont {V\'{\i}ctor}\
  \bibnamefont {Montenegro}}, \bibinfo {author} {\bibfnamefont {Alessandro}\
  \bibnamefont {Ferraro}}, \ and\ \bibinfo {author} {\bibfnamefont {Sougato}\
  \bibnamefont {Bose}},\ }\bibfield  {title} {\enquote {\bibinfo {title}
  {Nonlinearity-induced entanglement stability in a qubit-oscillator system},}\
  }\href {\doibase 10.1103/PhysRevA.90.013829} {\bibfield  {journal} {\bibinfo
  {journal} {Phys. Rev. A}\ }\textbf {\bibinfo {volume} {90}},\ \bibinfo
  {pages} {013829} (\bibinfo {year} {2014})}\BibitemShut {NoStop}%
\bibitem [{\citenamefont {Mukhopadhyay}\ \emph
  {et~al.}(2025{\natexlab{b}})\citenamefont {Mukhopadhyay}, \citenamefont
  {Bayat}, \citenamefont {Montenegro},\ and\ \citenamefont
  {Paris}}]{mukhopadhyay2025beatingjointquantumestimation}%
  \BibitemOpen
  \bibfield  {author} {\bibinfo {author} {\bibfnamefont {Chiranjib}\
  \bibnamefont {Mukhopadhyay}}, \bibinfo {author} {\bibfnamefont {Abolfazl}\
  \bibnamefont {Bayat}}, \bibinfo {author} {\bibfnamefont {Victor}\
  \bibnamefont {Montenegro}}, \ and\ \bibinfo {author} {\bibfnamefont {Matteo
  G.~A.}\ \bibnamefont {Paris}},\ }\href {https://arxiv.org/abs/2506.06075}
  {\enquote {\bibinfo {title} {Beating joint quantum estimation limits with
  stepwise multiparameter metrology},}\ } (\bibinfo {year}
  {2025}{\natexlab{b}}),\ \Eprint {http://arxiv.org/abs/2506.06075}
  {arXiv:2506.06075 [quant-ph]} \BibitemShut {NoStop}%
\bibitem [{\citenamefont {Rubio}\ and\ \citenamefont
  {Dunningham}(2020)}]{rubio2020bayesian}%
  \BibitemOpen
  \bibfield  {author} {\bibinfo {author} {\bibfnamefont {Jes\'us}\ \bibnamefont
  {Rubio}}\ and\ \bibinfo {author} {\bibfnamefont {Jacob}\ \bibnamefont
  {Dunningham}},\ }\bibfield  {title} {\enquote {\bibinfo {title} {Bayesian
  multiparameter quantum metrology with limited data},}\ }\href {\doibase
  10.1103/PhysRevA.101.032114} {\bibfield  {journal} {\bibinfo  {journal}
  {Phys. Rev. A}\ }\textbf {\bibinfo {volume} {101}},\ \bibinfo {pages}
  {032114} (\bibinfo {year} {2020})}\BibitemShut {NoStop}%
\bibitem [{\citenamefont {Rubio}\ and\ \citenamefont
  {Dunningham}(2019)}]{rubio2019quantummetrology}%
  \BibitemOpen
  \bibfield  {author} {\bibinfo {author} {\bibfnamefont {Jesús}\ \bibnamefont
  {Rubio}}\ and\ \bibinfo {author} {\bibfnamefont {Jacob}\ \bibnamefont
  {Dunningham}},\ }\bibfield  {title} {\enquote {\bibinfo {title} {Quantum
  metrology in the presence of limited data},}\ }\href {\doibase
  10.1088/1367-2630/ab098b} {\bibfield  {journal} {\bibinfo  {journal} {New
  Journal of Physics}\ }\textbf {\bibinfo {volume} {21}},\ \bibinfo {pages}
  {043037} (\bibinfo {year} {2019})}\BibitemShut {NoStop}%
\bibitem [{\citenamefont {Montenegro}(2025)}]{montenegro2025heisenberg}%
  \BibitemOpen
  \bibfield  {author} {\bibinfo {author} {\bibfnamefont {Victor}\ \bibnamefont
  {Montenegro}},\ }\bibfield  {title} {\enquote {\bibinfo {title}
  {Heisenberg-limited spin-mechanical gravimetry},}\ }\href {\doibase
  10.1103/PhysRevResearch.7.013016} {\bibfield  {journal} {\bibinfo  {journal}
  {Physical Review Research}\ }\textbf {\bibinfo {volume} {7}},\ \bibinfo
  {pages} {013016} (\bibinfo {year} {2025})}\BibitemShut {NoStop}%
\bibitem [{\citenamefont {Wang}\ \emph {et~al.}(2025)\citenamefont {Wang},
  \citenamefont {Dasari},\ and\ \citenamefont {Wrachtrup}}]{wang2025remote}%
  \BibitemOpen
  \bibfield  {author} {\bibinfo {author} {\bibfnamefont {Yang}\ \bibnamefont
  {Wang}}, \bibinfo {author} {\bibfnamefont {Durga Bhaktavatsala~Rao}\
  \bibnamefont {Dasari}}, \ and\ \bibinfo {author} {\bibfnamefont {J{\"o}rg}\
  \bibnamefont {Wrachtrup}},\ }\bibfield  {title} {\enquote {\bibinfo {title}
  {Remote cooling of spin-ensembles through a spin-mechanical hybrid
  interface},}\ }\href {\doibase 10.1038/s41534-025-00968-4} {\bibfield
  {journal} {\bibinfo  {journal} {npj Quantum Information}\ }\textbf {\bibinfo
  {volume} {11}},\ \bibinfo {pages} {24} (\bibinfo {year} {2025})}\BibitemShut
  {NoStop}%
\bibitem [{\citenamefont {G{\"u}hne}\ and\ \citenamefont
  {T{\'o}th}(2009)}]{guhne2009entanglement}%
  \BibitemOpen
  \bibfield  {author} {\bibinfo {author} {\bibfnamefont {Otfried}\ \bibnamefont
  {G{\"u}hne}}\ and\ \bibinfo {author} {\bibfnamefont {G{\'e}za}\ \bibnamefont
  {T{\'o}th}},\ }\bibfield  {title} {\enquote {\bibinfo {title} {Entanglement
  detection},}\ }\href {\doibase 10.1016/j.physrep.2009.02.004} {\bibfield
  {journal} {\bibinfo  {journal} {Physics Reports}\ }\textbf {\bibinfo {volume}
  {474}},\ \bibinfo {pages} {1--75} (\bibinfo {year} {2009})}\BibitemShut
  {NoStop}%
\bibitem [{\citenamefont {Polino}\ \emph {et~al.}(2020)\citenamefont {Polino},
  \citenamefont {Valeri}, \citenamefont {Spagnolo},\ and\ \citenamefont
  {Sciarrino}}]{polino2020photonic}%
  \BibitemOpen
  \bibfield  {author} {\bibinfo {author} {\bibfnamefont {Emanuele}\
  \bibnamefont {Polino}}, \bibinfo {author} {\bibfnamefont {Mauro}\
  \bibnamefont {Valeri}}, \bibinfo {author} {\bibfnamefont {Nicolò}\
  \bibnamefont {Spagnolo}}, \ and\ \bibinfo {author} {\bibfnamefont {Fabio}\
  \bibnamefont {Sciarrino}},\ }\bibfield  {title} {\enquote {\bibinfo {title}
  {Photonic quantum metrology},}\ }\href {\doibase 10.1116/5.0007577}
  {\bibfield  {journal} {\bibinfo  {journal} {AVS Quantum Science}\ }\textbf
  {\bibinfo {volume} {2}},\ \bibinfo {pages} {024703} (\bibinfo {year}
  {2020})}\BibitemShut {NoStop}%
\bibitem [{\citenamefont {Guo}\ and\ \citenamefont
  {Zhang}(2002)}]{guo2002scheme}%
  \BibitemOpen
  \bibfield  {author} {\bibinfo {author} {\bibfnamefont {Guang-Can}\
  \bibnamefont {Guo}}\ and\ \bibinfo {author} {\bibfnamefont {Yong-Sheng}\
  \bibnamefont {Zhang}},\ }\bibfield  {title} {\enquote {\bibinfo {title}
  {Scheme for preparation of the w state via cavity quantum electrodynamics},}\
  }\href {\doibase 10.1103/PhysRevA.65.054302} {\bibfield  {journal} {\bibinfo
  {journal} {Phys. Rev. A}\ }\textbf {\bibinfo {volume} {65}},\ \bibinfo
  {pages} {054302} (\bibinfo {year} {2002})}\BibitemShut {NoStop}%
\bibitem [{\citenamefont {de~Moraes~Neto}\ \emph {et~al.}(2017)\citenamefont
  {de~Moraes~Neto}, \citenamefont {Teizen}, \citenamefont {Montenegro},\ and\
  \citenamefont {Vernek}}]{neto2017steady}%
  \BibitemOpen
  \bibfield  {author} {\bibinfo {author} {\bibfnamefont {G.~D.}\ \bibnamefont
  {de~Moraes~Neto}}, \bibinfo {author} {\bibfnamefont {V.~F.}\ \bibnamefont
  {Teizen}}, \bibinfo {author} {\bibfnamefont {V.}~\bibnamefont {Montenegro}},
  \ and\ \bibinfo {author} {\bibfnamefont {E.}~\bibnamefont {Vernek}},\
  }\bibfield  {title} {\enquote {\bibinfo {title} {Steady many-body
  entanglements in dissipative systems},}\ }\href {\doibase
  10.1103/PhysRevA.96.062313} {\bibfield  {journal} {\bibinfo  {journal} {Phys.
  Rev. A}\ }\textbf {\bibinfo {volume} {96}},\ \bibinfo {pages} {062313}
  (\bibinfo {year} {2017})}\BibitemShut {NoStop}%
\bibitem [{\citenamefont {hossein Mehrinezhad~Chobari}\ \emph
  {et~al.}(2025)\citenamefont {hossein Mehrinezhad~Chobari}, \citenamefont
  {Aghababa},\ and\ \citenamefont
  {Kolahdouz}}]{chobari2025generationteleportationparticlew}%
  \BibitemOpen
  \bibfield  {author} {\bibinfo {author} {\bibfnamefont {Seyed~Amir}\
  \bibnamefont {hossein Mehrinezhad~Chobari}}, \bibinfo {author} {\bibfnamefont
  {Hossein}\ \bibnamefont {Aghababa}}, \ and\ \bibinfo {author} {\bibfnamefont
  {Mohammadreza}\ \bibnamefont {Kolahdouz}},\ }\href
  {https://arxiv.org/abs/2501.18743} {\enquote {\bibinfo {title} {Generation
  and teleportation of three and four particle w state},}\ } (\bibinfo {year}
  {2025}),\ \Eprint {http://arxiv.org/abs/2501.18743} {arXiv:2501.18743
  [quant-ph]} \BibitemShut {NoStop}%
\bibitem [{\citenamefont {Park}\ \emph {et~al.}(2025)\citenamefont {Park},
  \citenamefont {Hofmann}, \citenamefont {Okamoto},\ and\ \citenamefont
  {Takeuchi}}]{park2025entangled}%
  \BibitemOpen
  \bibfield  {author} {\bibinfo {author} {\bibfnamefont {Geobae}\ \bibnamefont
  {Park}}, \bibinfo {author} {\bibfnamefont {Holger~F.}\ \bibnamefont
  {Hofmann}}, \bibinfo {author} {\bibfnamefont {Ryo}\ \bibnamefont {Okamoto}},
  \ and\ \bibinfo {author} {\bibfnamefont {Shigeki}\ \bibnamefont {Takeuchi}},\
  }\bibfield  {title} {\enquote {\bibinfo {title} {Entangled measurement for w
  states},}\ }\href {\doibase 10.1126/sciadv.adx4180} {\bibfield  {journal}
  {\bibinfo  {journal} {Science Advances}\ }\textbf {\bibinfo {volume} {11}},\
  \bibinfo {pages} {eadx4180} (\bibinfo {year} {2025})}\BibitemShut {NoStop}%
\bibitem [{\citenamefont {Pu}\ \emph {et~al.}(2018)\citenamefont {Pu},
  \citenamefont {Wu}, \citenamefont {Jiang}, \citenamefont {Chang},
  \citenamefont {Li}, \citenamefont {Zhang},\ and\ \citenamefont
  {Duan}}]{pu2018experimental}%
  \BibitemOpen
  \bibfield  {author} {\bibinfo {author} {\bibfnamefont {Yunfei}\ \bibnamefont
  {Pu}}, \bibinfo {author} {\bibfnamefont {Yukai}\ \bibnamefont {Wu}}, \bibinfo
  {author} {\bibfnamefont {Nan}\ \bibnamefont {Jiang}}, \bibinfo {author}
  {\bibfnamefont {Wei}\ \bibnamefont {Chang}}, \bibinfo {author} {\bibfnamefont
  {Chang}\ \bibnamefont {Li}}, \bibinfo {author} {\bibfnamefont {Sheng}\
  \bibnamefont {Zhang}}, \ and\ \bibinfo {author} {\bibfnamefont {Luming}\
  \bibnamefont {Duan}},\ }\bibfield  {title} {\enquote {\bibinfo {title}
  {Experimental entanglement of 25 individually accessible atomic quantum
  interfaces},}\ }\href {\doibase 10.1126/sciadv.aar3931} {\bibfield  {journal}
  {\bibinfo  {journal} {Science Advances}\ }\textbf {\bibinfo {volume} {4}},\
  \bibinfo {pages} {eaar3931} (\bibinfo {year} {2018})}\BibitemShut {NoStop}%
\bibitem [{\citenamefont {Eibl}\ \emph {et~al.}(2004)\citenamefont {Eibl},
  \citenamefont {Kiesel}, \citenamefont {Bourennane}, \citenamefont
  {Kurtsiefer},\ and\ \citenamefont {Weinfurter}}]{eibl2004experimental}%
  \BibitemOpen
  \bibfield  {author} {\bibinfo {author} {\bibfnamefont {Manfred}\ \bibnamefont
  {Eibl}}, \bibinfo {author} {\bibfnamefont {Nikolai}\ \bibnamefont {Kiesel}},
  \bibinfo {author} {\bibfnamefont {Mohamed}\ \bibnamefont {Bourennane}},
  \bibinfo {author} {\bibfnamefont {Christian}\ \bibnamefont {Kurtsiefer}}, \
  and\ \bibinfo {author} {\bibfnamefont {Harald}\ \bibnamefont {Weinfurter}},\
  }\bibfield  {title} {\enquote {\bibinfo {title} {Experimental realization of
  a three-qubit entangled {W} state},}\ }\href {\doibase
  10.1103/PhysRevLett.92.077901} {\bibfield  {journal} {\bibinfo  {journal}
  {Physical Review Letters}\ }\textbf {\bibinfo {volume} {92}},\ \bibinfo
  {pages} {077901} (\bibinfo {year} {2004})}\BibitemShut {NoStop}%
\bibitem [{\citenamefont {H{\"a}ffner}\ \emph {et~al.}(2005)\citenamefont
  {H{\"a}ffner}, \citenamefont {H{\"a}nsel}, \citenamefont {Roos},
  \citenamefont {Benhelm}, \citenamefont {Chek-al kar}, \citenamefont
  {Chwalla}, \citenamefont {K{\"o}rber}, \citenamefont {Rapol}, \citenamefont
  {Riebe}, \citenamefont {Schmidt}, \citenamefont {Becher}, \citenamefont
  {G{\"u}hne}, \citenamefont {D{\"u}r},\ and\ \citenamefont
  {Blatt}}]{haefner2005scalable}%
  \BibitemOpen
  \bibfield  {author} {\bibinfo {author} {\bibfnamefont {H.}~\bibnamefont
  {H{\"a}ffner}}, \bibinfo {author} {\bibfnamefont {W.}~\bibnamefont
  {H{\"a}nsel}}, \bibinfo {author} {\bibfnamefont {C.~F.}\ \bibnamefont
  {Roos}}, \bibinfo {author} {\bibfnamefont {J.}~\bibnamefont {Benhelm}},
  \bibinfo {author} {\bibfnamefont {D.}~\bibnamefont {Chek-al kar}}, \bibinfo
  {author} {\bibfnamefont {M.}~\bibnamefont {Chwalla}}, \bibinfo {author}
  {\bibfnamefont {T.}~\bibnamefont {K{\"o}rber}}, \bibinfo {author}
  {\bibfnamefont {U.~D.}\ \bibnamefont {Rapol}}, \bibinfo {author}
  {\bibfnamefont {M.}~\bibnamefont {Riebe}}, \bibinfo {author} {\bibfnamefont
  {P.~O.}\ \bibnamefont {Schmidt}}, \bibinfo {author} {\bibfnamefont
  {C.}~\bibnamefont {Becher}}, \bibinfo {author} {\bibfnamefont
  {O.}~\bibnamefont {G{\"u}hne}}, \bibinfo {author} {\bibfnamefont
  {W.}~\bibnamefont {D{\"u}r}}, \ and\ \bibinfo {author} {\bibfnamefont
  {R.}~\bibnamefont {Blatt}},\ }\bibfield  {title} {\enquote {\bibinfo {title}
  {Scalable multiparticle entanglement of trapped ions},}\ }\href {\doibase
  10.1038/nature04279} {\bibfield  {journal} {\bibinfo  {journal} {Nature}\
  }\textbf {\bibinfo {volume} {438}},\ \bibinfo {pages} {643--646} (\bibinfo
  {year} {2005})}\BibitemShut {NoStop}%
\bibitem [{\citenamefont {Máttar}\ \emph {et~al.}(2017)\citenamefont
  {Máttar}, \citenamefont {Skrzypczyk}, \citenamefont {Aguilar}, \citenamefont
  {Nery}, \citenamefont {Ribeiro}, \citenamefont {Walborn},\ and\ \citenamefont
  {Cavalcanti}}]{mattar2017experimental}%
  \BibitemOpen
  \bibfield  {author} {\bibinfo {author} {\bibfnamefont {A}~\bibnamefont
  {Máttar}}, \bibinfo {author} {\bibfnamefont {P}~\bibnamefont {Skrzypczyk}},
  \bibinfo {author} {\bibfnamefont {G~H}\ \bibnamefont {Aguilar}}, \bibinfo
  {author} {\bibfnamefont {R~V}\ \bibnamefont {Nery}}, \bibinfo {author}
  {\bibfnamefont {P~H~Souto}\ \bibnamefont {Ribeiro}}, \bibinfo {author}
  {\bibfnamefont {S~P}\ \bibnamefont {Walborn}}, \ and\ \bibinfo {author}
  {\bibfnamefont {D}~\bibnamefont {Cavalcanti}},\ }\bibfield  {title} {\enquote
  {\bibinfo {title} {Experimental multipartite entanglement and randomness
  certification of the w state in the quantum steering scenario},}\ }\href
  {\doibase 10.1088/2058-9565/aa629b} {\bibfield  {journal} {\bibinfo
  {journal} {Quantum Science and Technology}\ }\textbf {\bibinfo {volume}
  {2}},\ \bibinfo {pages} {015011} (\bibinfo {year} {2017})}\BibitemShut
  {NoStop}%
\bibitem [{\citenamefont {Mitchell}\ \emph {et~al.}(2004)\citenamefont
  {Mitchell}, \citenamefont {Lundeen},\ and\ \citenamefont
  {Steinberg}}]{mitchell2004super}%
  \BibitemOpen
  \bibfield  {author} {\bibinfo {author} {\bibfnamefont {M.~W.}\ \bibnamefont
  {Mitchell}}, \bibinfo {author} {\bibfnamefont {J.~S.}\ \bibnamefont
  {Lundeen}}, \ and\ \bibinfo {author} {\bibfnamefont {A.~M.}\ \bibnamefont
  {Steinberg}},\ }\bibfield  {title} {\enquote {\bibinfo {title}
  {Super-resolving phase measurements with a multiphoton entangled state},}\
  }\href {\doibase 10.1038/nature02493} {\bibfield  {journal} {\bibinfo
  {journal} {Nature}\ }\textbf {\bibinfo {volume} {429}},\ \bibinfo {pages}
  {161--164} (\bibinfo {year} {2004})}\BibitemShut {NoStop}%
\bibitem [{\citenamefont {Walther}\ \emph {et~al.}(2004)\citenamefont
  {Walther}, \citenamefont {Pan}, \citenamefont {Aspelmeyer}, \citenamefont
  {Ursin}, \citenamefont {Gasparoni},\ and\ \citenamefont
  {Zeilinger}}]{walther2004debroglie}%
  \BibitemOpen
  \bibfield  {author} {\bibinfo {author} {\bibfnamefont {P.}~\bibnamefont
  {Walther}}, \bibinfo {author} {\bibfnamefont {J.-W.}\ \bibnamefont {Pan}},
  \bibinfo {author} {\bibfnamefont {M.}~\bibnamefont {Aspelmeyer}}, \bibinfo
  {author} {\bibfnamefont {R.}~\bibnamefont {Ursin}}, \bibinfo {author}
  {\bibfnamefont {S.}~\bibnamefont {Gasparoni}}, \ and\ \bibinfo {author}
  {\bibfnamefont {A.}~\bibnamefont {Zeilinger}},\ }\bibfield  {title} {\enquote
  {\bibinfo {title} {De broglie wavelength of a non-local four-photon state},}\
  }\href {\doibase 10.1038/nature02552} {\bibfield  {journal} {\bibinfo
  {journal} {Nature}\ }\textbf {\bibinfo {volume} {429}},\ \bibinfo {pages}
  {158--161} (\bibinfo {year} {2004})}\BibitemShut {NoStop}%
\bibitem [{\citenamefont {Afek}\ \emph {et~al.}(2010)\citenamefont {Afek},
  \citenamefont {Ambar},\ and\ \citenamefont {Silberberg}}]{afek2010highnoon}%
  \BibitemOpen
  \bibfield  {author} {\bibinfo {author} {\bibfnamefont {Itai}\ \bibnamefont
  {Afek}}, \bibinfo {author} {\bibfnamefont {Oron}\ \bibnamefont {Ambar}}, \
  and\ \bibinfo {author} {\bibfnamefont {Yaron}\ \bibnamefont {Silberberg}},\
  }\bibfield  {title} {\enquote {\bibinfo {title} {High-noon states by mixing
  quantum and classical light},}\ }\href {\doibase 10.1126/science.1188172}
  {\bibfield  {journal} {\bibinfo  {journal} {Science}\ }\textbf {\bibinfo
  {volume} {328}},\ \bibinfo {pages} {879--881} (\bibinfo {year}
  {2010})}\BibitemShut {NoStop}%
\bibitem [{\citenamefont {Zhang}\ and\ \citenamefont
  {Chan}(2018)}]{zhang2018scalable}%
  \BibitemOpen
  \bibfield  {author} {\bibinfo {author} {\bibfnamefont {Lu}~\bibnamefont
  {Zhang}}\ and\ \bibinfo {author} {\bibfnamefont {Kam Wai~Clifford}\
  \bibnamefont {Chan}},\ }\bibfield  {title} {\enquote {\bibinfo {title}
  {Scalable generation of multi-mode noon states for quantum multiple-phase
  estimation},}\ }\href {\doibase 10.1038/s41598-018-29828-2} {\bibfield
  {journal} {\bibinfo  {journal} {Scientific Reports}\ }\textbf {\bibinfo
  {volume} {8}},\ \bibinfo {pages} {11440} (\bibinfo {year}
  {2018})}\BibitemShut {NoStop}%
\end{thebibliography}%

\clearpage
\onecolumngrid
\widetext

\begin{center}
\textbf{\large Appendix: Stroboscopic Saturation of Multiparameter Quantum Limits in Distributed Quantum Sensing}
\end{center}

\begin{center}
Berihu Teklu$^{1,2}$ and Victor Montenegro$^{1,3,4}$\\
\vspace{0.2cm}
\textit{$^1$\small{College of Computing and Mathematical Sciences, Department of Applied Mathematics and Sciences, Khalifa University of Science and Technology, 127788 Abu Dhabi, United Arab Emirates}}\\
\textit{$^2$\small{Center for Cyber-Physical Systems (C2PS), Khalifa University of Science and Technology, 127788, Abu Dhabi, United Arab Emirates.}}\\
\textit{$^3$\small{Institute of Fundamental and Frontier Sciences, University of Electronic Science and Technology of China, Chengdu 611731, China.}}\\
\textit{$^4$\small{Key Laboratory of Quantum Physics and Photonic Quantum Information, Ministry of Education, University of Electronic Science and Technology of China, Chengdu 611731, China.}}
\end{center}

\setcounter{equation}{0}
\setcounter{figure}{0}
\setcounter{table}{0}
\setcounter{page}{1}
\setcounter{section}{0}
\makeatletter
\renewcommand{\theequation}{S\arabic{equation}}
\renewcommand{\thefigure}{S\arabic{figure}}
\renewcommand{\thesection}{\Alph{section}}

\textbf{OUTLINE:}

\begin{enumerate}
    \item[A.] Analytical Derivation of the Networked Quantum Probe State.
    \item[B.] Linear Entropy of the Reduced Probe State Across the Network.
    \item[C.] Quantum Fisher Information Matrix for General Distributed Quantum-Enhanced Sensing.
    \item[D.] Weak Commutativity Condition: Saturation of Ultimate Multi-Parameter Bound.
    \item[E.] Single-Parameter Estimation: Optimal Measurement Basis via the Symmetric Logarithmic Derivative Eigenbasis.
\end{enumerate}

\section{Analytical Derivation of the Networked Quantum Probe State}\label{sec_sm_analytical}

This section clarifies how the general wavefunction of Eq.~\eqref{eq_quantum_probe} is obtained in the main text.

Recall the Hamiltonian:
\begin{equation}
\frac{\hat{H}_j}{\Omega\hbar} = \hat{b}_j^\dagger \hat{b}_j - (k_j \hat{\Lambda}_j - \mathcal{E}_j)(\hat{b}_j^\dagger + \hat{b}_j),
\end{equation}

with unitary operator
\begin{equation}
\hat{u}_j(\tau) = e^{i (k_j \hat{\Lambda}_j - \mathcal{E}_j)^2 (\tau - \sin\tau)}e^{(k_j \hat{\Lambda}_j - \mathcal{E}_j)(\eta(\tau)\hat{b}_j^\dagger - \eta^*(\tau)\hat{b}_j)} e^{-i \hat{b}_j^\dagger \hat{b}_j \tau},
\end{equation}
which can be written as
\begin{equation}
\hat{u}_j(\tau) = e^{i (k_j \hat{\Lambda}_j - \mathcal{E}_j)^2 (\tau - \sin\tau)} D_j[(k_j \hat{\Lambda}_j - \mathcal{E}_j)\eta(\tau)] e^{-i \hat{b}_j^\dagger \hat{b}_j \tau},
\end{equation}
where the displacement operator is defined as
\begin{equation}
D_j[z] \equiv e^{z \hat{b}_j^\dagger - z^* \hat{b}_j},
\end{equation}
with $\eta(\tau) \equiv 1 - e^{-i\tau}$ and $\tau \equiv \Omega t$. Since $[\hat{H}_j, \hat{H}_{j'}] = 0$ for all $1 \leq (j,j') \leq N$, the total unitary operator for the network is given by the product
\begin{equation}
\hat{U}(\tau) = \prod_{j=1}^N \hat{u}_j(\tau).
\end{equation}

We considered the initial state:
\begin{equation}
|\psi(0)\rangle = |\text{W}_N\rangle \big(|\alpha\rangle^{\otimes N}\big).
\end{equation}

For simplicity, and without loss of generality, let us consider the first non-trivial network, namely $N=3$. In this case:
\begin{equation}
|\psi(0)\rangle = |\text{W}_3\rangle\big(|\alpha\rangle^{\otimes 3}\big) = \frac{1}{\sqrt{3}}\Big(|\lambda, \lambda', \lambda'\rangle + |\lambda', \lambda, \lambda'\rangle + |\lambda', \lambda', \lambda\rangle\Big)|\alpha\rangle|\alpha\rangle|\alpha\rangle.
\end{equation}

The corresponding evolved wavefunction reads:
\begin{eqnarray}
\nonumber |\psi(\tau)\rangle &=& \Big(\frac{1}{\sqrt{3}}e^{i (k_1 \lambda - \mathcal{E}_1)^2 (\tau - \sin\tau)}D_1[(k_1 \lambda - \mathcal{E}_1)\eta(\tau)]D_1[\alpha e^{-i\tau}]\\
\nonumber &\times&e^{i (k_2 \lambda' - \mathcal{E}_2)^2 (\tau - \sin\tau)}D_2[(k_2 \lambda' - \mathcal{E}_2)\eta(\tau)]D_2[\alpha e^{-i\tau}]e^{i (k_3 \lambda' - \mathcal{E}_3)^2 (\tau - \sin\tau)}D_3[(k_3 \lambda' - \mathcal{E}_3)\eta(\tau)]D_3[\alpha e^{-i\tau}]|\lambda, \lambda',\lambda'\rangle|vac\rangle|vac\rangle|vac\rangle\Big)\\
\nonumber &+&\Big(\frac{1}{\sqrt{3}}e^{i (k_1 \lambda' - \mathcal{E}_1)^2 (\tau - \sin\tau)}D_1[(k_1 \lambda' - \mathcal{E}_1)\eta(\tau)]D_1[\alpha e^{-i\tau}]\\
\nonumber &\times&e^{i (k_2 \lambda - \mathcal{E}_2)^2 (\tau - \sin\tau)}D_2[(k_2 \lambda - \mathcal{E}_2)\eta(\tau)]D_2[\alpha e^{-i\tau}]e^{i (k_3 \lambda' - \mathcal{E}_3)^2 (\tau - \sin\tau)}D_3[(k_3 \lambda' - \mathcal{E}_3)\eta(\tau)]D_3[\alpha e^{-i\tau}]|\lambda', \lambda,\lambda'\rangle|vac\rangle|vac\rangle|vac\rangle\Big)\\
\nonumber &+&\Big(\frac{1}{\sqrt{3}}e^{i (k_1 \lambda' - \mathcal{E}_1)^2 (\tau - \sin\tau)}D_1[(k_1 \lambda' - \mathcal{E}_1)\eta(\tau)]D_1[\alpha e^{-i\tau}]\\
\nonumber &\times&e^{i (k_2 \lambda' - \mathcal{E}_2)^2 (\tau - \sin\tau)}D_2[(k_2 \lambda' - \mathcal{E}_2)\eta(\tau)]D_2[\alpha e^{-i\tau}]e^{i (k_3 \lambda - \mathcal{E}_3)^2 (\tau - \sin\tau)}D_3[(k_3 \lambda - \mathcal{E}_3)\eta(\tau)]D_3[\alpha e^{-i\tau}]|\lambda', \lambda',\lambda\rangle|vac\rangle|vac\rangle|vac\rangle\Big).\\ \label{eq_wavefunctionSM}
\end{eqnarray}

In the above, $|vac\rangle$ denotes the field vacuum state. When acted upon by the product of two displacement operators on the same mode $j$, one obtains
\begin{equation}
D_j[z_1]D_j[z_2]|vac\rangle = e^{\tfrac{1}{2}(z_1 z_2^* - z_1^* z_2)}|z_1+z_2\rangle,
\end{equation}
where $|z_1+z_2\rangle$ is the coherent state with complex amplitude $z_1+z_2$. Eq.~\eqref{eq_wavefunctionSM} simplifies to:
\begin{eqnarray}
\nonumber |\psi(\tau)\rangle &=& \Big(\frac{1}{\sqrt{3}}e^{i (k_1 \lambda - \mathcal{E}_1)^2 (\tau - \sin\tau)} e^{i (k_2 \lambda' - \mathcal{E}_2)^2 (\tau - \sin\tau)} e^{i (k_3 \lambda' - \mathcal{E}_3)^2 (\tau - \sin\tau)}e^{-i\alpha(\mathcal{E}_1-k_1\lambda)\sin\tau}e^{-i\alpha(\mathcal{E}_2-k_2\lambda')\sin\tau}e^{-i\alpha(\mathcal{E}_3-k_3\lambda')\sin\tau}\\
\nonumber &\times& |\lambda, \lambda',\lambda'\rangle|\alpha e^{-i\tau} + (k_1 \lambda - \mathcal{E}_1)\eta(\tau)\rangle|\alpha e^{-i\tau} + (k_2 \lambda' - \mathcal{E}_2)\eta(\tau)\rangle|\alpha e^{-i\tau} + (k_3 \lambda' - \mathcal{E}_3)\eta(\tau)\rangle\Big)\\
\nonumber &+&\Big(\frac{1}{\sqrt{3}}e^{i (k_1 \lambda' - \mathcal{E}_1)^2 (\tau - \sin\tau)} e^{i (k_2 \lambda - \mathcal{E}_2)^2 (\tau - \sin\tau)} e^{i (k_3 \lambda' - \mathcal{E}_3)^2 (\tau - \sin\tau)}e^{-i\alpha(\mathcal{E}_1-k_1\lambda')\sin\tau}e^{-i\alpha(\mathcal{E}_2-k_2\lambda)\sin\tau}e^{-i\alpha(\mathcal{E}_3-k_3\lambda')\sin\tau}\\
\nonumber &\times& |\lambda', \lambda,\lambda'\rangle|\alpha e^{-i\tau} + (k_1 \lambda' - \mathcal{E}_1)\eta(\tau)\rangle|\alpha e^{-i\tau} + (k_2 \lambda - \mathcal{E}_2)\eta(\tau)\rangle|\alpha e^{-i\tau} + (k_3 \lambda' - \mathcal{E}_3)\eta(\tau)\rangle\Big)\\
\nonumber &+&\Big(\frac{1}{\sqrt{3}}e^{i (k_1 \lambda' - \mathcal{E}_1)^2 (\tau - \sin\tau)} e^{i (k_2 \lambda' - \mathcal{E}_2)^2 (\tau - \sin\tau)} e^{i (k_3 \lambda - \mathcal{E}_3)^2 (\tau - \sin\tau)}e^{-i\alpha(\mathcal{E}_1-k_1\lambda')\sin\tau}e^{-i\alpha(\mathcal{E}_2-k_2\lambda')\sin\tau}e^{-i\alpha(\mathcal{E}_3-k_3\lambda)\sin\tau}\\
&\times& |\lambda', \lambda',\lambda\rangle|\alpha e^{-i\tau} + (k_1 \lambda' - \mathcal{E}_1)\eta(\tau)\rangle|\alpha e^{-i\tau} + (k_2 \lambda' - \mathcal{E}_2)\eta(\tau)\rangle|\alpha e^{-i\tau} + (k_3 \lambda - \mathcal{E}_3)\eta(\tau)\rangle\Big).
\end{eqnarray}

Without loss of generality, we fix the global phase at the first site $j=1$, so that the wavefunction can be further simplified as:
\begin{eqnarray}
\nonumber |\psi(\tau)\rangle &=& \frac{1}{\sqrt{3}}|\lambda, \lambda',\lambda'\rangle |\alpha_{\lambda}\rangle_1 |\alpha_{\lambda'}\rangle_2 |\alpha_{\lambda'}\rangle_3\\
\nonumber &+&\frac{1}{\sqrt{3}}e^{-i (\lambda-\lambda')[(-2 \mathcal{E}_1 k_1 + 2 \mathcal{E}_2 k_2 + (k_1 - k_2)(k_1 + k_2)(\lambda+\lambda')) t 
+(2 \mathcal{E}_1 k_1 + \alpha (k_1 - k_2) - 2 \mathcal{E}_2 k_2 - (k_1 - k_2)(k_1 + k_2)(\lambda+\lambda'))\sin t]}|\lambda', \lambda,\lambda'\rangle|\alpha_{\lambda'}\rangle_1|\alpha_{\lambda}\rangle_2 |\alpha_{\lambda'}\rangle_3\\
\nonumber &+&\frac{1}{\sqrt{3}}e^{-i (\lambda-\lambda')[(-2 \mathcal{E}_1 k_1 + 2 \mathcal{E}_3 k_3 + (k_1 - k_3)(k_1 + k_3)(\lambda+\lambda')) t 
+ (2 \mathcal{E}_1 k_1 + \alpha (k_1 - k_3) - 2 \mathcal{E}_3 k_3 - (k_1 - k_3)(k_1 + k_3)(\lambda+\lambda'))\sin t]}|\lambda', \lambda',\lambda\rangle|\alpha_{\lambda'}\rangle_1|\alpha_{\lambda'}\rangle_2 |\alpha_{\lambda}\rangle_3,\\
\label{eq_sm_any_time}
\end{eqnarray}

where we have defined the coherent amplitudes as:
\begin{equation}
    |\alpha_{\Xi}\rangle_j=|\alpha e^{-i\tau} + (k_j \Xi - \mathcal{E}_j)\eta(\tau)\rangle.
\end{equation}

We can further simplify the notation by writing
\begin{equation}
|\alpha_j\rangle = |\alpha_{\lambda'}\rangle_1 \cdots |\alpha_{\lambda}\rangle_j \cdots |\alpha_{\lambda'}\rangle_N,
\end{equation}
which represents the tensor product of $N$ coherent states, with the $\lambda$-dependent displacement appearing at site $j$---all the remaining states are displaced states whose displacement explicitly depends on $\lambda'$. This is the notation used throughout the main text.

While Eq.~\eqref{eq_sm_any_time} gives the wavefunction at arbitrary times, in our analysis we focus primarily on the dynamics at $\tau = 2\pi$ (see main text). At this specific time, Eq.~\eqref{eq_sm_any_time} reduces to:
\begin{eqnarray}
\nonumber |\psi(\tau=2\pi)\rangle &=& \frac{1}{\sqrt{3}}\Big(|1\rangle+e^{-2\pi i (\lambda-\lambda')[-2 \mathcal{E}_1 k_1 + 2 \mathcal{E}_2 k_2 + (k_1 - k_2)(k_1 + k_2)(\lambda+\lambda')]}|2\rangle+e^{-2\pi i (\lambda-\lambda')[-2 \mathcal{E}_1 k_1 + 2 \mathcal{E}_3 k_3 + (k_1 - k_3)(k_1 + k_3)(\lambda+\lambda')]}|3\rangle\Big)|\alpha\rangle_1|\alpha\rangle_2|\alpha\rangle_3,\\ \label{eq_sm_2pi}
\end{eqnarray}
where all mechanical states return to their initial state $(|\alpha\rangle)^{\otimes 3}$ and
\begin{eqnarray}
    |1\rangle &=& |\lambda\rangle_1 |\lambda'\rangle_2 |\lambda'\rangle_3,\\
    |2\rangle &=& |\lambda'\rangle_1 |\lambda\rangle_2 |\lambda'\rangle_3,\\
    |3\rangle &=& |\lambda'\rangle_1 |\lambda'\rangle_2 |\lambda\rangle_3,
\end{eqnarray}
or, in general:
\begin{equation}
|j\rangle = |\lambda'\rangle_1 |\lambda'\rangle_2 \ldots |\lambda\rangle_j \ldots |\lambda'\rangle_{N-1} |\lambda'\rangle_N.
\end{equation}

\textbf{Case 1:} In Eq.~\eqref{eq_sm_2pi}, if we set $\mathcal{E}_j \neq \mathcal{E}_{j'} \neq 0$ while keeping $k_j = k$ for all $j$, we immediately obtain:
\begin{eqnarray}
|\psi(\tau=2\pi)\rangle &=& \frac{1}{\sqrt{3}}\Big(|1\rangle+e^{4\pi i (\lambda - \lambda')k(\mathcal{E}_1 - \mathcal{E}_2)}|2\rangle+e^{4\pi i (\lambda-\lambda')k(\mathcal{E}_1-\mathcal{E}_3)}|3\rangle\Big)|\alpha\rangle_1|\alpha\rangle_2|\alpha\rangle_3\label{eq_sm_case1}
\end{eqnarray}

\textbf{Case 2:} In Eq.~\eqref{eq_sm_2pi}, if we set $\mathcal{E}_j = 0$ and $k_j \neq k_{j'} \neq 0$, we get:
\begin{eqnarray}
|\psi(\tau=2\pi)\rangle &=& \frac{1}{\sqrt{3}}\Big(|1\rangle+e^{-2\pi i (\lambda-\lambda')(k_1 - k_2)(k_1 + k_2)(\lambda+\lambda')}|2\rangle+e^{-2\pi i (\lambda-\lambda')(k_1 - k_3)(k_1 + k_3)(\lambda+\lambda')}|3\rangle\Big)|\alpha\rangle_1|\alpha\rangle_2|\alpha\rangle_3.\label{eq_sm_case2}
\end{eqnarray}

Eqs.~\eqref{eq_sm_case1}–\eqref{eq_sm_case2} correspond to the cases analyzed in detail in the main text and match the scenarios presented in Table~\ref{tab_phases_cases}.

From the above expressions, it is clear we can ignore the coherent amplitudes and focus in the $\{|j\rangle\}$ subsystems. For completeness we will compute the QFIM for Eq.~\eqref{eq_sm_case1}. Using Eq.~\eqref{eq_QFI_multi_pure}, the QFIM reads as:
\begin{equation}
 Q=\frac{64\pi^2(\lambda-\lambda')^2 k^2}{9}
\begin{pmatrix}2 & -1\\[4pt]-1 & 2\end{pmatrix}, 
\end{equation}
with the trace of the inverse as:
\begin{equation}
    \mathrm{Tr}[Q^{-1}] = \frac{3}{[4\pi(\lambda-\lambda')k]^2}.
\end{equation}

\section{Linear Entropy of the Reduced Probe State Across the Network}\label{sec_sm_entanglement}

Recall the general pure state shown in Eq.~\eqref{eq_wavefunction_arbitrary_time}:
\begin{equation}
    |\psi(\tau)\rangle = \frac{1}{\sqrt{N}}\left(|1\rangle|\alpha_1(\tau)\rangle+\sum_{j=2}^N e^{ \xi_j(\tau)}|j\rangle|\alpha_j(\tau)\rangle\right),
\end{equation}
where
\begin{equation}
    \xi_j(\tau)=\frac{1}{2}e^{-i\tau}(\lambda-\lambda')[\alpha (1-e^{2 i \tau})(k_1-k_j)-2[2\mathcal{E}_1 k_1-2\mathcal{E}_j k_j-(k_1-k_j)(k_1+k_j)(\lambda+\lambda')](\tau-\sin\tau)(\sin\tau-i\cos\tau)].
\end{equation}

We aim to evaluate the linear entropy of the reduced probe state, defined as
\begin{equation}
S_L = 1 - \mathrm{Tr}\left[\rho_\mathrm{probe}^2\right],
\end{equation}
where $\rho_\mathrm{probe}$ is the reduced density matrix obtained by tracing out the mechanical (field) degrees of freedom from the full pure state, i.e.,
\begin{equation}
\rho_\mathrm{probe} = \mathrm{Tr}_\mathrm{fields} \left[ |\psi(\tau)\rangle \langle \psi(\tau)| \right].
\end{equation}

To compute $S_L$, we require the trace of the product of partial states involving overlaps of mechanical coherent states. Specifically, for any two coherent states $|\Upsilon_1\rangle$ and $|\Upsilon_2\rangle$, the trace over the mechanical subsystem reduces to the inner product:
\begin{equation}
\mathrm{Tr}_\mathrm{fields} \left( |\Upsilon_1\rangle \langle \Upsilon_2| \right) = \langle \Upsilon_2 | \Upsilon_1 \rangle = \exp\left( -\frac{1}{2}|\Upsilon_1|^2 - \frac{1}{2}|\Upsilon_2|^2 + \Upsilon_2^* \Upsilon_1 \right).
\end{equation}

Below we provide explicit analytical expressions for the linear entropy for small network sizes $N = 2, 3, 4, 5$.

For $N = 2$:
\begin{equation}
\mathcal{S}_L^{(2)} = \frac{1}{2} \left[1 - \exp\left(2 (k_1^2 + k_2^2) (\lambda-\lambda')^2 \left[\cos \tau - 1\right]\right)\right].
\end{equation}

For $N = 3$:
\begin{multline}
\mathcal{S}_L^{(3)} = \frac{2}{9} \Big[3 - \exp\left(2 (k_1^2 + k_2^2) (\lambda-\lambda')^2 \left[\cos \tau - 1\right]\right)
- \exp\left(2 (k_1^2 + k_3^2) (\lambda-\lambda')^2 \left[\cos \tau - 1\right]\right)
- \\
\exp\left(2 (k_2^2 + k_3^2) (\lambda-\lambda')^2 \left[\cos \tau - 1\right]\right)
\Big].
\end{multline}

For $N = 4$:
\begin{multline}
\mathcal{S}_L^{(4)} = \frac{1}{8} \Big[6 - \exp\left(2 (k_1^2 + k_2^2) (\lambda-\lambda')^2 \left[\cos \tau - 1\right]\right) - \exp\left(2 (k_1^2 + k_3^2) (\lambda-\lambda')^2 \left[\cos \tau - 1\right]\right) \\
- \exp\left(2 (k_2^2 + k_3^2) (\lambda-\lambda')^2 \left[\cos \tau - 1\right]\right)
- \exp\left(2 (k_1^2 + k_4^2) (\lambda-\lambda')^2 \left[\cos \tau - 1\right]\right) \\
- \exp\left(2 (k_2^2 + k_4^2) (\lambda-\lambda')^2 \left[\cos \tau - 1\right]\right) - \exp\left(2 (k_3^2 + k_4^2) (\lambda-\lambda')^2 \left[\cos \tau -1\right]\right)\Big].
\end{multline}

For $N = 5$:
\begin{multline}
\mathcal{S}_L^{(5)} = \frac{2}{25} \Big[
10 - \exp\left(2 (k_1^2 + k_2^2) (\lambda-\lambda')^2 \left[\cos \tau - 1\right]\right)
- \exp\left(2 (k_1^2 + k_3^2) (\lambda-\lambda')^2 \left[\cos \tau - 1\right]\right) \\
- \exp\left(2 (k_2^2 + k_3^2) (\lambda-\lambda')^2 \left[\cos \tau - 1\right]\right)
- \exp\left(2 (k_1^2 + k_4^2) (\lambda-\lambda')^2 \left[\cos \tau - 1\right]\right) \\
- \exp\left(2 (k_2^2 + k_4^2) (\lambda-\lambda')^2 \left[\cos \tau - 1\right]\right)
- \exp\left(2 (k_3^2 + k_4^2) (\lambda-\lambda')^2 \left[\cos \tau - 1\right]\right) \\
- \exp\left(2 (k_1^2 + k_5^2) (\lambda-\lambda')^2 \left[\cos \tau - 1\right]\right)
- \exp\left(2 (k_2^2 + k_5^2) (\lambda-\lambda')^2 \left[\cos \tau - 1\right]\right) \\
- \exp\left(2 (k_3^2 + k_5^2) (\lambda-\lambda')^2 \left[\cos \tau - 1\right]\right)
- \exp\left(2 (k_4^2 + k_5^2) (\lambda-\lambda')^2 \left[\cos \tau - 1\right]\right)\Big].
\end{multline}

From the explicit expressions derived for small network sizes, the structure of the linear entropy reveals a consistent and systematic pattern. This allows for a straightforward generalization to an arbitrary number of nodes $N$
\begin{eqnarray}
    \mathcal{S}_L^{(N)} &=& 1 - \frac{1}{N^2} \left[
N + 2 \sum_{1 \leq i < j \leq N} \exp\left(2(k_i^2 + k_j^2)(\lambda-\lambda')^2 [\cos\tau -1] \right)
\right],\\
&=& \frac{2}{N^2}\left[\frac{N(N-1)}{2} - \sum_{1 \leq i < j \leq N} \exp\left(2(k_i^2 + k_j^2)(\lambda-\lambda')^2 [\cos\tau -1] \right)
\right],
\end{eqnarray}
which gives a closed-form expression valid for any network size.

\section{Quantum Fisher Information Matrix for General Distributed Quantum-Enhanced Sensing}\label{sec_sm_qfim}

Consider the general state at $\tau = 2\pi$ from Eq.~\eqref{eq_quantum_probe}, derived in the main text, as
\begin{equation}
|\psi\rangle=\frac{1}{\sqrt{N}}\Big(|1\rangle+\sum_{j=2}^N e^{i\beta_j\Phi_j}|j\rangle\Big),
\end{equation}
where we have omitted the disentangled mechanical states $|\alpha\rangle^{\otimes N}$ for simplicity. To compute the QFIM entries using Eq.~\eqref{eq_QFI_multi_pure}, we first evaluate the partial derivatives
\begin{equation}
|\partial_j\psi\rangle
=\frac{i\beta_j}{\sqrt{N}}e^{i\beta_j\Phi_j}|j\rangle,
\end{equation}
the inner products
\begin{equation}
\langle\partial_j\psi|\partial_{j'}\psi\rangle
=\frac{\beta_j^2}{N}\delta_{j,j'},
\end{equation}
and the overlaps
\begin{equation}
\langle\psi|\partial_j\psi\rangle
=\frac{1}{N} i\beta_j.  
\end{equation}

Therefore, the QFIM is given by
\begin{equation}
Q
=\frac{4}{N}\mathrm{diag}(\boldsymbol{\beta}^2)-\frac{4}{N^2}\boldsymbol{\beta}\boldsymbol{\beta}^T,\label{eq_sm_QFIM}
\end{equation}
or in terms of its matrix entries
\begin{equation}
[Q]_{ij} = \frac{4}{N} \delta_{ij} \beta_i^2 - \frac{4}{N^2} \beta_i \beta_j,
\end{equation}
where $\delta_{ij}$ is the Kronecker delta, $\boldsymbol{\beta}=(\beta_2,\dots,\beta_N)^{T}$ (of length $N-1$), and let $\mathrm{diag}(\boldsymbol{\beta}^2)$ denote the diagonal matrix with entries $\beta_j^2$.

Eq.~\eqref{eq_sm_QFIM} can be rewritten as:
\begin{equation}
Q = A - u v^T,    
\end{equation}
where
\begin{equation}
A \equiv \frac{4}{N}\mathrm{diag}(\boldsymbol{\beta}^2),\qquad
u \equiv \frac{2}{N}\boldsymbol{\beta},\qquad
v \equiv \frac{2}{N}\boldsymbol{\beta}.
\end{equation}

We now use the Sherman–Morrison identity to obtain the inverse of the QFIM in Eq.~\eqref{eq_sm_QFIM}. This identity reads as:
\begin{equation}
    \left(A - u v^T\right)^{-1}
= A^{-1} + \frac{A^{-1} u v^T A^{-1}}{1 - v^T A^{-1} u},
\end{equation}
which holds for invertible $A$, i.e. $\beta_j\neq 0$ for all $j$, and when the scalar $(1 - v^T A^{-1} u) \neq 0$, i.e. $v^T A^{-1} u \neq 1$. The above terms explicitly are:
\begin{eqnarray}
A^{-1} &=& \frac{N}{4}\mathrm{diag}\big(\boldsymbol{\beta}^{-2}\big),\\
v^T A^{-1} u
&=& \frac{1}{N}\sum_{j=2}^N \beta_j\beta_j^{-1}= \frac{N-1}{N} \neq 1 \hspace{0.2cm} \forall N,\\
A^{-1} u v^T A^{-1}
&=& \left(\tfrac{1}{2}\boldsymbol{\beta}^{-1}\right)\!\left(\tfrac{1}{2}\boldsymbol{\beta}^{-1}\right)^T.
\end{eqnarray}

Replacing the above terms into Sherman–Morrison formula, one gets:
\begin{equation}
Q^{-1} = \frac{N}{4}\Big[ \mathrm{diag}(\boldsymbol{\beta}^{-2}) + \boldsymbol{\beta}^{-1}(\boldsymbol{\beta}^{-1})^T\Big],
\end{equation}
or, similarly as before, in terms of its entry elements
\begin{equation}
[Q^{-1}]_{ij} = \frac{N}{4} \left( \frac{\delta_{ij}}{\beta_i^2} + \frac{1}{\beta_i \beta_j} \right),
\end{equation}
which is a particularly useful representation for single-parameter distributed sensing and nuisance analysis.

Note that, in order to justify the invertibility of $Q$, one can also easily compute the determinant needed for its inverse as $\det Q=(1-v^T A^{-1} u )\det A=\frac{1}{N}\det A > 0$ for all $N>0$ and whenever all $\beta_j \neq 0$. Hence, $Q$ is positive definite with a well-defined inverse. If any $\beta_j = 0$, $\det A = 0$ and thus $Q$ becomes singular.

As stated in the main text, we focus on the scalar multi-parameter quantum Cram\'{e}r–Rao bound, which requires the trace of the inverse of the QFIM. To compute this, we simply recall that $\mathrm{Tr}[\boldsymbol{\beta}^{-1}(\boldsymbol{\beta}^{-1})^T]=\sum_j \beta_j^{-2}$. Hence:
\begin{equation}
\mathrm{Tr}[Q^{-1}] = \frac{N}{2}\sum_{j=2}^N \beta_j^{-2}.
\end{equation}
The above expression reproduces the results for the trace of the inverse of the QFIM presented in the main text.

\section{Weak Commutativity Condition: Saturation of Ultimate Multi-Parameter Bound}\label{sec_sm_weak}

Unlike the single parameter case, in the multi-parameter scenario the quantum Cram\'{e}r–Rao bound cannot always be saturated. A sufficient condition for saturation is the strong commutativity of the SLDs, $[\mathcal{L}_k, \mathcal{L}_{k'}]=0$, which guarantees the existence of a common eigenbasis, but this is not a necessary condition. A more general and less restrictive criterion is given within the framework of quantum local asymptotic normality from which a necessary and sufficient condition for the saturation of the multi-parameter quantum Cram\'{e}r–Rao bound is the so-called weak commutativity condition,
\begin{equation}
\langle \psi| [\mathcal{L}_k, \mathcal{L}_{k'}] |\psi\rangle = 0,    
\end{equation}
which ensures the asymptotic attainability of the bound even when the SLDs do not commute strongly~\cite{carollo2019quantumness}.

From Eq.~\eqref{SM_eq_Lk_explicit}, it is straightforward to compute the product
\begin{equation}
    \mathcal{L}_k\mathcal{L}_{k'}=-\frac{4\beta_k\beta_{k'}}{N}\left[\frac{e^{i\beta_k \Phi_k}}{\sqrt N}|k\rangle\langle\psi|-e^{i(\beta_k \Phi_k-\beta_{k'} \Phi_{k'})}|k\rangle\langle k'|+\frac{e^{-i\beta_{k'} \Phi_{k'}}}{\sqrt N}|\psi\rangle\langle k'|\right],
\end{equation}
from which one can readily compute $\langle \psi| [\mathcal{L}_k,\mathcal{L}_{k'}] |\psi\rangle = 0$. This means that saturation of the multi-parameter Cram\'{e}r-Rao bound is, in principle, possible.

To illustrate that all phases $\Phi_j$ can be estimated simultaneously using a projective measurement that is independent of the unknown parameters, and that this measurement saturates the quantum Cram\'{e}r-Rao bound, we now explicitly analyze the first non-trivial case $N=3$. As shown in the main text, using Eq.~\eqref{eq_basis}, we can compute the probabilities:
\begin{eqnarray}
p_0 &=& \frac{1}{9} \left(3 + 2\cos[\beta_2(\Phi_2 - \vartheta_2)] + 2\cos[\beta_3(\Phi_3 - \vartheta_3)] + 2\cos[\beta_2 \Phi_2 - \beta_3 \Phi_3 - \beta_2 \vartheta_2 + \beta_3 \vartheta_3] \right), \\
p_1 &=& \frac{2}{3} \sin^2\left(\frac{1}{2} \beta_2 (\Phi_2 - \vartheta_2) \right), \\
p_2 &=& \frac{1}{9} \left(3 + \cos[\beta_2(\Phi_2 - \vartheta_2)] - 2\cos[\beta_3(\Phi_3 - \vartheta_3)] - 2\cos[\beta_2 \Phi_2 - \beta_3 \Phi_3 - \beta_2 \vartheta_2 + \beta_3 \vartheta_3] \right).
\end{eqnarray}
From which the entries of the classical Fisher information matrix can be built.

\section{Single-Parameter Estimation: Optimal Measurement Basis via the Symmetric Logarithmic Derivative Eigenbasis}\label{sec_sm_optimal_measurement_single}

The symmetric logarithmic derivative (SLD) operator for pure states is given by:
\begin{equation}
    \mathcal{L}_k = 2 \left( |\partial_k \psi\rangle \langle \psi| + |\psi\rangle \langle \partial_k \psi| \right). \label{SM_eq_Lk}
\end{equation}

We aim to determine the optimal measurement basis for a single-parameter estimation scenario. In this context, single-parameter refers to the estimation process where only one phase $\Phi_k$ is estimated, while all other relative phases $\{\Phi_j\}_{j\neq k}$ are assumed to be known. For us to achieve the optimal measurement that saturates the single-parameter Cram\'{e}r-Rao bound, that is $Q = \mathcal{F}$, we need to compute the eigenbasis of the SLD operator. Recall the state described in Eq.~\eqref{eq_quantum_probe} and its derivative with respect to $\Phi_k$ as.
\begin{equation}
    |\psi\rangle = \frac{1}{\sqrt{N}}\left(|1\rangle + \sum_{j=2}^Ne^{i \beta_j\Phi_j}|j\rangle\right), \hspace{2cm} |\partial_k \psi\rangle = \frac{1}{\sqrt{N}}i\beta_k e^{i \beta_k \Phi_k}|k\rangle.
\end{equation}

Therefore, the SLD operator can be calculated explicitly as follows:
\begin{equation}
    \mathcal{L}_k=\frac{2i\beta_k}{N}\left[e^{i\beta_k \Phi_k}|k\rangle\langle 1|+\sum_{j=2}^N e^{i(\beta_k \Phi_k-\beta_{j} \Phi_{j})}|k\rangle\langle j|-e^{-i\beta_k \Phi_k}|1\rangle\langle k|-\sum_{j=2}^N e^{i(\beta_{j} \Phi_{j}-\beta_k \Phi_k)}|j\rangle\langle k|\right]. \label{SM_eq_Lk_explicit}
\end{equation}

From Eq.~\eqref{SM_eq_Lk}, it is straightforward to see that $\mathcal{L}_k$ has support only on $\mathrm{span}\{|\psi\rangle,|k\rangle\}$. All remaining vectors that are orthogonal to both $|\psi\rangle$ and $|k\rangle$ have a zero eigenvalue, making them uninformative when constructing the corresponding probability distributions for classical Fisher information functions. Let us construct an orthonormal pair:
\begin{eqnarray}
|e_1\rangle &:=& |\psi\rangle,\\
|e_2\rangle &:=& \sqrt{\frac{N}{N-1}}\Big(e^{i\beta_k \Phi_k}|k\rangle-\frac{1}{\sqrt N}|\psi\rangle\Big).
\end{eqnarray}
By construction $\langle e_a|e_b\rangle=\delta_{ab}$ and $\langle\psi|e_2\rangle=0$. Using these definitions, the SLD operator $\mathcal{L}_k$ in its matrix form spanned in $\{|e_1\rangle,|e_2\rangle\}$ basis is
\begin{equation}
[L_k]_{e_1,e_2}=\frac{2\beta_k\sqrt{N-1}}{N}\begin{pmatrix}
0 & -i\\ i& 0
\end{pmatrix}.
\end{equation}

From the above 2$\times$2 SLD operator in matrix form one can straightforwardly find its non-zero eigenvalues
\begin{equation}
    \lambda_k^{\pm}=\pm\frac{2\beta_k\sqrt{N-1}}{N},
\end{equation}
with normalized eigenvectors
\begin{eqnarray}
    |v_k^{\pm}\rangle&=&\frac{1}{\sqrt2}\Big(|e_1\rangle\ \pm\ i\,|e_2\rangle\Big),\\
    &=&\frac{1}{\sqrt2}\left[\Big(1\mp \frac{i}{\sqrt{N-1}}\Big)|\psi\rangle \pm i \sqrt{\frac{N}{N-1}}e^{i\beta_k \Phi_k}|k\rangle\right].
\end{eqnarray}

Finally, the optimal single‑parameter measurement (for estimating $\Phi_k$) is the optimal projective measurement
\begin{equation}
\Pi_\pm^{(k)}=|v_k^{\pm}\rangle\langle v_k^{\pm}|. \label{SM_eq_POVM}
\end{equation}

Using the optimal measurement for the single-parameter scenario described in Eq.~\eqref{SM_eq_POVM}, we can easily show that the classical Fisher information is equal to the quantum Fisher information. This means that the optimal measurement achieves the highest possible precision limit, as expected. To demonstrate this, we assume a fixed value of $\tilde{\Phi}_k$ and consider the optimal projective measurement. Under these conditions, the classical Fisher information can be expressed as follows:
\begin{equation}
\mathcal{F}(\tilde{\Phi}_k)=\left[\frac{1}{p_+(\Phi_k)[1-p_+(\Phi_k)]}\left(\frac{\partial p_+(\Phi_k)}{\partial \Phi_k}\right)^2\right]\Bigg|_{\Phi_k = \tilde{\Phi}_k},\label{SM_eq_CFI}
\end{equation}
where
\begin{equation}
    p_+(\Phi_k) = \frac{1}{2} + \frac{\sqrt{N-1}}{N}\sin \left( \beta_k (\Phi_k - \tilde{\Phi}_k) \right).
\end{equation}
Hence:
\begin{equation}
\mathcal{F}(\tilde{\Phi}_k)=
\frac{4 \beta_k^2 (N - 1) \cos^2\left[\beta_k(\tilde{\Phi}_{k} - \Phi_k)\right]}{N^2 - 4 (N - 1) \sin^2\left[\beta_k(\tilde{\Phi}_{k} - \Phi_k)\right]}\Bigg|_{\Phi_k = \tilde{\Phi}_k}=\frac{4\beta_k^2(N-1)}{N^2}.
\end{equation}
The above expression exactly matches the quantum Fisher information for the single-parameter estimation scenario, see Eq.~\eqref{eq_single_nonuisance}.

\end{document}